\documentclass[10pt,twocolumn,rmp,aps,nofootinbib,preprint]{revtex4} %

\usepackage{graphics}
\usepackage{dcolumn}
\usepackage{epsfig}

\begin{document}
\title{The Hierarchy of Incompressible Fractional Quantum Hall States}

\author{John J. Quinn}
\email{jjquinn@utk.edu} \affiliation{Department of Physics and
Astronomy, University of Tennessee, Knoxville, TN 37996, USA}

\author{Arkadiusz W\'{o}js}
\affiliation{Wroclaw University of Technology, 50-370 Wroclaw,
Poland} \affiliation{Department of Physics and Astronomy, University
of Tennessee, Knoxville, Tennessee 37996, USA}

\author{Kyung-Soo Yi}
\affiliation{Department of Physics, Pusan National University, Busan
609-735, Korea}

\author{George Simion}
\affiliation{ Department of Physics and Astronomy, University of
Tennessee, Knoxville, TN 37996, USA}

\date{\today}

\begin{abstract}

The correlations that give rise to incompressible quantum liquid
(IQL) states in fractional quantum Hall systems are determined by
the pseudopotential $V(\mathcal R)$ describing the interaction of a
pair of Fermions in a degenerate Landau level (LL) as a function of
relative pair angular momentum $\mathcal R$. $V(\mathcal R)$ is
known for a number of different Fermion systems, e.g. electrons in
the lowest Landau level (LL0) or the first excited Landau level
(LL1), and for quasiparticles of Laughlin-Jain IQL states. Laughlin
correlations, the avoidance of pair states with the smallest values
of $\mathcal R$, occur only when $V(\mathcal R)$ satisfies certain
conditions. We show that Jain's composite Fermion (CF) picture is
valid only if the conditions necessary for Laughlin correlations are
satisfied, and we present a rigorous justification of the CF picture
without the need of introducing an ``irrelevant" mean field energy
scale. Electrons in LL1 and quasielectrons in IQL states (e.g. QEs
in CF LL1) do not necessarily support Laughlin correlations.
Numerical diagonalization studies for small systems of Fermions
(electrons in LL0 or in LL1, and QEs in CF LL1), with the use
appropriate pseudopotentials $V(\mathcal R)$, show clear evidence
for different types of correlations. The relation between LL
degeneracy $g=2\ell+1$ and number of Fermions $N$ at which IQL
states are found is known for a limited range of $N$ values.
However, no simple intuitive models that we have tried
satisfactorily describe all of the systems we have studied.
Successes and shortcomings of some simple models are discussed, and
suggestions for further investigation are made.
\end{abstract}

\maketitle

\tableofcontents

\section{Introduction}
\label{sec:intro}

Solid state theory has developed from the realization
\cite{Sommerfeld} that simple metals could be described in terms of
free electrons that obeyed the Pauli exclusion principle
\cite{PauliPrinciple}. Very early work on the effect of the periodic
potential of the solid on the single electron eigenstates
\cite{Bloch} led to the concept of energy bands and bandgaps
\cite{WignerSeitz}, and to understanding of why some solids were
metals while others were insulators, semiconductors or semimetals
\cite{Wilson}. The early decades of solid state physics were
dominated by this ``single particle" description of electronic
states.

In the middle of the last century scientists began to worry about
why this single particle picture worked so well, since the
interaction between particles was not so small. Landau
\cite{Landau,Landau2} proposed the Fermi liquid theory to describe
the effect of short range many body interaction in liquid
$\rm{He}^3$. The concept of quasiparticles (QPs), elementary
excitations that satisfied Fermi-Dirac statistics and included a
``self-energy" (due to interaction with the ground-state) and a weak
interaction with one another, became a critical new concept in solid
state theory. Silin \cite{Silin} made use of Landau's idea to study
the properties of a "metallic" electron liquid with long range
Coulomb interactions. In microscopic studies of the electron liquid
many-electrons interactions were treated via diagrammatic
perturbation theory. The starting point, however, was still the
single electron eigenstates and the Fermi distribution function.

The BCS theory of superconductivity \cite{BCSpaper} demonstrates
that perturbation theory was not always adequate, even when
interactions were weak. However, even in BCS theory the
noninteracting electron states served as the starting point for
introduction of novel correlation effects via a generalized mean
field approximation.

During the past two decades novel systems have been discovered in
which many-body interaction appears to dominate over single particle
energies. Transition metal oxides displaying a metal-insulator
transition, magnetic phase transitions and high temperature
superconductivity are one technologically important class of such
``strongly interacting systems". When interactions dominate, the
standard technique of treating them as a perturbation on the single
particle spectrum is usually not adequate.

The paradigm for such systems is the fractional quantum Hall (FQH)
system. At very high values of the applied magnetic field the
massively degenerate single particle Landau levels (LLs) disappear
from the problem. The low energy spectrum is determined by a single
energy scale $e^2/\lambda$, where $\lambda=(\hbar c/eB)^{1/2}$ is
the magnetic length. The incompressible quantum liquid (IQL) states
discovered by Tsui et al. \cite{TsuiPRL82} result from the
interaction alone.

In this paper we present a review of the families of FQH states
observed experimentally and of how we understand them. Although a
lot of theoretical methods have been developed, we would limit
ourselves to those that are critical to our explanations, leaving
out for example some work rooted in field theories
\cite{BalatskyFradkinPRB91,FradkinSchaposnikPRL91,LopezFradkinPRB91,
LopezFradkinPRL69,LopezFradkinPRB93,LopezFradkinPRB95} and
Hamiltonian method \cite{MurthyShankarRMP03,ShankarMurthyPRL97,
MurthyShankarPRB99,MurthyShankarPRB02}. Laughlin's remarkable
insight \cite{LaughlinPRL83} into the nature of correlations giving
rise to an IQL state and the fractionally charged excitations:
quasielectrons (QEs) and quasiholes (QHs) are discussed. We consider
Haldane's idea \cite{HaldanePRL83} that a hierarchy of IQL daughter
states can be attributed to the fact that putting fractionally
charge QPs into a QP Landau level is essentially the same problem as
that of putting the original electrons in an electron Landau level.
We review Jain's remarkable composite Fermion (CF) picture
\cite{JainPRL89}. It predicts not only the filling factor $\nu$ at
which the most prominent IQL states are observed, but structure of
the lowest band of energy states for any value of the applied
magnetic field $B$. We emphasize the conditions under which the CF
picture is valid and discuss why it's valid. We give examples in
which the CF picture is not valid. We suggest that a useful approach
to many Fermion systems dominated by the interaction between pairs
is to study the antisymmetric eigenstates of a single pair and to
use them to construct an appropriate product over all pairs. For the
simplest case, this is exactly the Laughlin wavefunction, a better
starting point for a many Fermion system than a Slater determinant
of single particle wavefunctions. We propose novel correlations,
different from Laughlin's, when the pair interactions are different
from the Coulomb interaction in the lowest Landau level (LL0).

Our objective is to give a deeper intuitive understanding of all FQH
states in the hope that it may suggest novel ways to treat
correlations in other strongly interacting systems.

\section{Integral Quantum Hall Effect}

The integral quantum Hall (IQH) effect was discovered by von
Klitzing et al. \cite{vonKlitzingPRL80} who investigated the
magnetotransport properties of a quasi two dimensional (2D) electron
gas in a silicon surface inversion layer.

The Hamiltonian describing the motion of a single electron confined
to the $x-y$ plane in the presence of a dc magnetic field $\vec B =B
\hat z$ is simply $H=(2\mu)^{-1}[\vec p +(e/c) \vec A(\vec r)]^2$.
The vector potential $\vec A (\vec r)$ in the {\it{symmetric}} gauge
is given by $\vec A(\vec r)=(1/2)B(-y\hat x+x \hat y)$. We use
$\hat{x},~\hat{y}~,$ and $\hat{z}$ as unit vectors along the
Cartesian axes. The Schr\"odinger equation $(H-E)\Psi(\vec r)=0$
\cite{LandauLifshitz} has eigenstates:
\begin{eqnarray}
\label{eq:psiLL}
\Psi_{nm}(r,\phi)&=&e^{im\phi}u_{nm}(r)\\
\label{eq:EnergyLL} E_{nm}&=&\frac{1}{2}\hbar \omega_c(2n+1+m+|m|)~.
\end{eqnarray}

The radial wavefunction $u_{nm}(r)$ can be written as
\begin{equation}
u_{nm}(r)=x^{|m|}\exp \left[-\frac{x^2}{2}\right]L^{|m|}_n(x^2)~,
\end{equation}
where $x^2=1/2(r/\lambda)^2$, and $L_n^{|m|}$ is an associated
Laguerre polynomial. $L_0^{|m|}$ is independent of $x$, and
$L_1^{|m|}$ is proportional to $(|m|+1-x^2)$. From Eq.
\ref{eq:EnergyLL} it is apparent that the spectrum of
single-particle energies consists of highly degenerate levels; the
lowest LL has $n=0$ and $m=0,-1,-2,\ldots $, and its wavefunction
can be written $\Psi_{0m}=z^{|m|}\exp [-|z|^2/4/\lambda^2]$, where
$z$ stands for $re^{-i\phi}$. For a finite-size sample of area
$\mathcal A$, the number of single-particle states in the lowest LL
is given by $N_{\phi}=B\mathcal A /\phi_0$, where $\phi_0=hc/e$ is
the quantum of flux. The filling factor $\nu$ is defined as
$N/N_{\phi}$, so that $\nu^{-1}$ is simply equal to the number of
flux quanta of the magnetic field per electron.

When $\nu$ is equal to an integer, there is an energy gap (equal to
$\hbar \omega_c$) between the filled states and the empty states.
This makes the noninteracting electron system incompressible,
because an infinitesimal decrease in the area $\mathcal A$ can be
accomplished only at the expense of promoting an electron across the
energy gap and into the first unoccupied LL. This incompressibility
is responsible for the integral quantum Hall effect
\cite{vonKlizingRMP}.

\begin{figure}
\includegraphics[width=\linewidth]{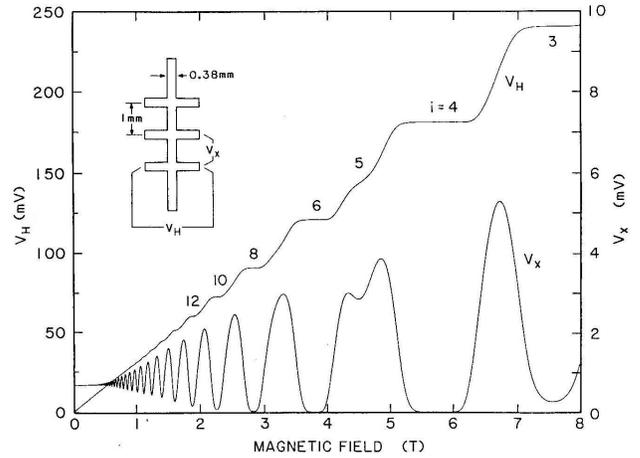}
\caption{$V_H$ and $V_x$ vs. B for a GaAs-AlGaAs heterostructure
cooled to 1.2 K. The source-drain current 25.5 $\rm{\mu A}$ and $n=
5.6 \times 10^{11}$ electrons/cm$^{2}$
\cite{Cage_book_chapter,IEEECage}.} \label{fig:IntegerQHE}
\end{figure}

In Fig. \ref{fig:IntegerQHE} we display typical results for $V_{X}$,
the voltage along the channel, and $V_H$, the Hall voltage. The
former contains zeros at the integral values of the filling factor
$\nu$ caused by the energy gap between the filled and empty LLs.
Both localized and extended states occur in the LLs. When the
chemical potential $\zeta$ resides in the localized states
$\sigma_{xx}$ vanishes (at T=0), and since localized states make no
contribution, the Hall conductivity $\sigma_{xy}$ remains constant
as $\zeta$ moves through the localized states. The integral value of
$\sigma_{xy}$ in units of $e^2/h$ is expected when $\nu$ is
precisely equal to an integer. The Hall plateaus depend on the
spectrum of the localized states which is related to the mobility of
the sample.

\section{Fractional Quantum Hall Effect}

\begin{figure}
\centerline{\includegraphics [width =1 \linewidth]{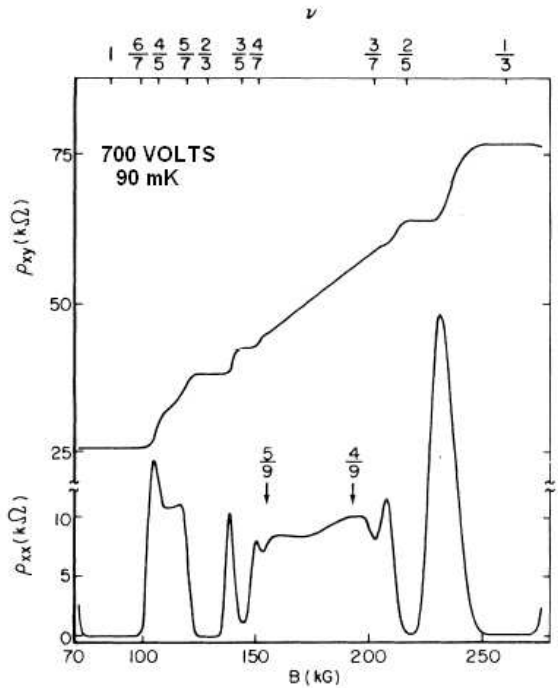}}
\caption{\label{Fig:FQHEgen} $\rho_{xx}$ and $\rho_{xy}$ at 90 mK,
for a sample which shows the fractional quantum Hall effect at
$\nu=1/3,~2/3,~2/5,~3/5,~3/7,~4/7,~4/9,$ and $5/9$
\cite{ChangPRL84}.}
\end{figure}

The observation of an incompressible quantum Hall liquid state in a
fractionally filled 2D Landau level by Tsui et al. \cite{TsuiPRL82}
was quite unexpected. The behavior of $\rho_{xx}$ and $\rho_{xy}$ as
a function of filling factor $\nu$ is displayed for a typical early
measurement in Fig. \ref{Fig:FQHEgen}. There are clear zeroes of
$\rho_{xx}$ at $\nu=1/3$ and $2/3$ and corresponding plateaus in
$\rho_{xy}$. At other fractions there are observable minima in
$\rho_{xx}$ and changes in slope in $\rho_{xy}$. The trace looks
like a continuation of Fig. \ref{fig:IntegerQHE} to higher magnetic
field or lower filling factor. Later, with significant improvement
of the quality of the sample, other filling fractions have been
observed in both lowest Landau level \cite{PanStormerTsuiPRL03}, and
higher LL \cite{EisensteinTsuiPRL87, PanStormerTsuiPRL99,
StormerTsuiPRL04,choiPRB08}. Unlike the IQH effect, the FQH effect
cannot be understood in terms of the single particle spectrum.
Coulomb correlations among electrons in the partially filled LL of
degenerate single particle states must be responsible for the
incompressibility (and the energy gap associated with it). Clearly,
this is a novel many-body state.

Laughlin \cite{LaughlinPRL83} correctly surmised that the FQH states
observed at filling factors $\nu=m^{-1}$, with $m$ being an odd
integer, resulted when the electrons were able to avoid pair states
with relative angular momentum smaller than $m$. These avoided pair
states have the smallest pair separation and the largest Coulomb
repulsion. Laughlin proposed a many-body wavefunction for the IQL
state at filling factor $\nu=m^{-1}$ given by
\begin{equation}
\label{eq:LaughlinPsi} \Psi_{m}(1,2,\ldots, N)=\prod\limits_{i<j}
z_{ij}^m \exp\left[-\frac{\sum_{k}|z_k|^2}{4 \lambda^2}\right]~.
\end{equation}
Here $z_i=r_ie^{-i\phi_i}$ is a complex coordinate for the position
of the $i^{\rm{th}}$ electron, $\lambda$ is the magnetic length and
$z_{ij}=z_i-z_j$. Clearly, in going from the filled $\nu=1$ state to
the $\nu=1/3$ state, the Laughlin wavefunction has introduced two
additional zeroes as a function of pair separation $|z_{ij}|$. The
relative pair angular momentum is simply $m$, the $z$-component of
the relative angular momentum of particles $i$ and $j$. Laughlin
also showed that the elementary excitations of the IQL state could
be described as fractionally charged QEs and QHs. Both localized and
extended states of the quasiparticles were required to understand
the observed behavior of $\rho_{xx}$ and $\rho_{xy}$.

The first explanation of FQH states at values of $\nu=n(1+2p)^{-1}$
with $n>1$ was given by Haldane \cite{HaldanePRL83}. He assumed that
the dominant interaction between quasiparticles was the short range
repulsive part of the pair interaction. Based on this assumption
Haldane suggested that the problem of filling the degenerate states
of the QP LL with $N_{\rm{QP}}$ Laughlin quasiparticles was similar
to that of filling the original $N_{\phi}$ states of the electron LL
with $N$ electrons. Because the number of QP states could not exceed
$N$, Haldane suggested the $N$ took place of $N_{\phi}$ and
$N_{\rm{QP}}$ the place of $N$ in the Laughlin's condition
$N_{\phi}=(2p+1)N$ for an IQL state. He proposed $N=2pN_{\rm{QP}}$
as the condition for new IQL states of the QPs. The even integer
$2p$ was chosen because, according to Haldane, the QPs were Bosons.
This ``Haldane hierarchy" of IQL states contained all odd
denominator fractions. Slightly different versions of Haldane's
hierarchy were independently suggested by Laughlin
\cite{LaughlinSurfSci84} and by Halperin
\cite{HalperinPRL84,HalperinHPA83}. The different versions differ in
the definition of the relative angular momentum of QPs, resulting in
different assignment of QP statistics. All of the versions depended
on the residual interactions between QPs (which were not well-known)
being sufficiently similar to the Coulomb interactions between
electrons in LL0.

\section{Numerical Diagonalization}
\label{sec:num_diag}

Confirmation of Laughlin's explanation of the correlations giving
rise to FQH states at $\nu=1/3,1/5,\ldots$ can be found through
numerical diagonalization of the Coulomb interaction within the
subspace of the lowest LL. Higher LLs play almost no role in the low
energy spectrum if the cyclotron energy $\hbar \omega_c$ is much
larger than the Coulomb energy $e^2/\lambda$. Exact numerical
diagonalization is limited to small systems, but it must give
qualitatively correct results as long as the correlation length is
much smaller than the radius of the system. Restricting the area of
the sample can be done in different ways, but probably the most
useful is to make the 2D surface on which the electrons reside a
sphere of radius $R$ \cite{HaldanePRL83,HaldaneRezayiPRB85}. In this
geometry a magnetic monopole of strength $2Q\phi_0$ (where $2Q$ is
an integer) at the center of sphere gives a radial magnetic field
$B=2Q\phi_0/4\pi R^2$. Boundary conditions are avoided and the
rotational invariance replaces the translational invariance of an
infinite plane.

The single particle eigenstates (called monopole harmonics
\cite{MonoploeHarmonics1,MonopoleHarmonics2}) are denoted by
$|Q,\ell,m>$, where the angular momentum $l$ and its $z$-component
$m$ must satisfy $|m|\leq l$. The single particle eigenvalues are
given by $E_l=(\hbar \omega_c/2Q)[\ell(\ell+1)-Q^2]$. Since $E_l$
cannot be negative, the minimum allowed value of $\ell$ must be $Q$.
We can write $\ell=Q+n$, with $n=0,1,\ldots$ playing the role of LL
index. For $\nu<1$ only the lowest LL (with $\ell=Q$) is relevant at
high magnetic fields. We can write $N$ electron basis states as:
$|m_1,m_2,\ldots,m_N>=c_{m_N}^{\dag}\ldots
c_{m_2}^{\dag}c_{m_1}^{\dag}|vac>$, where $|vac>$ is the vacuum
state and $c_m^{\dag}$ creates an electron in state $|Q,\ell,m>$
with $\ell=Q$. Of course the allowed values of $m$ must satisfy $|m|
\leq \ell$. Although the two body matrix elements of the Coulomb
interaction $<m_1,m_2|V|m_3,m_4>$ have a simple form in the lowest
Landau level \cite{FanoOrtolaniColomboPRB86}, the number of $N$
electron states $[N_{\phi}!/N!(N_{\phi}-N)!]$ grows rapidly with the
system size. In the lowest LL where $\ell=Q$ the $N$-electron states
can be written $|L,L_z,\alpha>$ with $L$ and $L_z$ being the total
angular momentum and its $z$ component, and $\alpha$ is an index
that distinguishes different multiplets  with the same value of $L$.
Because the Coulomb interaction Hamiltonian
$H=\sum_{i<j}V(|\vec{r}_i-\vec{r}_j|)$ is spherically symmetric, the
Wigner-Eckart theorem tells us that
$<L^{\prime},L^{\prime}_z,\alpha^{\prime}|H|L,L_z,\alpha>=\delta(L^{\prime},L)
\delta(L^{\prime}_z,L_z) V_{\alpha \alpha^{\prime}}(L)$, and the
reduced matrix element $V_{\alpha \alpha^{\prime}}$ is independent
of $L_z$. This fact can be used to reduce the size of matrices to be
diagonalized \cite{QuinnWojsPhysE98,QuinnVarenna}.

It is probably worth noting that on a plane \cite{QuinnWojsPhysE98}
the allowed values of $m$, the $z$-component of the single particle
angular momentum, are $0,1,\ldots, N_{\phi}-1$. $M=\sum_i m_i$ is
the total $z$-component of the angular momentum (the sum is over
occupied states). It can be divided into $M_{\rm{CM}}+M_{\rm{R}}$,
the center of mass and relative contributions. The connection
between the planar and spherical geometries is $M=N\ell+L_z$,
$M_R=N\ell-L$, and $M_{\rm{CM}}=L+L_z$. The interactions depend only
on $M_{\rm{R}}$ so $|M_{\rm{R}},M_{\rm{CM}}>$ acts just like
$|L,L_z>$. The absence of boundary conditions and the complete
rotational symmetry make the spherical geometry attractive to
theorists. Many experimentalists prefer using the
$|M_{\rm{R}},M_{\rm{CM}}>$ states of the planar geometry.

\begin{figure}
\centerline{\includegraphics [width =1 \linewidth]{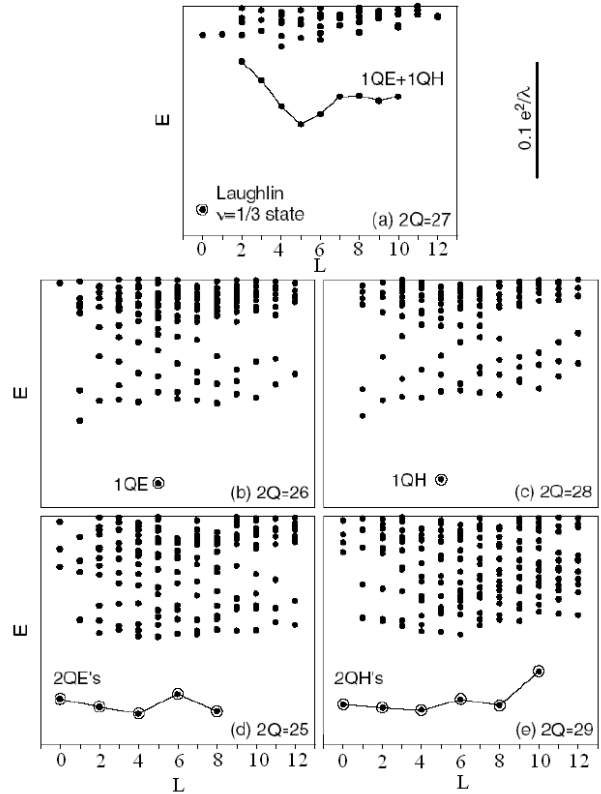}}
\caption{\label{fig:DiagExample} The spectra of 10 electrons in the
lowest Landau level calculated on a Haldane sphere with $2Q$ between
25 and 29. The open circles and solid lines mark the lowest energy
bands with the fewest composite Fermion quasiparticles
\cite{QuinnWoysJPhys00}.}
\end{figure}

Some exact diagonalization results ($E$ vs. $L$) for the ten
electron system are shown in Fig. \ref{fig:DiagExample}. The
Laughlin $L=0$ incompressible ground state occurs at $2Q=3(N-1)$ for
the $\nu=1/3$ state. States with larger values of $Q$ contain one or
two QHs ($2Q=28,29$), and states with smaller values of $Q$ contain
QEs in the ground states \cite{QuinnQuinnSSC06,QuinnVarenna,
QuinnWoysJPhys00}.

The energy of the multiplet $|L\alpha>$ can be expressed as
\begin{equation}
\label{eq:energyElalpha}
E_{\alpha}(L)=\left(
\begin{array} {c} N \\ 2\end{array}
\right) \sum\limits_{L^{\prime}}
P_{L\alpha}(L^{\prime})V(L^{\prime})~,
\end{equation}
where $P_{L\alpha}(L^{\prime})$ is the probability that $|L\alpha>$
contains pairs with pair angular momentum $L^{\prime}$, and
$V(L^{\prime})$ is the energy of interaction of a pair with angular
momentum $L^{\prime}=2\ell-\mathcal R$. Here $R=1,3,5,\ldots$ is
referred to as the relative pair angular momentum. We will sometimes
use the notation $V(\mathcal R)$ understanding this to mean
$V(2\ell-\mathcal R)$ i.e. the function $V(L^{\prime})$ with
$L^{\prime}$ expressed as $2\ell-\mathcal R$.

\begin{figure}
\centerline{\includegraphics
[width=1\linewidth]{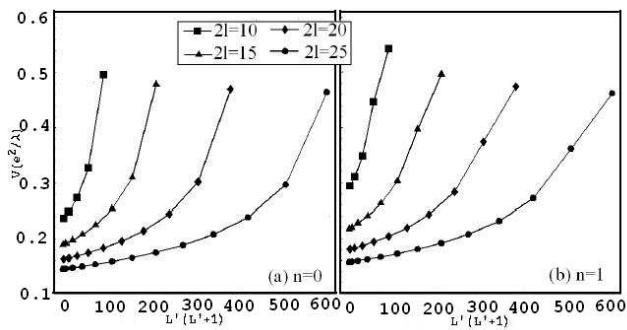}}
\caption{\label{Fig:Gen_Pseodpot} Pseudopotential $V(L ^{\prime})$
of the Coulomb interaction in the lowest (a) and the first excited
Landau level (b) as a function of squared pair angular momentum
$L^{\prime}(L^{\prime} + 1)$. Squares ($\ell = 5$), triangles ($\ell
= 15/2$), diamonds ($\ell = 10$), and circles ($\ell = 25/2$)
indicate data for different values of $Q = \ell +n$
\cite{QuinnVarenna}.}
\end{figure}

It is straightforward to evaluate the pseudopotential $V(\mathcal
R)$ describing the interaction of a pair of electrons in a shell of
angular momentum $l$ in the Haldane spherical geometry
\cite{FanoOrtolaniColomboPRB86}. It depends on the radius of the
sphere $R=(Q)^{1/2}\lambda$ and on the Landau level index
$n=\ell-Q=0,1,2,\ldots$. Simple results for $V^{(n)} (\mathcal R)$
are given in Fig. \ref{Fig:Gen_Pseodpot}.

\section{Chern-Simons Gauge Field and Jain's Composite Fermion Picture}

Jain \cite{JainPRL89} made the remarkable observation that the most
prominent IQL states observed experimentally could be understood in
terms of a simple composite Fermion (CF) picture. This picture made
use of a Chern--Simons (CS) transformation \cite{WilczekPRL82,
WilczekPRL82_1} and a CS gauge field familiar to field theorists.
The CS transformation can be described as attaching to the
$j^{\rm{th}}$ electron $(1\leq j \leq N)$ a flux tube carrying a
magnetic field $ \vec b =\alpha \phi_0 \delta (\vec r - \vec
r_j)\hat z$. Here $\phi_{0}=(hc)/e$ is the quantum of flux, $\alpha$
is a constant, and $\hat z$ a unit vector normal to the 2D layer. It
is well-known that when this CS flux is added via a gauge
transformation, no change in the classical equations of motion
results. Only the phase of the quantum mechanical wavefunction is
changed. However, the CS transformation results in a much more
complicated many-body Hamiltonian that includes a CS vector
potential $\vec a (\vec r)$ given by
\begin{equation}
\label{eq:gen_a_CS} \vec a (\vec r) = \alpha \phi_0 \int{\rm{d}^2
r_1 \frac{ \hat z \times (\vec r -\vec r_1)}{(\vec r- \vec
r_1)^{2}}\psi^{\dagger}(\vec r_1) \psi (\vec r_1)}~,
\end{equation}
in addition to the vector potential $\vec{A}(\vec{r})$ of the dc
magnetic field. Simplification results only when the mean field (MF)
approximation is made. This is accomplished by replacing the density
operator $\psi^{\dagger}(\vec r) \psi (\vec r) $ in the CS vector
potential and in the Coulomb interaction by its MF value $n_{s}$,
the uniform electron density. The resulting Hamiltonian is the sum
of single particle Hamiltonians in which an ``effective" magnetic
field $B^{\ast }=B-\nu \phi_0 n_s$ appears.

Jain introduced the idea of a CF to represent an electron with an
attached flux tube which carried an even number $\alpha (=2p)$ of
flux quanta \cite{JainPRB90}. In the MF approximation the CF filling
factor $\nu ^{\ast }$ is given by $\nu ^{\ast - 1}=\nu ^{ -
1}-\alpha $, i.e. the number of flux quanta per electron of the dc
field less the CS flux per electron. When $\nu ^{\ast }$ is equal to
an integer $n=\pm 1,\pm 2,\ldots$ , then $\nu =n(1+\alpha n)^{ - 1}$
generates (for \textit{$\alpha $}=2) quantum Hall states at $\nu
=1/3,2/5,3/7,\ldots $, and $\nu =1,2/3,3/5,\ldots $. These are the
most pronounced FQH states observed.

In the spherical geometry one can introduce an effective monopole
strength seen by one CF \cite{ChenQuinnSSC94}. It is given by
$2Q^{\ast }=2Q-\alpha (N-1)$ since the $\alpha $ flux quanta
attached to every other CF must be subtracted from the original
monopole strength 2$Q$. Then $|Q^{\ast}|=\ell_0^{\ast}$ plays the
role of the angular momentum of the lowest CF shell just as
$Q=\ell_{0}$ is the angular momentum of the lowest electron shell.
When $2Q$ is equal to an odd integer $(1+\alpha)$ times $(N-1)$, the
CF shell $\ell_0^{\ast}$ is completely filled, and an $L=0$
incompressible Laughlin state at filling factor $\nu=(1+\alpha)^{-
1}$ results. When $2|Q^{\ast }|+1$ is smaller (larger) than $N$, QEs
(QHs) appear in the shell
$\ell_{\rm{QE}}=\ell^{\ast}_0+1~(\ell_{\rm{QH}}=\ell_0^{\ast})$. The
low energy sector of the energy spectrum consists of the states with
the minimum number of QP excitations required by the value of
2$Q^{\ast }$ and $N$. The first excited band of states will contain
one additional QE-QH pair. The total angular momentum of these
states in the lowest energy sector can be predicted by addition of
the angular momenta ($\ell_{\rm{QH}}$ or $\ell_{\rm{QE}}$) of the
$n_{\rm{QH}}$ or $n_{\rm{QE}}$ quasiparticles treated as identical
Fermions. In Table \ref{tab:QE_QH_Laughlin_state} we demonstrated
how these allowed $L$ values are found for a ten electron system
with $2Q$ in the range $29 \geq 2Q \geq 25$. By comparing with
numerical results presented in Fig. \ref{fig:DiagExample}, we
readily observe that the total angular momentum multiplets appearing
in the lowest energy sector are always correctly predicted by this
simple MF CS picture
\cite{QuinnQuinnSSC06,QuinnVarenna,QuinnWoysJPhys00}.

\begin{table}[htbp]
\begin{tabular}
{|l|l|l|l|l|l|} \hline $2Q$& 29& 28& 27& 26&
25 \\
\hline $2Q^{\ast }$& 11& 10& 9& 8&
7 \\
\hline $n_{\rm{QH}}$& 2& 1& 0& 0&
0 \\
\hline $n_{\rm{QE}}$ & 0& 0& 0& 1& 2 \\
\hline $\ell_{\rm{QH}}$ & 5.5& 5& 4.5& 4&
3.5 \\
\hline $\ell_{\rm{QE}}$ & 6.5& 6& 5.5& 5&
4.5 \\
\hline $L$& 10,8,6,4,2,0& 5& 0& 5&
8,6,4,2,0 \\
\hline
\end{tabular}
\caption{\label{tab:QE_QH_Laughlin_state} The effective CF monopole
strength $2Q^{\ast }$, the number of CF quasiparticles (quasiholes -
$n_{\rm{QH}}$ and quasielectrons $n_{\rm{QE}}$), the quasiparticle
angular momenta and $\ell_{\rm{QH}}$, $\ell_{\rm{QE}}$ and the
angular momenta $L$ of the lowest lying band of multiplets for a ten
electron system at $2Q$ between 25 and 29.}
\end{table}

For example, the Laughlin $L=0$ ground state at $\nu =1/3$ occurs
when $2\ell_0^{\ast }=N-1$, so that the $N$ composite Fermions fill
the lowest shell (with angular momentum $\ell_0^{\ast}$). The CFQE
and CFQH states occur at $2 \ell_0^{\ast}=N-1\mp 1$ and have one too
many (for QE) or one too few (for QH) particles to give integral
filling. The single QPs have angular momentum $N/2$. The 2QE and 2QH
states occur at $2\ell_0^{\ast}=N-1\mp 2$. They have
$\ell_{\rm{QE}}=(N-1)/2$ and $\ell_{\rm{QH}}=(N+1)/2$. We expect,
for example, $\ell_{\rm{QE}}=4.5$ and $\ell_{\rm{QH}}=5.5$ for a ten
electron system, leading to low energy bands with $L=0 \oplus 2
\oplus 4\oplus 6\oplus 8$ for 2 QEs and to $L=0\oplus 2\oplus
4\oplus 6\oplus 8\oplus 10$ for 2 QHs. In the MF picture, which
neglects QP-QP interactions, these bands are degenerate. Of course,
numerical results in Fig. \ref{fig:DiagExample} show that two QP
states with different $L$ have different energy. From this numerical
data we obtain, up to an overall constant, $V_{\rm{QP}}$ the
residual interaction of a QP pair as a function of the pair angular
momentum $L^{\prime}$ \cite{QuinnQuinnSSC06,QuinnVarenna,
QuinnWoysJPhys00,WojsQuinnPRB00}.

In addition to the lowest energy band of multiplets, first excited
bands which contain one additional QE-QH pair can be observed in
Fig. \ref{fig:DiagExample}. The ``magnetoroton'' band lying between
the $L$=0 Laughlin IQL ground state and a continuum of higher energy
states can be observed in Fig. \ref{fig:DiagExample}(a). This band
contains one QH with $\ell_{\rm{QH}}=9/2$ and one QE with
$\ell_{\rm{QE}}=11/2$. By adding the angular momenta of these
distinguishable particles, a band with $1 = \ell_{\rm{QE}}-
\ell_{\rm{QH}} \leq L \leq  \ell_{\rm{QE}}+ \ell_{\rm{QH}} =10$
would be predicted. The state with $L=1$ does not appear in
Fig.\ref{fig:DiagExample} (a) suggesting that QE-QH pairs with $L=1$
are forbidden (or at least pushed into the higher energy continuum
by interactions). Furthermore, the states in the band are not
degenerate indicating residual interactions that depend on the
angular momentum of the pair. Other bands that are not quite so
clearly defined can be observed in other frames. For example, in
frame (b) between the single QE state at $L=5$ and the higher energy
continuum, there is a 2QE-1QH band. The allowed $L$ values can be
estimated by taking $\ell_{\rm{QE}}=5$ and $\ell_{\rm{QH}}=4$ and
adding angular momenta (treating the QEs as identical Fermions).
Interactions cause the predicted multiplets to overlap the bottom of
the continuum for $3 \leq L \leq 6$ but outside this range they are
separate from it \cite{QuinnQuinnSSC06,QuinnVarenna,
QuinnWoysJPhys00}.


\section{Beyond Mean Field}

\begin{figure}
\centerline{\includegraphics [width =1 \linewidth]
{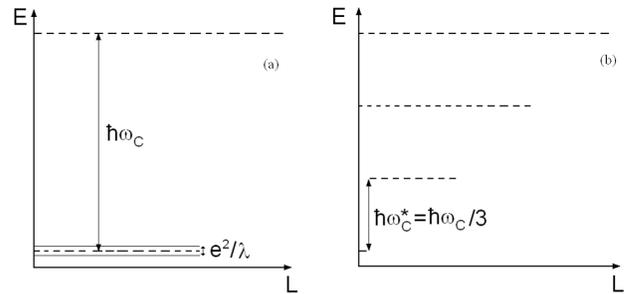}} \caption{Comparison of spectrum of $N$
noninteracting electrons (a) with that of $N$ noninteracting CFs (b)
the electron Landau levels are separated by $\hbar \omega_C$; the CF
levels $\hbar \omega_C^{\star}=\nu \hbar \omega_C$. For $\hbar
\omega_C \gg e^2/ \lambda$, the Coulomb energy scale, the degenerate
single electron levels are split by the Coulomb energy. This
splitting is much smaller than $\hbar \omega_C$ (or $\hbar
\omega_C^{\star}$). The higher electron LLs are not involved in
determining the interacting spectrum, so $\hbar \omega_C$ and $\hbar
\omega_C^{\ast}$, both proportional to $B$, are irrelevant.}
\label{fig:nint_CF}
\end{figure}

Despite the satisfactory description of the allowed angular momentum
multiplets, the magnitude of the MFCF energies is completely wrong.
The magnetoroton energy does not occur at the effective cyclotron
frequency $\hbar \omega_C^{\ast}=eB^{\star}/mc$. . This MF energy is
irrelevant at large values of $B$ (if we keep $m_{\rm{CF}}=m_{e}$),
so it is a puzzle why the CF picture does so well at predicting the
structure of the energy spectrum. It is interesting to compare the
energy spectrum of $N$ noninteracting electrons with that of $N$
noninteracting CFs as done in Fig. \ref{fig:nint_CF}.

For large values of $B$ the MF energy $\hbar \omega_C^{\ast}$ is
much larger than the Coulomb scale $e^{2}/\lambda $. Therefore the
low lying multiplets of interacting electrons will be contained in a
band of width $e^{2}$/\textit{$\lambda $} about the lowest electron
LL. The noninteracting CF spectrum contains a number of bands
separated by $\hbar \omega_C^{\ast}$. Interactions (Coulomb and CS
gauge interactions) among fluctuations beyond the MF essentially
have to restore the original noninteracting electron spectrum when
$B\rightarrow \infty$. Halperin et al.\cite{HalperinLeeReadPRB93}
and Lopez and Fradkin
\cite{LopezFradkinPRB91,LopezFradkinPRL69,LopezFradkinPRB93} have
used conventional manybody perturbation theory to treat
fluctuations. However, there is no small parameter to guarantee
convergence or to justify simple approximations like the random
phase approximation (RPA). The standard many-body perturbation
theory gives reasonable results probably because it can be thought
of as a Silin-Landau theory \cite{Landau,Silin} of an electron
liquid. Long range correlations are handled by RPA; short range
correlations by adding Landau Fermi liquid interactions
\cite{SimonHalperinPRB93}. What is clear is that the success of the
CF picture does not result from a cancelation between CS gauge
interactions and Coulomb interactions beyond MF.

Jain proposed a trial wavefunction which included a Jastrow factor
$\prod_{i<j}z_{ij}^2$, and he projected it onto the lowest Landau
level. He then diagonalized the Coulomb interaction using the
projected trial function \cite{JainPRB90}. Though the technique
seems to give reasonably good results, it is not obvious why it
works.


\section{Adiabatic Addition of CS Flux}

The CS magnetic field $\vec {b}(\vec r) =\alpha \phi_0 \sum_{j}
\delta(\vec r- \vec r_j) \hat z$ is usually introduced via a gauge
transformation. Then, it is a Bohm-Aharonov \cite{AharonovBohmPR59}
type field, having no effect on the classical equation of motion.
The Lorentz force on the $i^{\rm{th}}$ electron is given by
$(-e/c)\vec v_i \times \vec {b}(\vec r)$ with $\vec r=\vec r_i$. No
electron senses its own CS flux, and since $\vec r_i$ and $\vec r_j$
cannot have the same value for a pair of Fermions, there is no
effect from $\vec b (\vec r)$ on the classical motion of the
electrons. However, the CS flux does introduce a phase factor into
the quantum mechanical wavefunction.

Let's define $\psi_m(\vec r)=\exp (i m\phi) u_m(r)$ as the
wavefunction for the relative coordinate $\vec r=\vec r_i-\vec r_j$
of pair electrons in the lowest LL. For Fermions $m$, the
$z$-component of the relative angular momentum, must be odd so that
under exchange $(\phi \rightarrow \phi+\pi)$ the phase factor
changes sign. If a CS flux $\phi =\alpha \phi_0$ is introduced on
each electron via a gauge transformation, then $\phi_m \rightarrow
\exp [i(m-\alpha)\phi] u_m(r)$. The phase factor is changed by
$\exp(-i\alpha \pi)\phi_m$ under exchange. If $\alpha$ is not an
even integer this leads to the famous transmutation of statistics,
since $\phi_m \rightarrow \exp (-i\alpha \pi) \phi_m$ under exchange
\cite{LeinaasMyrheimNC77,Wilczekbook}.

A gauge transformation is not the only way by which CS flux can be
introduced. We can start with some initial state of the relative
coordinates of pair, e.g. one with $\alpha=0$, and adiabatically
increase the CS flux attached to each particle up to its final
value. In this case $\phi_m$ evolves adiabatically into $\exp(i m
\phi)u_{m+\alpha}(r)$. There is no change in phase (and therefore no
change in statistics for any value of $\alpha$). However, in the
semiclassical orbit (described by maximum in the density
$\rho(r)=|\phi_m|^2$) $u_m(r)$ is replaced by $u_{m+\alpha}(r)$. The
orbit size changes in such a way that the total flux (due to both
the applied field $B$ and the CS flux) is conserved. The change in
orbit size results from the Faraday emf acting on the relative
motion in the presence of perpendicular magnetic field $B$. If the
pair was initially in the smallest allowed pair orbit (with $m=-1$)
and two CS flux quanta opposite to $\vec B$ were added
($\alpha=-2$), then the resulting new orbit will have $m=-3$. This
is exactly what we mean by Laughlin correlations. The adiabatic
addition of CS flux has altered the orbit to avoid the most
repulsive pair state with $m=-1$. However, in the absence of Coulomb
interactions all negative values of $m$ belong to the lowest LL. No
change in energy occurs without Coulomb repulsion. No MF
approximation or MF energy scale is needed \cite{QuinnQuinnPRB03}.

If we write the pair wavefunction as a product of center of mass
(CM) and relative motion we find $\psi(\vec r_i, \vec
r_j)=\exp(im\phi_{ij})u_m (r_{ij})u_0(R_{ij})$ can be written as
$z_{ij}^{|m|}\exp[-(r_i^2+r_j^2)/(4\lambda ^2)]$. Here
$z_i=r_i\exp(-i\phi_i)$ is the complex coordinate of the $i^{th}$
particle, and $\lambda^2=\hbar c/eB=2 \lambda_{CM}^2=\lambda_r^2/2$.
For an $N$ electron system the straightforward generalization of
this pair function is the Laughlin wavefunction
$\Phi_m(1,2,\ldots,N)=\prod_{i<j}z_{ij}^{|m|}\exp[-\sum_kr_k^2/(4
\lambda^2)]$ where $|m|$ is an odd integer. Small values of $|m|$
correspond to small pair orbits, with $|m|=1$ having the largest
Coulomb repulsion. Adiabatic addition of CS flux to every electron
forces each pair to be Laughlin correlated by avoiding pair orbits
with $|m|=1$. This is accomplished without the necessity of a MF
approximation or the introduction of a MF energy scale.

From our previous discussion we know that we can form total angular
momentum multiplets $|\ell^N;L\alpha>$ by addition of the angular
momenta $\hat {\ell}_i=\hat {\ell}$ ($i=1,2,\ldots, N $) of $N$
Fermions. In the absence of Coulomb repulsion, $E_{\alpha}(L)$ is
the same for every value of $L$ formed from $N$ electrons, each with
angular momentum $\ell$ in the lowest LL (with $\ell=Q$, the
monopole strength in the Haldane spherical geometry). Let's define
$G_{N\ell}(L)$ as the number of multiplets of total angular momentum
$L$. If we adiabatically add two CS flux quanta to each electron,
the $N$ particle multiplets that can be formed belong to a subset
$G_{N\ell}(L)$ with $\ell$ replaced by $\ell^{\ast}=\ell-(N-1)$. The
multiplets belonging to $G_{N\ell^{\ast}}(L)$ all avoid, to the
maximum extend possibly, pair states with $\mathcal R=1$. This
result is obviously true for a pair of Laughlin correlated
electrons. The smallest allowed pair angular momentum would be
$L^{\prime}=2\ell^{\ast}-1=2\ell-3$, completely avoiding $\mathcal
R=1$. In addition our numerical results, (Fig.
\ref{fig:DiagExample}) show that the allowed values of
$\ell_{\rm{QE}}$ and $\ell_{\rm{QH}}$ [frames (b) and  (c)] are
$\ell_{\rm{QE}}=\ell^{\ast}+1=5$ and $\ell_{\rm{QH}}=\ell^{\ast}=5$.
This was easily understood in terms of ``effective monopole
strength", but the result does not depend on the MF approximation.
From frame (e) it is clear that $L_{2\rm{QH}}=2\ell^{\ast}-j$ where
$j$ is an odd integer [and from (d) that
$L_{2\rm{QE}}=2(\ell^*+1)-j$]. Thus, the adiabatic addition of CS
flux introduces Laughlin correlations (avoiding $\mathcal R=1$) and
selects \cite{QuinnQuinnCombTh01,QuinnWQuinnBPhysE01} from
$G_{N\ell}(L)$ the subset $G_{N\ell^*}(L)$ that avoids the smallest
(and most repulsive) pair orbit with $\mathcal R=1$. The proof that
$G_{N\ell^*}(L)$ is a subset of $G_{N\ell}(L)$ has been given in
\textcite{QuinnQuinnCombTh01}.

When $N>2\ell^{\ast}+1$, there are more particles than can be
accommodated in the lowest CF LL. An integral number of filled CF
levels occurs when $N=n(2\ell^*+n)$, where $n=1,2,\ldots$. Then, the
only state belonging to $G_{N\ell^*}(L)$ is the $L=0$ incompressible
Jain state with filling factor $\nu=n(2n\pm 1)^{-1}$. This
completely explains the Jain sequences $1/3,2/5,3/7,\ldots$ and
$1,2/3,3/5, \ldots$ (though for simplicity we have considered $p=1$
instead of addition of $2p$ CS flux quanta). The gap between the
lowest band of states (containing the minimum number of QPs required
by the values of $2Q$ and $N$) and the first excited band is
proportional to $V(\mathcal R)$, the pseudopotential describing the
interaction of a pair as a function of the relative pair angular
momentum $\mathcal R$, for the value of $\mathcal R$ avoided in the
Laughlin correlated state. Note that the only energy scale is the
Coulomb scale, and even though no extraneous MF energy has been
introduced, the occurrence of Jain states, the form of the low
energy spectrum, and the size of gaps has been determined
qualitatively.

\begin{figure}
\centerline{\includegraphics [width =1 \linewidth] {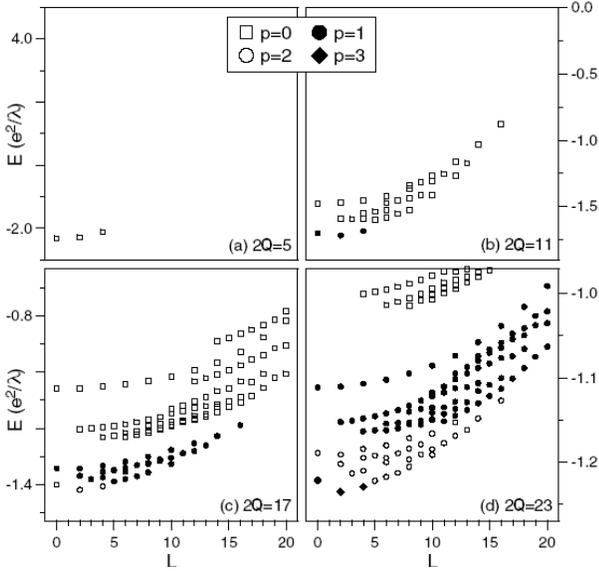}}
\caption{The energy spectra of 4 electrons in the lowest Landau
level at different monopole strength (a) $2Q=5$, (b) $2Q=11$, (c)
$2Q=17$, (d) $2Q=23$. All those $2Q$ values are equivalent in
mean-field CF picture (CS transformation with $p=0,1,2$ and 3,
respectively). (Solid diamonds: states with $\mathcal R \geq 7$,
that is $P(1)\approx P(3)\approx P(5)\approx 0$ and $P(7)>0$; open
circles: states with $\mathcal R \geq 5$, that is $P(1)\approx
P(3)\approx 0$ and $P(5)>0$;  solid circles: states with $\mathcal R
\geq 3$, that is $P(1)\approx 0$ and $P(3)>0$; open squares: states
with $\mathcal R \geq 1$, that is $P(1)>0$)
\cite{QuinnWoysJPhys00}.} \label{fig:bands_CF}
\end{figure}

Fig. \ref{fig:bands_CF} is a simple illustration of this for a
system of four electrons. If we start at $2\ell=23$ [frame (d)] we
find four bands. The highest band contains pairs with the largest
values of $L^{\prime}$ (i.e. the largest pair repulsion). When we
consider $2\ell^*=2\ell-2(N-1)=17$ [frame (c)] we eliminate the
largest $L^{\prime}$ and there are only three bands. Ultimately at
$2\ell^{*}=2\ell-6(N-1)=5$ [frame (a)] there is only a single band
(with low $L$ values and low pair repulsion). If we had chosen
$2\ell=21$ instead of 23, we would have had a Laughlin $L=0$ IQL
state for frame (a) since $2\ell^{\ast}=21-6\times 3=3$ and the
level is filled by four electrons.


\section{The Composite Fermion Hierarchy}
\label{sec:CFHierarchy}
Haldane \cite{HaldanePRL83} introduced the idea of a hierarchy of
condensed states in which Laughlin QPs of a condensed electron state
could form daughter states. The new daughter states have their
excitations (a second generation of QPs) which, in turn, could form
new IQL daughter states with their own QPs, ad infinitum. Haldane
assumed the problem of partial filling of a Landau level of QPs (or
a QP angular momentum shell) was essentially the same as the
original problem of putting $N$ electrons into $N_{\Phi}$ single
particle states of the lowest LL. Because the maximum allowed value
of the number of QP states, was equal to $N$, the number of
electrons in the Laughlin condensed state, he replaced the electron
LL degeneracy $N_{\Phi}$ by $N$, and replaced the number of
electrons by $N_{\rm{QP}}$ in the Laughlin condition
$N_{\Phi}=(2p+1)N$ for an IQL state. Because he treated the
excitations as Bosons, Haldane's condition for a daughter state was
taken as $N=2pN_{\rm{QP}}$, with the even integer $2p$ replacing
Laughlin's odd integer $2p+1$ appropriate to Fermions. This
hierarchy picture implicitly assumed that residual interactions
between QPs would give rise to Laughlin correlations among them.

Slightly different versions of the hierarchy were later
independently proposed by Halperin \cite{HalperinPRL84} and by
Laughlin \cite{LaughlinPRL88, LaughlinSurfSci84}. They differed
primarily in the statistics (anyon, Fermion, Boson) satisfied by the
QPs. These hierarchy schemes suggested that all odd denominator
filling fractions should be IQL states.

Sitko et al. \cite{SitkoYiYiQuinnPRL96,SitkoYiQuinnPRB97} introduced
a very simple CF hierarchy picture in an attempt to understand the
connection between Haldane's hierarchy of Laughlin correlated QP
daughter states and Jain's sequence of IQL states with integrally
filled CF Landau levels. Jain's CF picture neglected interactions
between QPs. The gaps causing incompressibility were energy
separations between the single particle CFLLs. Not all odd
denominator fractions occurred in Jain's sequence $\nu=n(2pn \pm
1)^{-1}$ where $n$ and $p$ are integers. The missing IQL states,
which occurred for partially filled QP shells (or CFQP Landau
levels), had to depend on ``residual interactions" between QPs,
neglected in Jain's mean field CF picture.

In the CF hierarchy picture
\cite{SitkoYiYiQuinnPRL96,SitkoYiQuinnPRB97,YiSitkoKhuranaQuinn,
YiQuinnSSC97,WojsQuinnPRB00}
an initial electron filling factor $\nu_0$ was related to an
effective CF filling factor $\nu_0^{\ast}$ by the relation
\begin{equation}
\label{eq:nu_0_gen_CF} {\nu_0^{\ast}}^{-1}=\nu_0^{-1}-2p_0~.
\end{equation}
This says that the total number of flux quanta (of both the dc
magnetic filed and CS gauge field) seen by one CF was equal to the
dc flux per electron minus the CS flux per electron subtracted in
the CF transformation. If $\nu_0^{\ast}$ were an integer $n$, then
the IQL state of the CFs would occur at $\nu_0=n(2p_0n \pm 1)^{-1}$.
This is the Jain sequence of integrally filled CF LLs.

What happens if $\nu_0^{\ast}$ is not equal to an integer? Sitko et
al. \cite{SitkoYiYiQuinnPRL96,SitkoYiQuinnPRB97} suggested that one
writes $\nu_0^{\ast}$ as $\nu_0^{\ast}=n_1+\nu_1$, where $n_1$ was
an integer and $\nu_1$ represented the filling factor of the
partially filled CFQP level. If, as Haldane \cite{HaldanePRL83}
suggested, the residual interactions between QPs were sufficiently
similar to the Coulomb interaction between electrons in the lowest
LL, one could assume Laughlin correlations among QPs. By reapplying
the CF transformation to them and writing
${\nu_1^{\ast}}^{-1}=\nu_1^{-1}-2p_1$, $\nu_1^{\ast}$ could be an
integer $n_2$ resulting in $\nu_1=n_2(2p_1n_2 \pm 1)^{-1}$ and an
IQL daughter state at
\begin{equation}
\frac{1}{\nu_0}=2p_1+\left[n_1+n_2(2p_1n_2+1)^{-1}\right]^{-1}~.
\label{eq:nu_gen_CF_1}
\end{equation}
This is a new odd denominator fraction not belonging to the Jain
sequence. If $\nu_1^{\ast}$ is not an integer, simply set
$\nu_1^{\ast}=n_2+\nu_2$ and reapply the CF transformation to the
new QPs in the shell of filling factor $\nu_2$. In general one finds
\begin{equation}
\frac{1}{\nu_l}=2p_l+\frac{1}{n_{l+1}+\nu_{l+1}}~.
\label{eq:nu_gen_CF}
\end{equation}
at the $l^{\rm{th}}$ level of the hierarchy. When $\nu_{l+1}=0$,
there is a filled shell of CFs at the $l^{\rm{th}}$ level of the
hierarchy. The procedure generates Haldane's continued fraction
leading to IQL states at all odd denominator fractions. It gives the
Jain sequence as a special case in which integral CF filling
$\nu_0^{\ast}=n$ of the CFQP shell is found at the first level of
the CF hierarchy. No residual interactions are needed to obtain the
Laughlin-Jain sequence of IQL states; it arises from the gap between
the last filled CF level and the empty ones. Haldane's result
assumes QP interactions are responsible for the incompressibility
gap, and that the interactions cause Laughlin correlations among the
QPs.

\begin{figure}
\centerline{\includegraphics [width =1 \linewidth]
{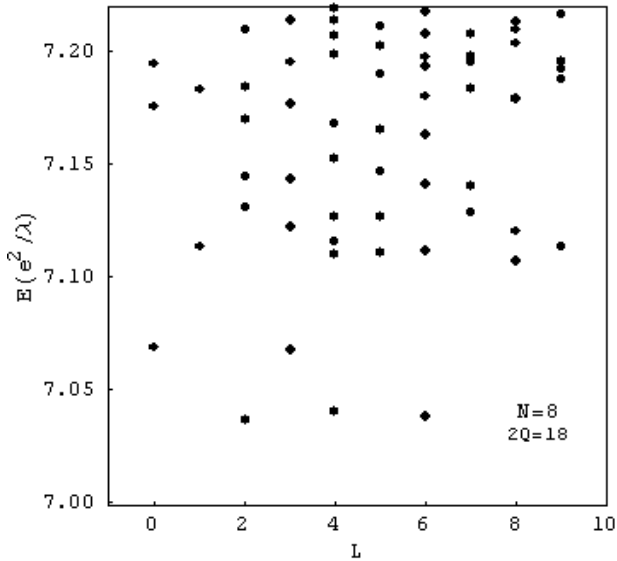}} \caption{Low energy spectrum of 8
electrons at $2\ell=18$. The lowest band contains 3 QEs each with
$\ell_{\rm{QE}}=3$. Reapplying the CS mean field approximation to
these QEs would predict an $L=0$ daughter state corresponding to
$\nu=4/11$. The data makes it clear that this is not valid.}
\label{fig:En_spectrum_8e_2l_18.eps}
\end{figure}

It is not difficult to show by numerical diagonalization that
hierarchy picture can't be correct in general. The reason, as
suggested by Sitko et al. \cite{SitkoYiYiQuinnPRL96} has to do with
the residual QP interactions. Consider, for example, the electron
system with $(N,2\ell)$ given by $(8,18)$. Applying the CF
transformation with $2p_0=2$ gives $2\ell_1^{\ast}=18-2(8-1)=4$.
Thus, the lowest CF shell has $\ell_1^{\ast}=2$; it can accommodate
five CFs. The remaining three CFs must go into the first excited CF
shell with $\ell_{\rm{QE}}=3$. The five CFs in the lowest shell
would give an IQL state if three CFQEs were not present. Only the
CFs in the partially filled CF shell are considered to be QPs. Three
Fermions each with $\ell_{\rm{QE}}=3$ give the multiplets $L=0\oplus
2\oplus 3\oplus 4\oplus 6$. If the CF hierarchy were correct,
applying a second CF transformation with $2p_1=2$ to the three CF
QEs would give $2
\ell_{\rm{QE}}^{\ast}=2\ell_{\rm{QE}}-2(N_{\rm{QE}}-1)=2$. The new
level of second generation CFs would exactly accommodate three QEs
and give an $L=0$ IQL ground state. Numerical diagonalization of the
$(N,2\ell)=(8,18)$ system gives the spectrum shown in Fig.
\ref{fig:En_spectrum_8e_2l_18.eps}.

The low lying multiplets are exactly as predicted at the first CF
level, giving three QEs each with $\ell_{\rm{QE}}=3$. However, the
$L=0$ multiplet is clearly not the ground state as predicted by
reapplying the CF transformation. It should be emphasized that the
numerical results are obtained for a spin polarized system (with
total spin $S=N/2=4$). The reason of this failure [the
``subharmonic" behavior of the CFQE pseudopotential
\cite{WojsQuinnPRB00}] will be explained later (see Sections
\ref{sec:res_int}, \ref{sec:QE}).

\section{Residual Interactions}
\label{sec:res_int}

The QEs and QHs have residual interactions that are more complicated
than simple Coulomb interactions. They are difficult to calculate
analytically, but if we look at an $N$ electron system at a value of
$2\ell=3(N-1)\pm 2$, we know that the lowest band of states in the
spectrum will correspond to 2 QEs or 2 QHs for the minus and plus
signs respectively. Fig. \ref{fig:DiagExample} gives the spectrum
for $N=10$ electrons at $2\ell=25$ (2 QE case) and $2\ell=29$ (2 QH
case). It is clear that the low energy bands are not degenerate, but
that the energy $E$ depends on $L$, which (as we have seen) can be
understood as the total angular momentum of the QP pair. For QEs,
$E(L)$ has a maximum at $L=2\ell_{\rm{QE}}-3$ and a minima at
$L=2\ell_{\rm{QE}}-1$ and $2\ell_{\rm{QE}}-5$. For QHs, $E(L)$ has a
maximum at $L=2\ell_{\rm{QH}}-1$ and $L=2\ell_{\rm{QH}}-5$, and a
minimum at $L=2\ell_{\rm{QH}}-3$. This is quite different from the
pseudopotentials for electrons (i.e. the energy of interaction as a
function of total pair angular momentum), and it is undoubtedly the
reason why the CF picture fails when it is reapplied to QEs.

\begin{figure}
\centerline{\includegraphics [width =1
\linewidth]{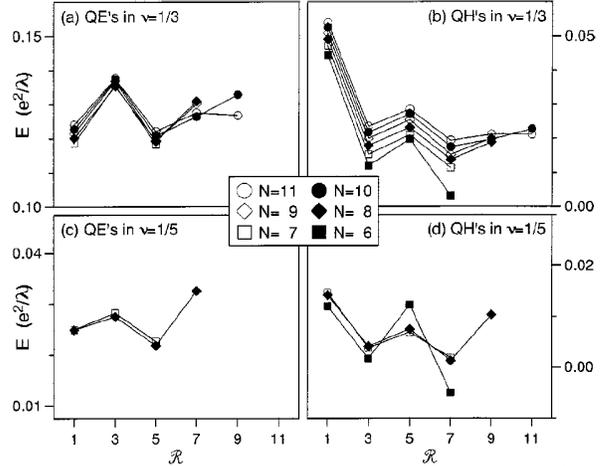}} \caption{The pseudopotentials of
a pair of quasielectrons (left) and quasiholes (right) in Laughlin
$\nu=1/3$ (top) and $\nu=1/5$ (bottom) states, as a function of
relative angular momentum $\mathcal R$. Different symbols mark data
obtained in the diagonalization of between 6 and 11 electrons
\cite{WojsQuinnPRB00}.} \label{fig:Calc_Pseodpot}
\end{figure}

More careful estimates of $V_{\rm{QE}}(\mathcal R)$ and
$V_{\rm{QH}}(\mathcal R)$ (where $\mathcal R=2\ell-L^{\prime}$ and
$L^{\prime}$ is the pair angular momentum) are shown for QPs of the
Laughlin $\nu=1/3$ and $\nu=1/5$ IQL states in Fig.
\ref{fig:Calc_Pseodpot}. The values of $V_{\rm{QP}}(\mathcal R)$ are
determined (up to an overall constant) by diagonalization of $N$
electron systems with $6 \leq N \leq 11$.

\begin{figure}
\centerline{\includegraphics [width =1
\linewidth]{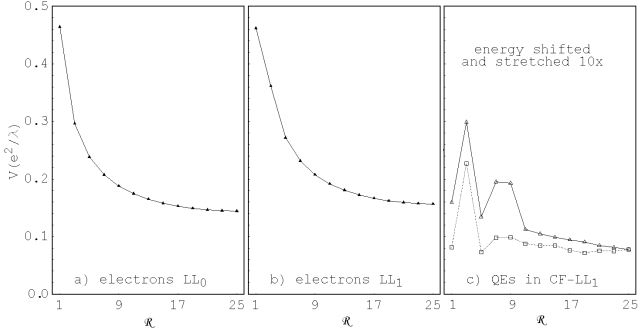}} \caption{Pair interaction
pseudopotentials as a function of relative angular momentum
$\mathcal R$ for electrons in LL0 (a), LL1 (b) and for the QEs of
the Laughlin $\nu=1/3$ state calculated by Lee et al.
\cite{LeeScarolaJainPRL01,LeeScarolaJainPRB02} (squares) and by
W\'ojs et al. \cite{WojsWodzinskiQuinnPRB06} (triangles).}
\label{fig:Pseudopot}
\end{figure}

In Fig. \ref{fig:Pseudopot} we display the pseudopotentials for
electrons in LL0 and LL1 with that for QEs of the Laughlin $\nu=1/3$
IQL state in CF LL1. The electron pseudopotentials are the same ones
presented in Fig. \ref{Fig:Gen_Pseodpot} but are presented here as a
function of $\mathcal R =2 \ell-L ^{\prime}$, the relative angular
momentum of a pair, for large systems.

We define a pseudopotential to be harmonic if it increases with
$L^{\prime}$ as $V_H(L^{\prime})=A+BL^{\prime}(L^{\prime}+1)$, where
$A$ and $B$ are constants. The superharmonic behavior of
$V^{(0)}(\mathcal R)$ (i.e. it increases faster than
$V_H(L^{\prime})$ everywhere) is clear from the increasing slope
with decreasing $\mathcal R$. For $V^{(1)}(\mathcal R)$, only at
$\mathcal R=1$ is the pseudopotential harmonic (the slope for
$1<\mathcal R<3$ is the same as that for $3<\mathcal R<5$). The QE
pseudopotentials in frame (c) were taken from the calculations of
Lee et al. \cite{LeeScarolaJainPRL01,LeeScarolaJainPRB02} and from
the diagonalization of small electron systems done by W\'ojs et al.
\cite{WojsWodzinskiQuinnPRB06,wojs2006}, and are known up to a
constant. The magnitude of interaction of CFQEs is much smaller, and
has a sharp maximum at $\mathcal R=3$ and minima at $\mathcal R=1$
and $5$.

These pseudopotentials have been obtained for 2D electron layers of
zero width. It is well known \cite{WojsQuinnPRB07,
arxivPetersonDasSarma,HeZhangXieDasSarmaPRB90} that the finite
extent of the subband wavefunction in the direction perpendicular to
the layer introduces a correction to the electron pseudopotentials.
The QP pseudopotentials are also sensitive to the layer width since
they are obtained from the energy of the two QP band obtained by
exact diagonalization of the appropriate electron system and the
specific form of the (lowest) subband wave function.

\section{Pair Angular Momentum Theorem and Coefficients of
Fractional Parentage} \label{sec:ang_mom_theorems}

 We can define the total angular
momentum operator $L=\sum_i \hat \ell_i$ for an $N$ electron system
in a shell of angular momentum $\ell$, and $\hat L_{ij}=\hat \ell_i+
\hat \ell_j$, the angular momentum operator for the pair $<i,j>$.
The operator identity
\begin{equation}
\hat L^2+N(N-1) \hat \ell^2-\sum_{<i,j>} \hat L_{ij}^2=0 ~,
\label{eq:L_identity}
\end{equation}
where the summation is over all pairs $<i,j>$, can be obtained
simply \cite{WojsQuinnSSC99} by writing out $\hat L^2$ and
$\sum_{<i,j>}L^2_{ij}$, and eliminating $\hat \ell_i\cdot \hat
\ell_j$. We consider the $N$ electron multiplet $|\ell^N;L\alpha>$
of total angular momentum $L$. The index $\alpha$ is used to
distinguish independent multiplets with the same total angular
momentum $L$. Taking the expectation value of Eq.
\ref{eq:L_identity} for the state $|\ell^N;L\alpha>$ we obtain
\begin{equation}
\label{eq:first_sum_rule_L} L(L+1)+N(N-2)\ell (\ell+1)=\left<
\sum_{<i,j>}\hat L^2_{ij}\right>~.
\end{equation}

This relates the expectation value of the sum over all pairs of the
squared pair angular momenta to $L$ and $\ell$.

The antisymmetric angular momentum multiplet $|\ell^N;L\alpha>$ can
be written
\begin{equation}
\label{eq:Fractional_Parentage}
|\ell^N;L\alpha>=\sum_{L_{12}}\sum_{L^{\prime}\alpha^{\prime}}G_{L\alpha
L^{\prime}\alpha^{\prime}}|\ell^2,L_{12};\ell^{N-2},L^{\prime}\alpha^{\prime};L>~.
\end{equation}
Here $|\ell^2,L_{12};\ell^{N-2},L^{\prime}\alpha^{\prime};L>$ is an
$N$ electron multiplet of total angular momentum $L$ formed from an
$N-2$ electron multiplet $|\ell^{N-2};L^{\prime}\alpha^{\prime}>$
and a pair wavefunction $|\ell^2;L_{12}>$. It is antisymmetric with
respect to the exchange of indices $i,j$ when both $i$ and $j$
belong to the set $(1,2)$ or when both belong to the set
$(3,4,\ldots N)$. It is not antisymmetric if $i$ belongs to one set
and $j$ to the other. However, the coefficient $G_{L\alpha
L^{\prime}\alpha^{\prime}}$, called the coefficient of fractional
parentage, can be chosen so that $|\ell^N;L\alpha>$ is totally
antisymmetric. Fractional parentage has been widely used in atomic
and nuclear physics \cite{NuclearShellBook}, but all that we need to
know is that
\begin{equation}
\sum_{L^{\prime}\alpha^{\prime}}|G_{L\alpha,L^{\prime}\alpha^{\prime}}(L_{12})|^2=P_{L\alpha}(L_{12})~.
\label{eq:Pr_Coeff_Fract_Parentage}
\end{equation}
This says that the probability $P_{L\alpha}(L_{12})$ that the
multiplet $|\ell^N,L\alpha>$ has pairs with pair angular momentum
$L_{12}$ is equal to the sum over all $N-2$ particle multiplets
$|\ell^{N-2};L^{\prime}\alpha^{\prime}>$ of the square of the
magnitude of $G_{L\alpha,L^{\prime}\alpha^{\prime}}(L_{12})$. Since
$|\ell^N;L\alpha>$ is totally antisymmetric, we can select a single
pair $<i,j>=<1,2>$ and multiply by the number of pairs. The right
hand side of Eq. \ref{eq:Fractional_Parentage} is a linear
combination of $\hat L_{12}^2$ whose coefficients are
$G_{L\alpha,L^{\prime}\alpha^{\prime}}(L_{12})$. The net result is
that
\begin{equation}
\left<\sum_{<i,j>}\hat
L_{ij}^2\right>=\frac{N(N-1)}{2}\sum_{L_{12}}L_{12}(L_{12}+1)P_{L\alpha}(L_{12}).
\label{eq:sum_L_ij_square}
\end{equation}

The summation on the right hand side is over all the allowed values
of the pair angular momentum $L_{12}$, and $P_{L\alpha}(L_{12})$ was
given in Eq. \ref{eq:Pr_Coeff_Fract_Parentage}. This leads to two
useful sum rules:
\begin{eqnarray}
   \sum_{L_{12}} P_{L\alpha}(L_{12})&=&1~, \label{eq:Sum_rule_1}\\
   \frac{1}{2}N(N-1) \sum_{L_{12}} L_{12} (L_{12}+1
)P_{L\alpha}(L_{12})&=&L(L+1)\nonumber\\
+N(N-2)\ell (\ell+1)~.\label{eq:Sum_rule_2}
\end{eqnarray}

It is interesting to note that the expectation value of
$\sum_{<i,j>}\hat L_{ij}^2$ in the multiplet $|L,\alpha>$ is
independent of $\alpha$ since the right hand side of Eq.
\ref{eq:Sum_rule_2} is independent of $\alpha$.

\section{Harmonic Pseudopotential and Absence of Correlations}
\label{sec:harmonic_pp}

The two sum rules allow us to make use of the concept of a harmonic
pseudopotential. In Fig. \ref{Fig:Gen_Pseodpot} we plotted the
pseudopotential for the Coulomb interaction of electrons in the LL0
and LL1 as a function of the eigenvalues of the square of the pair
angular momentum $L^{\prime}$. For LL0 $V^{(0)}(L^{\prime})$
increases with increasing $L^{\prime}$ faster than
$L^{\prime}(L^{\prime}+1)$; for LL1 this is true only for
$L^{\prime}>2\ell-5$. Between $L^{\prime} = 2\ell-5$ and $L^{\prime}
=2 \ell-1$, $V^{(1)}(L^{\prime})$ increases approximately as a
linear function of $L^{\prime}(L^{\prime}+1)$. Let's define
\begin{equation}
V_H(L^{\prime})=A+BL^{\prime }(L^{\prime }+1)~,
\label{eq:Harmonic_Pot_def}
\end{equation}
as a harmonic pseudopotential, with $A$ and $B$ being constants.
From Eqs. \ref{eq:energyElalpha} and \ref{eq:Sum_rule_2}, we can
write, for a harmonic pseudopotential, the energy of the multiplet
$|\ell^N;N\alpha>$ as
\begin{equation}
E_{\alpha}(L)=N\left[\frac{1}{2}(N-1)A+B(N-2)\ell(\ell+1)\right]+BL(L+1)~.
\label{eq:Energy_Harmonic_potential}
\end{equation}

We note that for a harmonic pseudopotential $E_{\alpha}(L)$ is
totally independent of the multiplet index $\alpha$. Every multiplet
with the same angular momentum $L$ has the same energy. As long as
the constant $B$ is positive, the energy increases with $L$ as
$BL(L+1)$, but the degeneracy of the myriad multiplets of a given
value of $L$ is not removed, implying the absence of correlations
for the harmonic potential.

\section{The Simplest Anharmonicity and Laughlin Correlations}
\label{sec:anharmonicity}
We define $\Delta V(L^{\prime})=V(L^{\prime})-V_H(L^{\prime})$ as
the anharmonic part of the pseudopotential. $\Delta V(L^{\prime})$
is responsible for lifting the degeneracy of different multiplets
having the same value of the total angular momentum $L$. We suggest
that the simplest anharmonic contribution to the pseudopotential be
taken as
\begin{equation}
\label{eq:simple_anharmonic} \Delta
V(L^{\prime})=k\delta(L^{\prime},2\ell-1)~.
\end{equation}

If $k>0$, it is apparent that the lowest energy multiplet for each
value of $L$ will be the one with the smallest value of
$P_{L\alpha}(L^{\prime}=2\ell-1)$ [or $P_{L\alpha}(\mathcal R=1)$].
This is exactly what is meant by Laughlin correlations. Complete
avoidance of $\mathcal R=1$ pairs (or $m=1$ pairs in the planar
geometry) cannot occur unless $2\ell \geq 3(N-1)$. In the limit of
large systems this corresponds to a filling factor $\nu\geq 1/3$.

If $k<0$ in Eq.\ref{eq:simple_anharmonic}, then the lowest energy
state for each $L$ will have the largest value of
$P_{L\alpha}(\mathcal R=1)$. This suggests a tendency to form
$\mathcal R=1$ pairs rather than Laughlin correlations.

It is important to emphasize that Laughlin correlations (e.g.
maximum avoidance of pairs with $\mathcal R=2\ell-L^{\prime}$ equal
to unity) occur only when $V(\mathcal R)$ is ``superharmonic" at
$\mathcal R=1$. From Fig. \ref{fig:Pseudopot}, we can see that
electrons in LL0 (a) satisfy this condition, while QEs of the
Laughlin $\nu=1/3$ state (c) do not. This means that at
$\nu_{\rm{QE}}=1/3$, the quasielectrons in CF LL1 will not be
Laughlin correlated. This is in agreement with the numerical results
of Sitko et al. \cite{SitkoYiYiQuinnPRL96}. Now, however, we
understand why the CF hierarchy picture fails for a spin polarized
system. The QE pseudopotential is subharmonic at $\mathcal R=1$ and
does not support Laughlin correlations. There have been a number of
papers suggesting that the IQL states observed by Pan et al.
\cite{PanStormerTsuiPRL03}, like the $\nu=4/11$ IQL, can be
understood as a second generation of CFs
\cite{SmetNature,GoerbigPhysicaE,GoerbigPRB69,LopezFradkinPRB04}.
This suggestion cannot be correct. As previously shown in Sec.
\ref{sec:CFHierarchy}, the idea is not new
\cite{SitkoYiYiQuinnPRL96,SitkoYiQuinnPRB97}, and it had already
been shown numerically to fail. The theorem on pair angular momentum
and the harmonic potential make it clear
\cite{WojsQuinnPRB00,WojsYiQuinnPRB04,QuinnVarenna,QuinnWQuinnBPhysE01,
QuinnWPhysE00,QuinnJPSK,QuinnQuinnSSC06} why the second generation
of CFs can't be correct for fully spin polarized states like
$\nu=4/11$: $V_{\rm{QE}}(\mathcal R)$ will not support Laughlin
correlations at $\nu_{\rm{QE}}=1/3$.

If QEs of a spin polarized electron system can't be Laughlin
correlated at $\nu_{\rm{QE}}=1/3$, how will these QEs be correlated?
Before considering this problem in detail, it is worthwhile looking
at the problem of electrons in LL1. For electrons confined to a 2D
surface, Fig. \ref{fig:Pseudopot} (b) shows that the pseudopotential
is very close to harmonic for $\mathcal R<3$. In such a case,
Laughlin correlations (avoidance of $\mathcal R=1$) will not produce
the lowest energy state. There is no reason to avoid $\mathcal R=1$
in favor of $\mathcal R=3$ in the lowest band of energy states.
Let's study the problem by numerical diagonalization and attempt to
understand the results in terms of simple intuitive pictures.


\section{Incompressible Quantum Liquids in the First Excited Landau Level}
\label{sec:LL1}

\subsection{The $\nu=5/2$ Incompressible Quantum Liquid}
\label{sec:5_2_IQL}

It has been known for some time
\cite{EisensteinTsuiPRL87,PanStormerTsuiPRL99,EisensteinPRL02} that
at filling factor $\nu=2+\nu_1=5/2$ (half-filling of one spin state
of the LL1), an IQL state with a robust energy gap occurs. This is
in stark contrast to the compressible state found at $\nu=1/2$ (half
filling of the lower spin state of LL0). The compressible state at
$\nu=1/2$ can be described in terms of CFs which experience a ``mean
magnetic field " $B^{\ast}$ equal to $B-\nu n \phi_0$, where $n$ is
the electron concentration, and $\phi_0=hc/e$ is the quantum of flux
\cite{HalperinLeeReadPRB93}. $B^{\ast}$ vanishes at $\nu=1/2$.
Shubnikov-de Haas oscillations in the magnetoconductivity are
observed as a function of $B^{\ast}$ for small deviations away from
filling factor $\nu=1/2$ \cite{StromerTsuiPRL93,SdHPRB96}. For
$\hbar \omega_C \gg e^2/\lambda$, the difference between the
behavior of electrons in LL0 and LL1 must be related to their
pseudopotentials. In LL0 Laughlin correlations occur because
$V_0(L^{\prime})$ is ``superharmonic". Jain's CF picture can be
applied resulting in the Laughlin-Jain sequence of "filled CF"
shells in the mean-field approximation. The Halperin, Lee, and Read
(HLR) picture \cite{HalperinLeeReadPRB93} treats the interactions
between the CFs beyond the mean field approximation (both Coulomb
and Chern-Simons gauge interactions) by standard many-body
perturbation theory. HLR gives surprisingly good agreement with the
qualitative features of $\nu=1/2$ state that are observed
experimentally.

For the electrons in LL1 the pseudopotential $V_1(\mathcal R)$ is
not superharmonic at $\mathcal R =2\ell-L_2 =1$. Therefore,
electrons in LL1 will not support Laughlin correlations and cannot
be described in terms of weakly interacting CFs. Finite well width
changes $V_1(\mathcal R)$ through form factors associated with the
subband wavefunction of the quantum well. It is possible that the
effect can lead to a change in the ratio of $V_1(\mathcal R=1)$ to
$V_1(\mathcal R=3)$ that will support Laughlin correlations within a
certain range of well widths \cite{RezayiHaldanePRL00}. Only then
can the $\nu=5/2$ state be thought of as a CF state at $B^{\ast}=0$,
which might undergo a ``Cooper pairing" instability and form the
gapped IQL state observed in some experiments.

For the moment, let's concentrate on the case of zero well width
where $V_1(\mathcal R)$ is given by Fig. \ref{fig:Pseudopot} (b). By
standard numerical diagonalization within LL1 (i.e. neglecting
Landau level mixing) we can obtain the energy spectra for $N$
electrons in a shell of angular momentum $\ell$ interacting via the
pseudopotential $V_1(\mathcal R)$. We have carried out such
diagonalizations for $N\leq 16$ and for different values of $2\ell$
\cite{WojsPRB01,WojsQuinnPRB05,WojsQuinnPRB06,prlarxiv}.
Incompressible $L=0$ ground states are found to fall into families.
The most prominent ones occur at $2\ell=2N-3$ for even values of
$N$, and at $2\ell=3N-7$ (and by electron-hole symmetry at their
e--h conjugate states $2\ell=2N+1$ and $2\ell=3N/2+2$). The
conjugate states are obtained with the replacement of $N$ by
$2\ell+1-N$. The energy gap for the $\nu_1=1/2$ state is less than
1/3 of the gap for the $\nu=1/3$ state in LL0. The behavior of the
gap with increasing particle number $N$ suggests that this IQL state
at $\nu_1=1/2$ will persist for macroscopic systems.

There has been a considerable amount of theoretical work on the
$\nu=5/2$ state (the half filled LL1 lower spin state). Moore and
Read \cite{MooreReadNuclPhys91} proposed a Pfaffian wavefunction for
this state based on ideas from conformal field theory. Greiter et
al. \cite{GreiterWenWilczekPRL91,GreiterWenWilczekNuclPhys92} showed
that the Pfaffian state is an exact solution to a special
Hamiltonian which is large and repulsive when three electrons form a
single droplet (with the total three particle angular momentum
$L_3=3\ell-3$ or $\mathcal R_3=3 \ell -L_3=3)$ and zero otherwise.
For the Pfaffian state at $\nu_1=1/2$ in LL1, $2\ell$ is given by
$2N-3$ (or its conjugate $2N+1$) in agreement with numerical
diagonalization.

It should be noted that Laughlin correlated states at $\nu=1/m$ in
LL0 occur at $2\ell=mN-m$, where $m$ is an odd integer. States in
the Jain sequence \cite{JainPRB90} $\nu=n(2pn\pm 1)^{-1}$, where $n$
and $p$ are positive integers, occur at $2\ell=\nu^{-1}N \pm n-2p$
(and their e-h conjugate values). No even denominator fractional
fillings are IQL states in the Laughlin-Jain sequence. How then can
we understand the IQL state observed experimentally at
$\nu_1=\nu-2=1/2$ in LL1 and found in numerical diagonalization of
small systems at $2\ell=2N-3$?


\subsection{Heuristic Picture of the $\nu=5/2$ State}
\label{sec:5_2_picture}

As we have already noted, the pseudopotential $V_1(\mathcal R)$ is
not superharmonic at $\mathcal R=1$, and the probability $P
(\mathcal R)$ of finding pairs with $\mathcal R=1$ in the ground
state will not be a minimum as it is for the Laughlin correlated
case in LL0. Let's make the assumption that $\mathcal R=1$ pairs
form. Of course, a state consisting of only $N/2$ pairs, each with
pair angular momentum $L_2=2\ell-1$ (or relative angular momentum
$\mathcal R=1$) is not an eigenstate of the interacting system. The
electrons can scatter, breaking up the pairs, as long as both the
total angular momentum of the system $L$ and it $z$-component are
conserved. However, we can think of this state as a ``parent state"
which will generate the exact ground state when Coulomb interactions
admix different configurations with the same $L$ and $L_z$.

\begin{figure}
\centerline{\includegraphics [width =1
\linewidth]{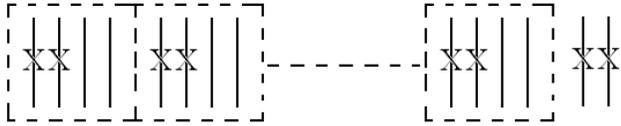}} \caption{Simple picture of the
$\nu_1=1/2$ paired state. The vertical lines represent single
particle states of different $\ell_z$, going form $-\ell$ to $\ell$.
Occupied states are marked by an X on the vertical line. The ``unit
cell" is shown by the dashed rectangles. Occupancy is chosen so that
$L_z=0$. The number of single particle states satisfies the relation
$2\ell+1=4(N/2-1)+2$, or $2\ell=2N-3$, corresponding to $\nu_1=1/2$
state (of LL1). It conjugate state at $2\ell=2N+1$ should, by e-h
symmetry, also be an IQL state. } \label{fig:HeuristicPicture}
\end{figure}

A simple heuristic picture of the parent state with $L=0$,
containing $N/2$ pairs each with $\mathcal R=1$ in an angular
momentum $\ell=(2N-3)/2$ is shown in Fig.
\ref{fig:HeuristicPicture}. It corresponds to the maximum number of
electrons in a $\nu_1=1/2$ filled state which has $L_z$, the
$z$-component of the total angular momentum equal to zero. The
$\mathcal R=1$ pairs have total pair angular momentum $\ell_{\rm
{P}}=2\ell-1$. The pairs of electrons might normally be thought of
as Bosons. However, in 2D, they can be treated as either Fermions of
angular momentum $\ell_{\rm{F}}$, or as Bosons with
$\ell_{\rm{B}}=\ell_{\rm{F}}-\frac{1}{2}(N-1)$, where $N$ is the
number of particles \cite{QuinnQuinnCombTh01,QuinnWQuinnBPhysE01,
XieHeDasSarmaPRL91}. Let's assume that $N$ is even and that we form
$N/2$ pairs. The pairs cannot approach one another too closely
without violating the Pauli exclusion principle with respect to
exchange of identical constituent Fermions belonging to different
pairs. We can account for this effect by introducing an effective
Fermion pair (FP) angular momentum defined by
\begin{equation}
2\ell_{\rm{FP}}=2(2\ell-1)-\gamma_F(N_{\rm{P}}-1)~. \label{eq:l_FP}
\end{equation}
For a single pair $\ell_{\rm{FP}}=2\ell-1$. As $N_{\rm{P}}$
increases, the allowed values of the total angular momentum of two
pairs is restricted to the values less than or equal to
$2\ell_{\rm{FP}}$. The value of $\gamma_{\rm{F}}$ is determined by
requiring that the FP filling factor $\nu_{\rm{FP}}$ be equal to
unity when the single Fermion filling factor has the electron
filling factor corresponding to the appropriate FP filling. For the
pair having $\ell_{\rm{P}} =2\ell-1$, this corresponds to $\nu=1$.
\cite{WojsQuinnPRB00,WojsYiQuinnPRB04}. Remembering that
$\nu_{\rm{FP}}^{-1}=(2\ell_{\rm{FP}}+1)/N_{\rm{P}}$ and that
$\nu^{-1}=(2\ell+1)/N$, then we find in the large $N$ limit that
$\gamma_{\rm{F}}=3$ and
\begin{equation}
\label{eq:nu_FP_nu} \nu_{\rm{FP}}^{-1}=4\nu^{-1}-3~.
\end{equation}

If we treated pairs as Bosons, $\gamma_{\rm{F}}$ would be replaced
by $\gamma_{\rm{B}}=\gamma_{\rm{F}}+1$. The factor of $4$ in Eq.
\ref{eq:nu_FP_nu} results from having half as many pairs
($N_{\rm{P}}=N/2$) filling as twice as many states of the pair LL
(since the pairs have charge $-2e$ giving the degeneracy of the pair
Landau level $g_{\rm{P}}=2g$). The pairs form not because there is
an attractive interaction between electrons, but because the
anharmonic contribution to the pseudopotential, which determines the
correlations, is attractive at $\mathcal R=1$. By forming $N/2$
pairs that can be more widely separated than $N$ electrons, the
slightly stronger anharmonic part of the e-e repulsion at $\mathcal
R=3$ can be avoided. In fact, the pairs can become  Laughlin
correlated. For electrons in LL1 at filling factor $\nu=1/2$, Eq.
\ref{eq:nu_FP_nu} gives $\nu_{\rm{FP}}=1/5$. Fermions in a  Laughlin
correlated $\nu_{\rm{FP}}=1/5$ state must have
$2\ell_{\rm{FP}}=5(N_{\rm{P}}-1)$. This, together with Eq.
\ref{eq:l_FP} in which $\gamma_{\rm{F}}$ is set equal to 3, gives
$2\ell=2N-3$, the relation between $2\ell$ and $N$ appropriate for
the $\nu_1=1/2$ filling of LL1 (i.e. for the total filling factor
$\nu=2+1/2=5/2$).

It is worth noting that the heuristic picture of Fig.
\ref{fig:HeuristicPicture} has been used before for Laughlin-Jain
states in LL0 \cite{QuinnGiulianiPRB85}. For example, at $\nu=3/7$,
the unit cell contains seven single particle states, the first three
of which are filled. The number of unit cells is $(N/3)-1$, three
electrons being reserved to fill three states after the last unit
cell to give an $L_z=0$ state. This picture suggests that the
``parent state" produces IQL states for $2\ell=(7/3) N-5$. The minus
five is the appropriate finite size correction for the Laughlin
correlated $\nu=3/7$ state. The finite size corrections obtained in
the families of IQL states found in numerical studies appear to
contain important information about correlations in the IQL state.
It should be noted that our pair-state is different from the
Moore-Read Pfaffian state since the square of the overlap of the two
wavefunctions is not so close to the unity 
for a 14-electron system. As a consequence, the wavefunctions
describing the ground state and excited states are different from
those predicted by Greiter at al. \cite{GreiterWenWilczekNuclPhys92}
and T\"{o}ke at al. \cite{TokeJainPRL06, TokeJRegnaultJainPRL07}.
The work of T\"{o}ke and Jain \cite{TokeJainPRL06} describes the IQL
state at $\nu=5/2$ as a  result of residual CF interaction, showing
that a realistic Coulomb interaction would produce a  wavefunction
which is somehow different from the Pfaffian one. A numerical study,
made by T\"{o}ke at al. \cite{TokeJainPRL06, TokeJRegnaultJainPRL07}
shows that excited states of the $\nu=5/2$ state in the presence of
the Coulomb potential differ from those expected when a Pfaffian
wavefunction is used. The absence of a degenerate band of
quasiparticle states might suppress the expected non-Abelian
behavior.


\subsection{Excitations of $\nu=5/2$ State}
\label{sec:5_2_Excitations}

\begin{figure}
\centerline{\includegraphics [width =1
\linewidth]{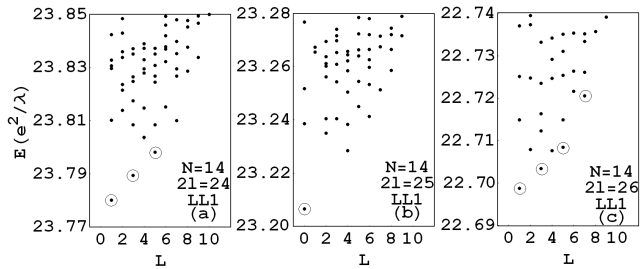}} \caption{Spectra of fourteen
electrons in the first excited LL of a zero width quantum well. The
values of $2\ell$ are 24 (a), 25 (b), 26 (c). Frame (b) has an $L=0$
IQL ground state. Frames (a)  and (c) contain at least two
elementary excitations [two FP quasiparticles in (a) and two FP
quasiholes in (b)] \cite{prlarxiv}.} \label{Fig:14_electrons_LL1}
\end{figure}

\begin{figure}
\centerline{\includegraphics [width =1
\linewidth]{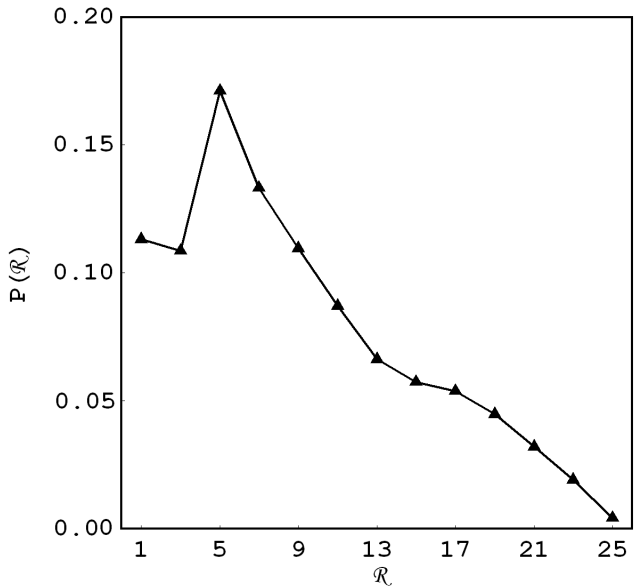}} \caption{$P(\mathcal R)$
vs $\mathcal R$ for the $L=0$ ground state of case (b) in Fig.
\ref{Fig:14_electrons_LL1} . The profile is very different from that
of a Laughlin correlated electron state in LL0 \cite{prlarxiv}.}
\label{Fig_Pr_5_2_14_electrons}
\end{figure}

In Fig. \ref{Fig:14_electrons_LL1} we display the spectra for $14$
electrons in LL1 at values of $2\ell$ equal to $24$ (a), $25$ (b),
and $26$ (c). In each case, the lowest band of states can be
interpreted using a simple picture which assumes that the $14$
electrons give rise to a ``parent" state with seven pairs, each pair
having pair angular momentum $\ell_{\rm{P}}=2\ell-1$. We treat the
pairs as Fermions with
$2\ell_{\rm{FP}}=2\ell_{\rm{P}}-3(N_{\rm{P}}-1)$. For case (b)
$\ell_{\rm{P}}=24$, giving $2\ell_{\rm{FP}}=30$. Then, by assuming
that the seven pairs are Laughlin correlated with
$2\ell_{\rm{FP}}^{\ast}=2\ell_{\rm{FP}}-2p(N_{\rm{P}}-1)$ and $p=2$,
we obtain $2\ell_{\rm{FP}}^{\ast}=6$. The shell of Laughlin
correlated pairs (LCPs) can accommodate $2\ell_{\rm{FP}}^{\ast}+1=7$
pairs giving an $L=0$ IQL ground state. For case (a)
$\ell_{\rm{P}}=23$ giving $2\ell_{\rm{FP}}=28$ and
$2\ell_{\rm{FP}}^{\ast}=4$. The lowest shell of FPs can accommodate
only five pairs; the remaining two become FP quasiparticles with
$\ell_{\rm{QP}}=3$. The allowed values of the total angular momentum
$L$ of two FP quasiparticles each with $\ell_{\rm{QP}}=3$ is
$L=2\ell_{\rm{QP}}-j$, where $j$ is an odd integer. This gives the
band $1\oplus 3\oplus 5$ as seen in frame (a). For $2\ell=26$,
$2\ell_{\rm{FP}}^{\ast}=8$, and we find two FP quasiholes each with
$\ell_{\rm{QH}}=4$ giving the band $1\oplus 3\oplus 5\oplus 7$ as
suggested in frame (c). The simple picture of $N_{\rm{P}}(=N/2)$
pairs for even values of $N$, correctly predicts the lowest band of
states for all even $N$ at $2\ell=2N-3$ or $2\ell=2N-3\pm 1$ that we
have tested. In Fig. \ref{Fig_Pr_5_2_14_electrons} we show
$P(\mathcal R)$, the probability of electron pairs with relative
angular momentum $\mathcal R$ for the $L=0$ ground state in (b).
Because $P(\mathcal R)$ is a maximum for $\mathcal R=1$ and a
minimum for $\mathcal R=3$, this IQL ground state is not a Laughlin
correlated state of electrons.

In LL0, excitations of the $\nu=m^{-1}$ Laughlin IQL states obtained
by changing $2\ell=m(N-1)$ by one unit consist of single QPs of
angular momentum $\ell_{\rm{QP}}=N/2$. For LL1, changing $2\ell$
from the $\nu_1=1/2$ value (of $2N-3$) by unity must produce two QPs
and a low lying band of excitations with angular momentum
$L=2\ell_{\rm{QP}}-j$, where $j$ is an odd integer. This is a very
strong indication that the IQL state consists of pairs with pair
angular momentum $\ell_{\rm{P}}=2\ell-1$. The angular momentum of
the pair changes by two units when the electron angular momentum
$\ell$ changes by one. The variation with total angular momentum $L$
of the energy in these bands can be interpreted as a pseudopotential
$V_{\rm{QP}}(L_2)$ describing the interaction of two Fermion pair
QPs. Unfortunately, the dispersion of these bands is rather
sensitive to the electron pseudopotential $V_1(\mathcal R)$. Small
changes like $\delta V_1(\mathcal R)=xV_1(\mathcal R)\delta
(\mathcal R,1)$ have noticeable effect on $V_{\rm{QP}}(L_2)$ even
for $x\lesssim 0.1$. In addition, the bands (especially the QH
bands) are not well separated from the quasicontinuum of higher
excitations.


\subsection{Other Incompressible Quantum Liquid States in the First Excited Landau Level}
\label{sec:other_LL1}

\begin{figure}
\centerline{\includegraphics [width =1
\linewidth]{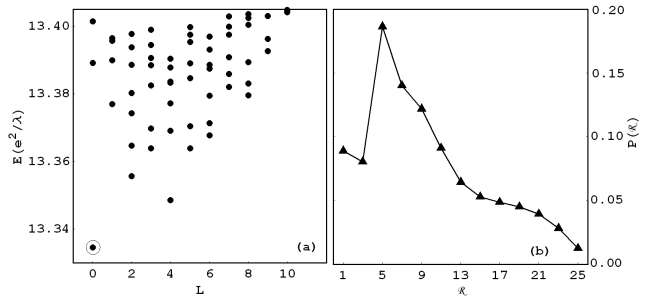}} \caption{(a) Spectrum of eleven
electrons at $2\ell=26$ in LL1. The ground state is an $L=0$ IQL
state. (b) $P(\mathcal R)$ vs. $\mathcal R$ for the IQL ground
state. It is clearly not a Laughlin correlated electron state
\cite{prlarxiv}.} \label{Fig:11el_2l_26_LL1}
\end{figure}

In Fig. \ref{Fig:11el_2l_26_LL1} (a) we display the spectrum of an
$N=11$ electron system at $2\ell=3N-7=26$ in LL1. The $L=0$ ground
state is separated from higher states by a clearly observable energy
gap. In frame (b) we show $P(\mathcal R)$ versus $\mathcal R$ for
this ground state. Again $P(\mathcal R)$ is neither a minimum at
$\mathcal R=1$ nor a maximum at $\mathcal R=3$, indicating that it
is not a Laughlin correlated electron state. Unfortunately, the
energy gap of the $\nu_1=1/3$ state for $6 \leq N \leq 12$ electron
system is not a smooth function of $N^{-1}$. Therefore we cannot
extrapolate to the macroscopic limit with any certainty. In
addition, no simple heuristic picture seems to describe the
correlations at $\nu_1=1/3$ for all values of $N$. Mixed clusters
(single electrons, pairs, triplets, etc.) treated by generalized CF
picture \cite{WojsSzYiQuinnPRB99} may be necessary for an intuitive
understanding of the correlations and elementary excitations at
$\nu_1=1/3$.

\begin{figure}
\centerline{\includegraphics [width =1
\linewidth]{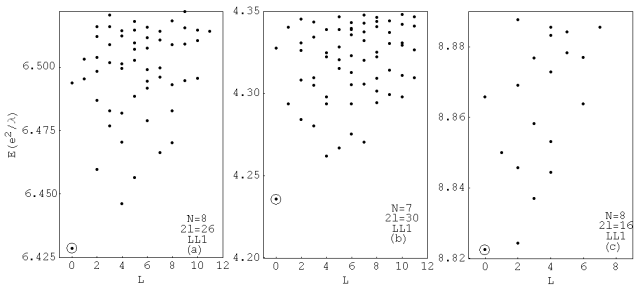}} \caption{Spectra for
$\nu_1=2/7$ (a), $1/5$ (b), and  $2/5$ (c) obtained using
$(N,2\ell)=(8,26)$, $(7,30)$, and $(8,16)$ respectively. Case (c)
has a very small gap and is not a robust IQL state. Case (a) and (b)
have bigger gaps and could persist in the macroscopic limit
\cite{prlarxiv}.} \label{Fig:Other_states_LL1}
\end{figure}

Because $V_1(\mathcal R)$ is not superharmonic at $\mathcal R=1$,
but it is at $\mathcal R=3$, we do not expect Laughlin correlated
electron (LCE) states for $1/2 \geq \nu_1 \geq 1/3$ where LCEs in
LL0 can form Laughlin-Jain states with $\nu=n(1+2n)^{-1}$ and $n$ an
integer. However, we do expect LCE states for $1/3 > \nu_1 \geq 1/5$
where electrons will avoid pair states with $\mathcal R=1$ and
$\mathcal R=3$, forming Laughlin-Jain states with
$\nu_1=n(1+4n)^{-1}$. In Fig. \ref{Fig:Other_states_LL1} we show
spectra obtained using the pseudopotential $V_1(\mathcal R)$
appropriate for a quantum well of zero width. The IQL states at
$\nu_1=1/5$ and $\nu_1=2/7$ are LCE states that can be understood
using Jain's CF$^4$ picture. $P(\mathcal R)$ is a minimum for
$\mathcal R=1$ and a maximum at $\mathcal R=5$ for each of these
states. For $\nu_1=2/5$ there is an extremely small gap between the
$L=0$ ground state and the lowest excited state. For this state
$P(\mathcal R)$ is a minimum at $\mathcal R=3$ and a maximum at
$\mathcal R=1$ and $\mathcal R=5$, implying 'pairing' rather than
Laughlin correlation between electrons.

\begin{figure}
\begin{center}
\centerline{
\includegraphics[width=0.5 \linewidth]{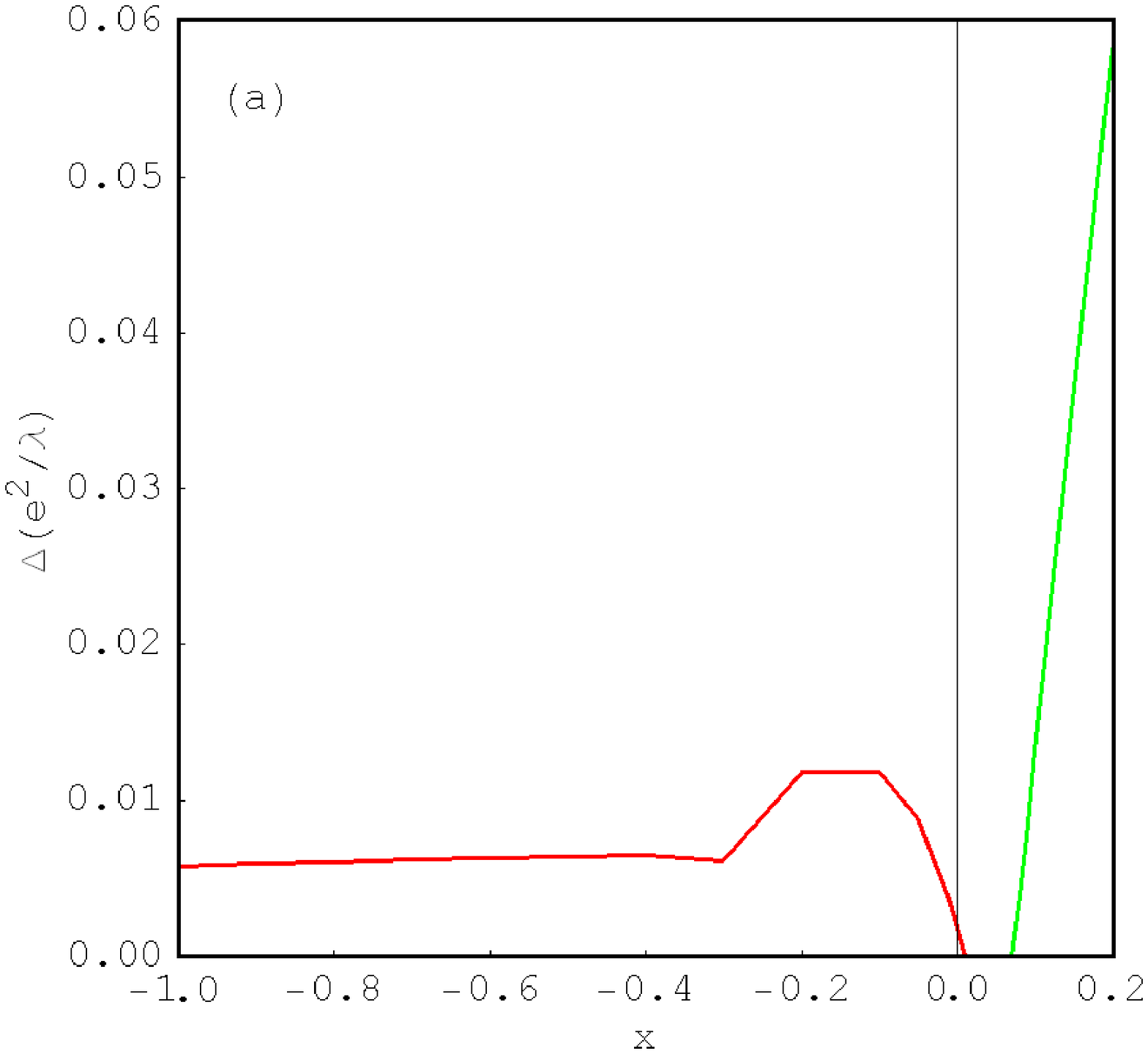}
\includegraphics[width=0.5 \linewidth]{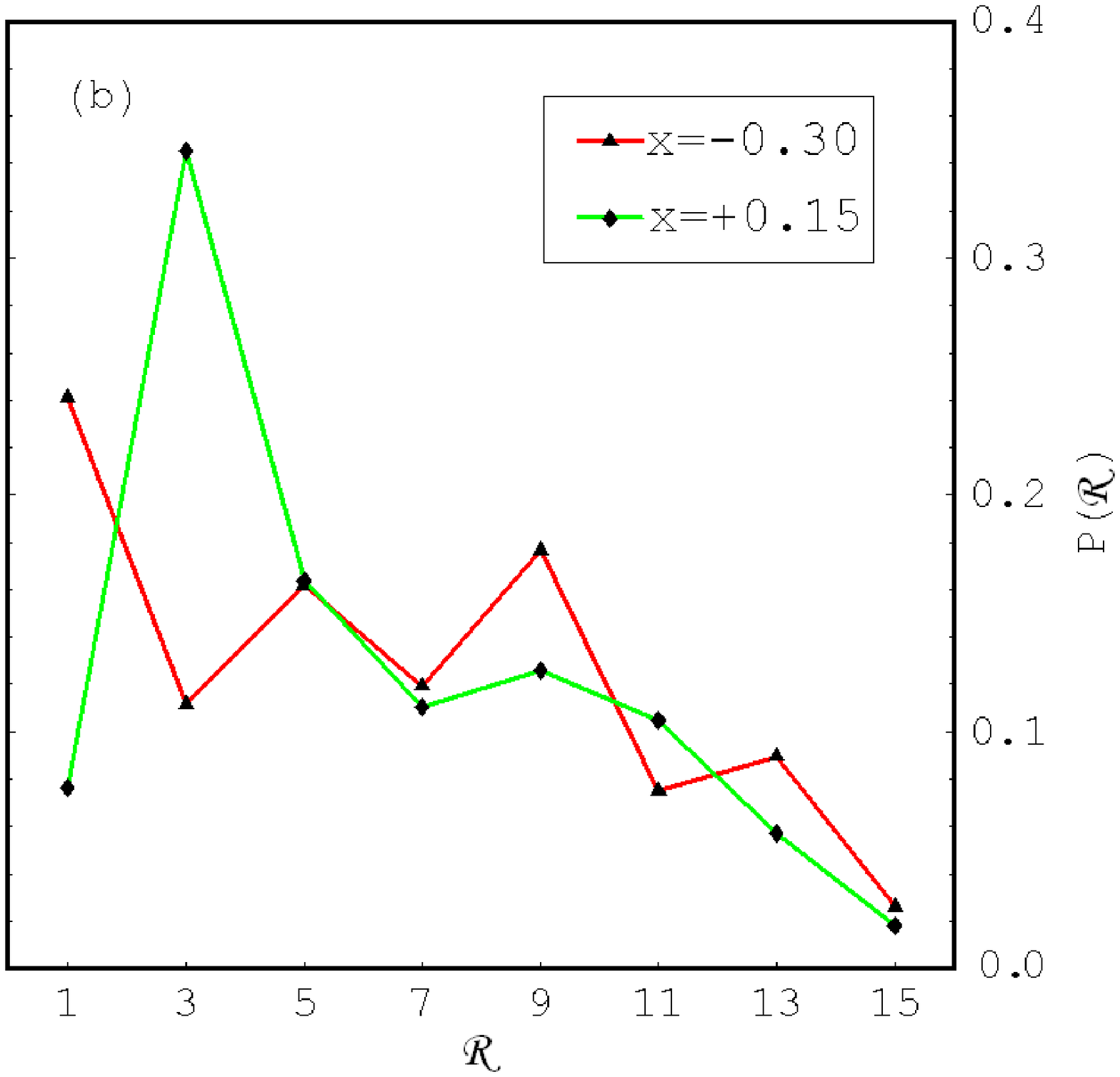}}
\caption{\label{Fig:2_5_gap_pr} (color online) (a) Energy gap vs.
$\delta V_1/V_1=x$. Remainder of pseudopotential $V_1(\mathcal R)$
(for $\mathcal R=3,5,\ldots$) is unchanged. (b) Sketch of pair
probability for $x=-0.3$ (red) and $x=0.15$(green) \cite{prlarxiv}.}
\end{center}
\end{figure}

We have studied the $\nu_1=2/5$ state (in the case $N=8$ and
$2\ell=16$) for the situation in which the pseudopotential
$V_1(\mathcal R)$ for a well of zero width is changed by an amount
$\delta V_1(\mathcal R)=xV_1(\mathcal R) \delta (\mathcal R,1)$
\cite{prlarxiv}. As shown in Fig. \ref{Fig:2_5_gap_pr} (a), a very
small gap $\Delta$ between $L=0$ ground state and the lowest excited
state is found for $x<-0.35$. The gap increases slightly with
increasing $x$, but begins to decrease for $x>-0.1$. It disappears
at $x \simeq +0.01$, but reappears at $x \gtrsim +0.08$ and then
increases roughly linearly with $x$. A plot of $P(\mathcal R)$
versus $\mathcal R$ is shown in Fig. \ref{Fig:2_5_gap_pr} (b) for
$x=-0.3$ (red) and $x=+0.15$ (green). Clearly the latter case is an
LCE state, while the former must contain $\mathcal R=1$ pairs. For
$x=0$, corresponding to the Coulomb pseudopotential in LL1, at most
a very small gap (associated with Laughlin correlations among
$\mathcal R=1$ pairs) can occur.

Our simple picture suggests that when the pseudopotential is
superharmonic at the value of  relative pair angular momentum
$\mathcal R$ to be avoided in a Laughlin correlated electron state,
Laughlin correlations occur and give rise to robust IQL ground
states at special values of $\nu$. When the pseudopotential is not
superharmonic, LCE states do not occur. Other kinds of correlations
(like formation of electron pairs or electron triplets) can occur,
but they result in weaker IQL states than LCE states. It is
well-known that Laughlin-Jain states at $\nu=n(1\pm 2pn)^{-1}$ are
the most robust FQH states in LL0, when $V_0(\mathcal R)$ is
superharmonic for $\mathcal R=1,3,5 \ldots$. For LL1, $V_1(\mathcal
R)$ is not superharmonic at $\mathcal R=1$. FQH states at
$\nu=1/3,1/2$, and $2/3$ can't be LCE states. They must involve
formation of clusters (pairs, triplets, etc.) despite the repulsive
nature of Coulomb interaction. Gaps are smaller than for the LCE
states. FQH states at $\nu_1=1/5$ and $2/7$ (and their e-h
conjugates at $4/5$ and $5/7$) are LCE states quite similar to
states of the same filling in LL0. The $\nu=2/5$ state cannot be an
LCE state. At most a very small gap, associated with correlations
between pairs of electrons, can occur. This picture is in excellent
qualitative agreement with the size of the energy gap determined
from thermally activated conductivity of the IQL states in LL0 and
LL1 \cite{choiPRB08}.

\subsection{Other Elementary Excitations of IQLs of $\nu=5/2$ IQL }
\label{sec:El_exc_others LL1}

It is clear that the correlations and the elementary excitations are
better understood for LL0 than for LL1 and higher Landau levels. In
LL0 the CF picture allows us to introduce $\ell^{\ast}=
|\ell-p(N-1)|$, where $p$ is an integer. Integral filling
$\nu^{\ast}=n$ ($n=1,2,3\ldots$) of the CF angular momentum shells
gives $L=0$ ground states at $\nu=n(2pn \pm 1)^{-1}$. The lowest
band of states will contain the minimum number of QP excitations
required by the values of $N$ and $2\ell$ \cite{ChenQuinnPRB93} The
QHs reside in the angular momentum shell
$\ell_{\rm{QH}}=\ell^{\ast}+n$; the QEs are in the shell
$\ell_{\rm{QE}}=\ell_{\rm{QH}}+1$. The CF picture describes the
lowest band of states for any value of the applied magnetic field.
The band containing two QEs (or two QHs) can be used to determine
(up to an overall constant) the pseudopotential
$V_{\rm{QP}}(L^{\prime})$ describing the pairwise interaction
between QPs of the Laughlin-Jain IQL states at $\nu=n(2pn\pm
1)^{-1}$. Higher bands of excitations contain one or more additional
QE-QH pairs. They are not as well defined as the lowest band,
overlapping at intermediate values of the allowed angular momentum.
However, most of the states predicted by simple CF picture are found
via numerical diagonalization.

For LL1, we do understand the correlations for $\nu_1=1/2$. They can
be described in terms of the formation of $N_{\rm{P}}=N/2$ pairs
when $N$ is even. The pair Landau level has a degeneracy
$g_{\rm{P}}$ twice that of the original electron LL. This increase
in degeneracy and decrease in particle number can lead to Laughlin
correlations among the pair giving rise to an IQL state of LCPs at
$2\ell=2N-3$. This was illustrated in Fig.
\ref{Fig:14_electrons_LL1} (a) and (c) where the lowest bands of
states contain two QP excitations in a Fermion pair excited LL of
angular momentum $\ell_{\rm{FPQP}}=3$ in frame (a) and two quasihole
excitations in a FP Landau level with $\ell_{\rm{FPQH}}=4$ in frame
(c). Some of these states were discussed by Greiter et al.
\cite{GreiterWenWilczekPRL91, GreiterWenWilczekNuclPhys92} but not
in terms of a generalized CF picture capable of predicting the
allowed values of $L$ in the lowest band of energy levels.

\begin{figure}
\centerline{\includegraphics[width=
\linewidth]{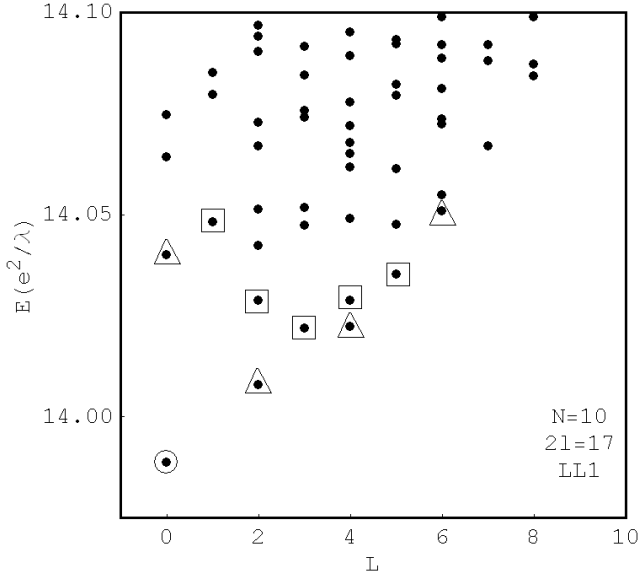}}  \caption{Energy spectrum
for $N=10$ electrons and $2\ell=17$ in LL1 for $V_1(\mathcal R)$
corresponding to zero width quantum well.}
\label{Fig:excitations_LL1}
\end{figure}

Not all the elementary excitations are QP pairs (occupying an
excited state FP LL) or QH pairs in the IQL state of Laughlin
correlated FPs. We have attempted to interpret spectra which contain
other kinds of excitations (e.g. unpaired electrons). In Fig.
\ref{Fig:excitations_LL1} we show the energy spectrum of a system
containing ten electrons in a shell of angular momentum $\ell=17/2$,
interacting through the Coulomb pseudopotential appropriate for LL1
in an ideal quantum well. In addition to $L=0$ ground state
corresponding to the IQL with $\nu=5/2$, there appear to be two low
lying bands with $L=0\oplus 2 \oplus 4 \oplus 6$ and $L=1 \oplus 2
\oplus 3 \oplus 4 \oplus 5$ respectively. We suggest that these
excitations can be identified using a slight generalization of the
composite Fermion picture applied to an intuitive guess at the
nature of excitations.

\subsection{Generalized CF Picture}

As we discussed earlier, the ground state in Fig.
\ref{Fig:excitations_LL1} should contain $N_{\rm{P}}=N/2$ pairs. In
the absence of correlations the pairs have a charge of $-2e$. If we
treat the pairs as Fermions, then the FP angular momentum is given
by $\ell_{\rm{FP}}=(2\ell-1)-3(N_{\rm{P}}-1)/2$. In low lying
excited states it is possible that one of the ground state pairs
breaks up into two unpaired electrons, each with charge $-e$ and
angular momentum $\ell$. We propose that the FPs and the unpaired
electrons have correlations among themselves and with one another.
We introduce the correlations in the standard CF way, by attaching
CS flux quanta (opposite to the dc magnetic field) to each particle
(both CF pairs and unpaired electrons).

We propose a generalized CF approximation to describe the
correlations using the following equations:
\begin{equation}
2\ell_{\rm{FP}}^{\ast}=2\ell_{\rm{FP}}-2p_{\rm{P}} (N_{\rm{P}}-1)-
2\gamma N_e~. \label{eq:FP_FP_e_corr}
\end{equation}
\begin{equation}
2\ell_{e}^{\ast}=2\ell_{e}-2p_e (N_e-1) - \gamma N_{\rm{P}}~.
\label{eq:e_FP_e_corr}
\end{equation}

It is straightforward to understand these correlations using the
following simple picture.
\begin{enumerate}
\item the effective CS charge on the composite Fermion pairs is thought of as ``red"
in color and that on the unpaired electrons as ``blue".
\item in Eq. \ref{eq:FP_FP_e_corr} $2p_{\rm{P}}$ ``red" and
$2\gamma$ ``blue" CS flux quanta are attached to each CF pair.
\item in Eq. \ref{eq:e_FP_e_corr} $2p_{e}$ ``blue" and
$\gamma$ ``red" CS flux quanta are attached to each unpaired
electron.
\item the CS charges sense only the CS flux quanta of the same
color, and no particle senses the flux attached to itself
\end{enumerate}
Thus Eq. \ref{eq:FP_FP_e_corr} tells us that the effective angular
momentum of one FP is decreased from $\ell_{\rm{FP}}$ by
$p_{\rm{P}}$ times the  number of other FPs and by $\gamma$ times
the number of unpaired electrons. Eq. \ref{eq:e_FP_e_corr} tells us
that the effective angular momentum of one unpaired electron is
decreased by $p_e$ times the number of other unpaired electrons and
by $\gamma /2$ times the number of CF pairs.

We know that this generalization of Jain's  mean field CF picture
results in exactly the same correlations as the adiabatic addition
of the CS flux, but that the latter approach needs no mean field
approximation. Note that $2p_{\rm{FP}}$ and $2p_e$ are even, and
that $\gamma$ can be odd or even. Adding $2\gamma$ ``blue" fluxes to
the CF pair causes the unpaired electron of the ``blue" charge to
have exactly the same $e-$CF pair correlations as adding $\gamma$
``red" fluxes sensed by the CF pair of ``red" charge  $-2e$ to the
unpaired electron. The CS charge times the CS flux must be the same
in step 2) and 3) to obtain the same correlations. Equations
\ref{eq:FP_FP_e_corr} and \ref{eq:e_FP_e_corr} define the
generalized CF picture in which different types of Fermions,
distinguishable from one another, experience correlations which
leave them as Fermions (since $2p$ is even) and give the same
correlations between members of two different species since the
product of CS charge and the CS flux added are the same (i.e $-e
\cdot 2\gamma = -2 e \cdot \gamma$).

If we apply the generalized composite Fermion (GCF) picture to Fig.
\ref{Fig:excitations_LL1} we know that the ground state has
$N_{\rm{P}}=5$ and $N_e=0$. Using GCF equation with $2p_e=4$ gives
$2\ell_{\rm{FP}}^*=4$, so that the $N_{\rm{P}}=5$ FPs fill the
$\ell_{\rm{FP}}^*=2$ shell giving $L=0$ IQL ground state. We can
think of two kinds of elementary excitations. First, one FP might be
promoted from $\ell_{\rm{FP}}^*=2$ shell (leaving an FP quasiholes
in this shell) into the $\ell_{\rm{FP}}^*+1$ shell (i.e we can
excite QEFP-QHFP with $\ell_{\rm{QEFP}}=3$ and
$\ell_{\rm{QHFP}}=2$). This would produce a band of states with
$1\leq L \leq 5$. Second, we could have an excited state with
$N_{\rm{FP}}=4$ and $N_e=2$ (i.e. one broken FP). This gives
$2\ell_{\rm{FP}}=23$, $2\ell_{\rm{FP}}^{\ast}=3$ and
$2\ell_{\rm{QE}}=7$ (when $p_{\rm{P}}=\gamma=2$ and $p_e=1$). The
four FPs fill the shell $\ell_{\rm{FP}}^{\ast}=3/2$ giving
$L_{\rm{FP}}=0$. The two QEs each with $\ell_{\rm{QE}}=7/2$ produce
the band $L=0\oplus 2\oplus 4\oplus 6$. This band is marked by
triangle in Fig. \ref{Fig:excitations_LL1}, while the FPQE-FPQH band
is marked by open squares going from $L=1$ to $L=5$. This
interpretation is suggestive, but not completely certain because we
know neither the QEFP-QHFP interaction nor the pseudopotential
$V_{\rm{QE}}(L')$ describing the interaction of a QE pair embedded
in an IQL state of four FPS. However the assignment of $L$ values
fits the numerical results for the low energy excited states with
$L\leq 6$.

It is worth noting that for the generalized CF picture
\cite{WojsSzYiQuinnPRB99} the correlations between a pair of
particles can be thought of as resulting from adiabatic addition of
fictitious CS flux quanta to one particle that is sensed by
fictitious charge on the other. The correlations among the particles
cause pairs to avoid the smallest pair orbits by introducing an
effective FP angular momentum $\ell_{\rm{FP}}^{\ast}$ and an
effective electron angular momentum $\ell_e^{\ast}$ given by Eqs.
\ref{eq:FP_FP_e_corr} and \ref{eq:e_FP_e_corr}. The allowed values
of the total angular momentum are obtained by addition of the
angular momenta of $N_{\rm{P}}$ identical correlated FPs, each with
angular momentum $\ell_{\rm{FP}}^{\ast}$ to obtain $L_{\rm{FP}}$,
the total FP angular momentum, and of $N_e$ identical correlated
electrons, each with angular momentum $\ell_e^{\ast}$ to obtain the
total electron angular momentum $L_e$. Then $L_{\rm{FP}}$ and $L_e$
are added as the angular momenta of distinguishable systems to
obtain the allowed total angular momentum values $L$ of the system.

Our interpretation is an attempt to understand some of the low lying
excitations of the $\nu_1=1/2$ state in a simple CF type picture. We
present the ideas here, even though they are not firmly established,
to motivate additional work on this important topic. We suggest
investigating other values of $N$ and $2\ell$ hoping that the
generalized CF type picture might fit numerical data and give us
better insight. The spectrum is more sensitive to small changes in
the pseudopotential $V_1(\mathcal R)$ than the spectrum in LL0 is to
small changes in $V_0(\mathcal R)$. Not understanding the
correlations at $\nu_1=1/3$ gives us, at the moment, no hope of
understanding the low energy excitations. We are still a long way
from knowing anything about the interactions between the elementary
excitations in that case.


\section{Model Pseudopotentials and Clusters of $j$ Particles}
\label{sec:model_pp_j}

\subsection{Energy of Clusters of $j$ Particles}

Thus far we have used the actual Coulomb pseudopotential describing
the interaction energy of a pair of electrons in the LL0 and LL1.
The pseudopotentials depend on the total pair angular momentum $L_2$
(or on $\mathcal R_2=2\ell-L_2$, where $\ell$ is the angular
momentum of the shell in which the electrons reside). They also
depend on Landau level index since the antisymmetric wavefunction
describing the relative motion of the pair is different for
different Landau levels. We have already noted that the energy of
the multiplet $|\ell^N;L\alpha>$ is given by

\begin{equation}
\label{eq:en_2p_cluster} E_{\alpha}(L)=\left(
\begin{array}{c}
  N \\
  2 \\
\end{array}
\right) \sum_{L_2} V(L_2) P_{L\alpha}(L_2)~.
\end{equation}
where $V(L_2)$ is the pair pseudopotential as a function of pair
angular momentum $L_2$, and $P_{L\alpha}(L_2)$ is the probability
that $|\ell^N;L\alpha>$ contains pairs with pair angular momentum
$L_2$. The sum in Eq. \ref{eq:en_2p_cluster} is over all allowed
values of $L_2=2\ell-\mathcal R_2$, where $\mathcal R=1,3,5 \cdots$.

For a cluster of $j$ particles, we can define $V(L_j,\beta_j)$ as
the interaction energy of the multiplet $|\ell^j;L_j\beta_j>$. It is
given by Eq. \ref{eq:en_2p_cluster} with $(N,L,\alpha)$ replaced by
$(j,L_j,\beta_j)$. Clearly one can write for the energy of
$|\ell^N;L\alpha>$
\begin{equation}
E_{\alpha}(L)=a_j\left(%
\begin{array}{c}
  N \\
  j \\
\end{array}%
\right) \sum_{L_j\beta_j}P_{L\alpha}(L_j,\beta_j)V(L_j,\beta_j)~.
\label{eq:E_j_part_int_a_j}
\end{equation}
Here $V(L_j,\beta_j)$ is the interaction energy of the electrons in
the multiplet $|\ell^j;L_j\beta_j>$ and $P_{L\alpha}(L_j\beta_j)$ is
the probability that the multiplet $|\ell^j;L_j\beta_j>$ appears in
the eigenfunction $|\ell^N;L\alpha>$
\cite{WojsQuinnPRB05,SimonRezayiCooperPRB07}. The coefficient $a_j$
is introduced to avoid overcountimg of the number of pairs. We can
use Eq. \ref{eq:en_2p_cluster} for $V(E_j,\beta_j)$  with
$|N,L,\alpha>$ replaced by $|j,L_j,\beta_j>$. Making use of identity
$P_{L\alpha}(L_2)=\sum_{L_j\beta_j}
P_{L_j\beta_j}(L_2)P_{L\alpha}(L_j\beta_j)$, and requiring Eq.
\ref{eq:E_j_part_int_a_j} to reduce to the results given by
\ref{eq:en_2p_cluster} gives us the value of $a_j$;
$a_j=(N-j)!(j-2)!/(N-2)!$. Thus we find:
\begin{equation}
E_{\alpha}(L)=\frac{N(N-1)}{j(j-1) }\sum_{L_j \beta_j}
P_{L\alpha}(L_j \beta_j)V(L_j,\beta_j)~.
\label{eq:E_j_part_cluster_pp}
\end{equation}
This result gives us the energy of $|\ell^N;L\alpha>$ in terms of
the energies $V(L_j\beta_j)$ of $j$ particle multiplets
$|\ell^j;L_j\beta_j>$.

It is worth recalling that when the pair pseudopotential $V(L_2)$ is
``harmonic" (i.e. $V(L_2)=A+BL_2(L_2+1)$, where $A$ and $B$ are
constants) the energies of the states $|\ell^j;L_j\beta_j>$ are
independent of the multiplet index $\beta_j$. Every state with the
same value of the $j$ particle angular momentum $L_j$ has the same
energy. In addition, the energy increases with $L_j$ as
$BL_j(L_j+1)$. This means that a harmonic $V_H(L_2)$ leads to a
harmonic $V_H(L_j)$ given by:
\begin{equation}
\label{eq:3p_harmonic_pseudopot} V_H(L_j)=A_j+BL_j(L_j+1)~.
\end{equation}
The coefficient $B$ is independent of $j$, and the constant $A_j$
gives an unimportant overall shift in the energy spectrum. Just as
$V_H(L_2)$, the harmonic pair pseudopotential causes no
correlations, $V_H(L_j)$, the harmonic pseudopotential of a
$j$-particle cluster also causes no correlations.

\subsection{Model Pseudopotentials}
\label{sec:LL1_Model Pseudopot}

Many authors
\cite{GreiterWenWilczekPRL91,GreiterWenWilczekNuclPhys92,
RezayiHaldanePRL00,WojsPRB01, WojsQuinnPRB05} have noted that the
most important pseudopotential coefficients are those with small
values of $\mathcal R$ ($\mathcal R=1,3,5,\ldots$) corresponding to
small pair separations. For example, if the ``superharmonic"
pseudopotential for LL0 is approximated by $V_0(\mathcal
{R}_2)=k\delta (\mathcal {R}_2,1)$ where $k>0$,the energy spectra
obtained in numerical diagonalization for $1/2 > \nu \geq 1/3$
filling factors are in excellent qualitative agreement with those
obtained using the full Coulomb pseudopotential
\cite{WojsPRB01,QuinnVarenna}. By this we mean that IQL states with
gaps proportional to $k$ occur at the values of $2\ell$ predicted by
Jain's CF picture, in agreement with the numerical results for both
Coulomb and model pseudopotentials.

This fact and the behavior of the leading pseudopotential
coefficients for  electrons in LL0 and LL1, and for CFQEs in CFLL1
have led to the introduction of a model two particle pseudopotential
\cite{WojsPRB01,WojsQuinnPRB05}
\begin{equation}
V_{\alpha}(\mathcal R)=(1-\alpha) \delta (\mathcal
R,1)+\frac{\alpha}{2}\delta (\mathcal R,3)~. \label{eq:model_pp}
\end{equation}
This model pseudopotential mimics the short range behavior of the
Coulomb pseudopotential in LL0 if $\alpha=0$, and in LL1 if $\alpha$
is approximately equal to $1/2$. It also mimics the QE-QE
pseudopotential of QEs of the Laughlin $\nu=1/3$ state (i.e CFLL1
where these QEs reside) if $\alpha$ is approximately equal to unity.
Of course, any harmonic contribution to the model potential can be
added to the Eq. \ref{eq:model_pp} without any effect on the
correlations.

Greiter et al. \cite{GreiterWenWilczekPRL91,
GreiterWenWilczekNuclPhys92} introduced a model three particle
pseudopotential that is equivalent to
\begin{equation}
\label{eq:pp_GWW_eq} V(\mathcal R_3)=\delta(\mathcal R_3,3)
\end{equation}
Here $L_3$, the total angular momentum of a three particle cluster,
is given by $L_3=3\ell-\mathcal R_3$. $\mathcal R_3$ is called the
three particle relative angular momentum. GWW showed that their
three particle pseudopotential, which forbids occurrence of compact
three particle droplets with $\mathcal R_3=3$, had the Moore-Read
\cite{MooreReadNuclPhys91} Pfaffian state as an exact solution. MR
proposed the Pfaffain wavefunctions based on correlations functions
used in conformal field theory in order to look for an explanation
for the IQL state observed at filling factor $\nu=5/2$. The proposed
Moore-Read Pfaffian wavefunction is to the GWW model pseudopotential
exactly what the Laughlin $\nu=1/3$ wavefunction is to the short
range two particle pseudopotential obtained by taking $\alpha=0$ in
Eq. \ref{eq:model_pp}. The $L=0$ IQL ground state of the Moore-Read
wavefunction occurs at $2\ell=2N-3$ (and its e-h conjugate
$2\ell=2N+1$). If the value of $2\ell$ is increased, QHs of zero
excitation energy appear in the IQL state. When a number of
degenerate QH states occur at the same value of total angular
momentum $L$ (e.g if $N_{\rm{QH}}=4$ and $\ell_{\rm{QH}}=5$ there
will be 3 degenerate zero energy states at $L=4 \oplus 2 \oplus 0$),
any normalized linear combination of degenerate states of
$N_{\rm{QH}}$ quasiholes is a perfectly good eigenstate. According
to Read and Rezayi \cite{ReadRezayiPRB99} this can lead to new
linear combinations of states under permutation of quasiholes that
can give rise to non-Abelian statistics. Non-Abelian quasiparticles
are of great current interest. Their existence appears to depend on
the very special form of the GWW three particle potential (or on
similar pseudopotential that vanish for relative angular momentum
larger than some value). Neither the actual Coulomb pair
pseudopotential nor the model pseudopotential given by
\ref{eq:model_pp} with $\alpha$ approximately equal to one half give
rise to sets of zero energy eigenstates with the same total angular
momentum. Although the idea of non-Abelian quasiparticles is very
intriguing and of potential value in quantum computing, it isn't yet
clear that such quasiparticles occur in systems with realistic
pseudopotentials.

\subsection{Model Three Body Pseudopotential}

We know that Eq. \ref{eq:model_pp} mimics the behavior of the short
range part of the two particle pseudopotentials for the lowest
Landau level when $\alpha=0$, for the first excited Landau level
when $\alpha\simeq 1/2$, and for the quasielectron of a Laughlin
$\nu=1/3$ state when $\alpha\simeq 1$. The energy spectra obtained
using these short range pseudopotentials give reasonably good
agreement with those obtained using the full pseudopotentials when
filling factor $\nu$ is between $1/3$ and $1/2$.

According to Section \ref{sec:model_pp_j} A
\begin{equation}
\label{eq:model_3p_pp} V(L_j,\beta_j)
=\frac{j(j-1)}{2}\sum_{L_12}V(L_{12})P_{L_j\beta_j}(L_{12})
\end{equation}
gives the energy of a $j$-particle cluster in the multiplet
$|\ell^j;L_j\beta_j>$ in terms of the pair pseudopotential and the
probability that $|\ell^j;L_j\beta_j>$ has pairs with pair angular
momentum $L_{12}$. If short range interactions dominate in
determining the correlations, then we can hope that the small values
of $\mathcal R_3=2\ell-L_3$ in $V(L_3,\beta_3)$ are the important
ones in determining the correlations. We have used
\ref{eq:model_pp}, the model short range pair pseudopotential, to
determine the $V(L_3,\beta_3)$ that it produces through Eq.
\ref{eq:model_3p_pp}. For $\mathcal R\leq 8$, there is only a single
allowed multiplet for each value of $\L_3=3\ell-\mathcal R_3$. For
$\mathcal R_3 \leq 8$, we can think of $V(\mathcal R_3)$ as a short
range three particle pseudopotential produced by the pair potential
given by Eq. \ref{eq:model_pp}. If short range interactions are the
important ones for determining correlations, we can simply set
$V(L_3)=0$ for $\mathcal R_3>8$, expecting this portion of the
interaction to have little effect on the correlations.
\begin{figure}
\centerline{\includegraphics[width=
\linewidth]{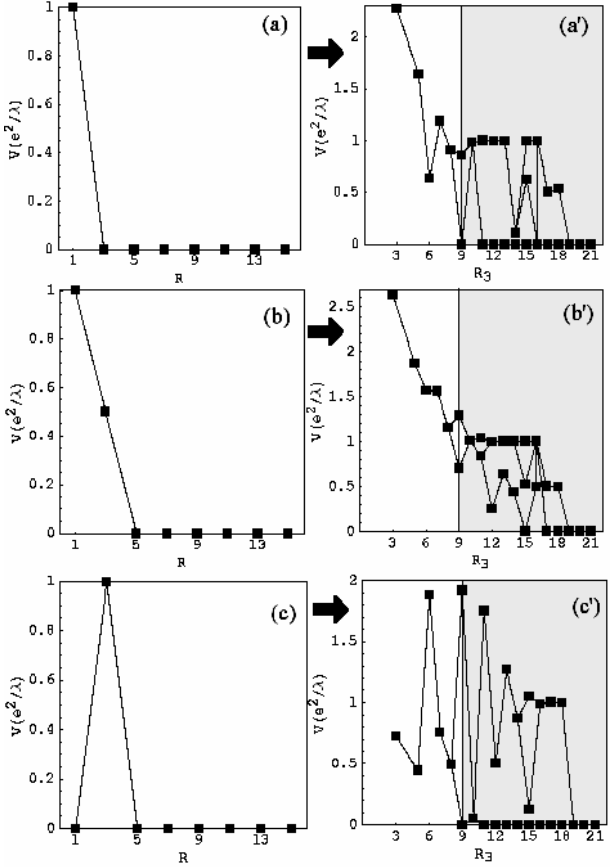}} \caption{The three particle
pseudopotentials $V(\mathcal R_3)$ (a', b', c') produced from the
model pair pseudopotentials $V(\mathcal R)$ (a,b,c)}
\label{Fig:Model2b_3b_pp}
\end{figure}

Fig. \ref{Fig:Model2b_3b_pp} shows the short range model pair
potential for $\alpha=0, 1/2$, and $1$ (frames a,b,c respectively).
Frames a', b', c' show the short range pseudopotential resulting
from Eq. \ref{eq:model_3p_pp} with $j=3$. It is worth noting that in
case b', $V(\mathcal R_3)$ is slightly superharmonic at $\mathcal
R_3=3$ and has no strong maxima excepts at $\mathcal R=3$ and no
minima for $\mathcal R_3<9$. In contrast, in c' $V(\mathcal R_3)$
has maxima at $\mathcal R_3=3$ and $6$, and a deep minima at
$\mathcal R_3=5$. These two three-body pseudopotentials are quite
different, from one another and from frame a'. It seems unlikely
that the electrons in LL1 and QEs of the Laughlin $\nu=1/3$ state
have the same correlations. The fact that frame b' has weak
superharmonic behavior at $\mathcal R_3=3$ might support the idea of
using the GWW pseduopotential as the simplest (one parameter
$V(\mathcal R_3=3) \neq 0$) three particle pseudopotential for LL1.

One can easily generalize the very special GWW three particle
pseudopotential to larger clusters. For $j$ Fermions in a shell of
angular momentum $\ell$, the maximum allowed value of the total $j$
particle angular momentum is $L_j^{MAX}=j(\ell-\frac{j-1}{2})$. This
means that the minimum allowed value of $\mathcal R_j=j\ell-L_j$,
the relative angular momentum of a $j$ particle cluster is,
$\mathcal R_j=j\ell-L_j^{MAX}=j(j-1)/2$. A model $j$ particle
pseduopotential given by $V(\mathcal R_j)=k\delta(\mathcal R_j,
\mathcal R_j^{MIN})$ eliminates compact $j$ particle clusters, just
as the three particle GWW pseudopotential eliminated three particle
compact droplets. The eigenstates of this model $j$ particle
pseudopotential are referred as 'parafermion' states
\cite{ReadRezayiPRB99}. It is not clear whether these simple
parafermion states give a reasonable approximation to those obtained
with a realistic pseudopotential.


\section{Spin Polarized Quasiparticles in a Partially Filled Composite Fermion Shell}
\label{sec:QE}

\subsection{Heuristic Picture}

In Section \ref{sec:anharmonicity} we demonstrated that the simplest
repulsive anharmonic pseudopotential $V(\mathcal R_2)=V_H(\mathcal
R_2)+k \delta (\mathcal R_2,1)$ caused the lowest energy state for
each value of the total angular momentum $L$ to be Laughlin
correlated. For a spin polarized LL0 with $1/3 \leq \nu \leq 2/3$
such a potential (superharmonic at $\mathcal R=1$) gives rise to the
Laughlin-Jain sequence of integrally filled CF levels with
$\nu_{\pm}=n(2n \pm 1)^{-1}$, where $n$ is an integer. No
interaction between the CFs is required. The gaps causing the IQL
state are associated with the energy needed to create a QE-QH pair
in the CF angular momentum shells. Haldane \cite{HaldanePRL83}
suggested that if the highest occupied CF level is only partially
filled, a gap could result from the residual interactions between
the QPs, in the same way that the original gap resulted from the
electron interactions. However, this would require
$V_{\rm{QP}}(\mathcal R)$ to be ``superharmonic" at $\mathcal R=1$
to give rise to Laughlin correlations. In Section \ref{sec:res_int}
we showed that in a Laughlin $\nu=1/3$ or $1/5$ state
$V_{\rm{QE}}(\mathcal R)$ was not superharmonic at $\mathcal R=1$
and $\mathcal R=5$, and that $V_{\rm{QH}}$ was not at $\mathcal
R=3$. This means that many of the novel IQL states observed by Pan
et al. \cite{PanStormerTsuiPRL03} have to result from correlations
among the electrons that are quite different from the Laughlin
correlations.

\begin{figure}
\centerline{\includegraphics[width=
\linewidth]{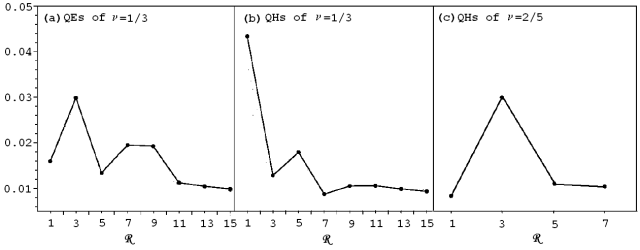}} \caption{$V_{\rm{QE}}(\mathcal
R)$ and $V_{\rm{QH}}(\mathcal R)$ for (a) QEs of $\nu=1/3$ state (b)
QHs of $\nu=1/3$ state, and (c) QHs of $\nu=2/5$ state.}
\label{Fig:QE_QH_psedopotentials}
\end{figure}

Just as electrons in LL1 tend to form clusters (pairs with pair
angular momentum $\ell _P=2 \ell -1$ or larger clusters), we expect
QPs in CF LL1 to tend to form pairs or larger clusters. The major
differences between electrons in LL1 and QPs in CF LL1 are: (i) the
pseudopotential $V_1(L^{\prime})$ for electrons in LL1 (shown in
Fig. \ref{fig:Pseudopot}) is an increasing function of $L^{\prime}$,
but it is not superharmonic at $\mathcal R=1$, while $V_{\rm{QE}}(L
^{\prime})$ is strongly subharmonic having a maximum at $\mathcal R
=2\ell-L ^{\prime}=3$ and minima at $\mathcal R=1$ and 5 and (ii)
the e-h symmetry of LL1 is not applicable to QEs and QHs in CF LL1
\cite{WojsPRB01_1}. The QEs are quasiparticles of the Laughlin
$\nu=1/3$ IQL state, while QHs in CF LL1 are actually quasiholes of
the Jain $\nu=2/5$ state. The QE and QH pseudopotentials in frames
a) and c) are similar, but not identical as shown in Fig.
\ref{Fig:QE_QH_psedopotentials}. The QEs of $\nu=1/3$ state and QHs
of the $\nu=2/5$ state reside in CF LL1. The QHs of the $\nu=1/3$
state reside in CF LL0.

The experimental results of Pan et al. \cite{PanStormerTsuiPRL03}
suggest that the novel $\nu=4/11$ IQL ground state is fully spin
polarized. In numerical studies testing their CF hierarchy picture,
Sitko et al. \cite{SitkoYiYiQuinnPRL96,SitkoYiQuinnPRB97} found that
this spin polarized IQL state did not occur at $\nu_{\rm{QE}}=1/3$.
They suggested that the reason for this was related to the
difference between $V_{\rm{QE}}(L^{\prime})$, the QE pseudopotential
describing the residual interactions between the CFQEs, and
$V_0(L^{\prime})$, the QE pseudopotential describing the
interactions between electrons in LL0. It was later shown that
because $V_{\rm{QE}}(L^{\prime})$ is not superharmonic at
$R^{\prime} \equiv 2\ell-L^{\prime}=1$, the CF picture could not be
reapplied to interacting QEs in the partially filled CF shell
\cite{WojsQuinnPRB00}. This led to the suggestion
\cite{ParkJainPRB00} that the QEs forming the daughter state had to
be spin reversed and reside in CF LL0 as quasielectrons with reverse
spin (QERs). Szlufarska et al. \cite{SzWojsQuinnPRB01_1} evaluated
$V_{\rm{QER}}(L^{\prime})$, the pseudopotential of QERs. They showed
that $V_{\rm{QER}}(L^{\prime})$ was superharmonic at $\mathcal R=1$,
so that unlike majority spin QEs, they could support Laughlin
correlations at $\mathcal R=1$.

This leaves at least two possible explanations of the $\nu=4/11$ IQL
state. It could be a Laughlin correlated daughter state of spin
reversed QEs (i.e. QERs), or it could be a spin polarized state in
which the QEs form pairs or larger clusters. In a later section we
compare the energies of these two states. The total energies involve
the QE (or QER) energies, the interaction energies of the QEs (or
QERs), and the Zeeman energy. For the moment let's look at the
completely polarized case.

The simplest idea is exactly that used for electrons in LL1. There,
the $\nu_1=1/2$ state could be attributed to the formation of pairs
with $\ell_{\rm{P}}=2\ell-1$, where $\ell$ is the angular momentum
of the shell occupied by electrons. If we assume that the QEs form
pairs and treat them as Fermions, then Eq. \ref{eq:FP_FP_e_corr} and
Eq. \ref{eq:e_FP_e_corr} give us the relation between the
``effective FP angular momentum" $\ell_{\rm{FP}}$, and the QE
angular momentum $\ell$, and the relation between the ``effective FP
filling factor $\nu_{\rm{FP}}$, and the QE filling factor $\nu$. If
we take $\nu_{\rm{FP}}$ equal to $m^{-1}$, where $m$ is an odd
integer, we can obtain the value of $\nu_{\rm{QE}}$ corresponding to
the Laughlin correlated state of FPs (pairs of quasielectrons with
$\ell_{\rm{P}}=2\ell-1$). Exactly the same procedure can be applied
to QHs in CF LL1 since $V_{\rm{QE}}(\mathcal R)$ and
$V_{\rm{QH}}(\mathcal R)$ are qualitatively similar at small values
of $\mathcal R$. Here we are assuming that $V_{\rm{QE}}(\mathcal R)$
and $V_{\rm{QH}}(\mathcal R)$ are dominated by their short range
behavior $\mathcal R \leq 5$.  The QH pseudopotential is not as well
determined for $\mathcal R>5$ because it requires larger $N$
electron systems then we can treat numerically. The electron filling
factor is given by $\nu^{-1}=2+(1+\nu_{\rm{QE}})^{-1}$ or by
$\nu^{-1}=2+(2-\nu_{\rm{QH}})^{-1}$. This results in the values of
$\nu$ shown in the Table \ref{tab:nu_QE_nu} for $2/3 \geq
\nu_{\rm{QP}} \geq 1/3$.

\begin{table}
 \caption{Values of $\nu_{\rm{FP}}=m^{-1}$ for $m=3,5,7$, and 9 and
 the resulting values of
 $\nu_{\rm{QE}}$, $\nu_{\rm{QH}}$ and the electron filling factor that they
generate.}
 \label{tab:nu_QE_nu}
\begin{tabular}{|c|c|c|c|c|}
  \hline
  $\nu_{\rm{FP}}^{-1}$ & 3 & 5 & 7 & 9 \\
  \hline
  $\nu_{\rm{QE}}$ & 2/3 & 1/2 & 2/5 & 1/3 \\
  \hline
  $\nu$ & 5/13 & 3/8 & 7/19 & 4/11 \\
 \hline\hline
  $\nu_{\rm{QH}}$(CF LL1) & 2/3 & 1/2 & 2/5 & 1/3 \\
 \hline
  $\nu$ & 4/11 & 3/8 & 8/21 & 5/13 \\
  \hline
\end{tabular}
\end{table}

The states generated at the values of $\nu_{\rm{QE}}$ and
$\nu_{\rm{QH}}$ equal to $2/5$ have not been observed. Clear IQL
states were observed by Pan et al. at $\nu=3/8$ and $\nu=4/11$. A
somehow weaker IQL state at $\nu=5/13$ is also observed.

The daughter states generated by QPs in CF LL1 (QEs of the parent
$\nu=1/3$ Laughlin state or QHs of the parent $\nu=2/5$ Jain state)
give rise to filling factors for the electron system with $\nu>1/3$.
QHs of the $\nu=1/3$ Laughlin state (residing in CF LL0) can also
form daughter states, and they result in electron filling factor
$\nu$ in the range $1/3 > \nu \geq 1/5$. The pseudopotential for
these QHs is superharmonic at $\mathcal R=1$ and has a strong
minimum at $\mathcal R=3$. Because of this, if they form pairs, the
pairs must have angular momentum $\ell_{\rm{P}}=2\ell-3$ (instead of
$\ell_{\rm{P}}=2\ell-1$ for QE pairs). Eq. \ref{eq:FP_FP_e_corr} and
\ref{eq:e_FP_e_corr}must be modified. We then replace $2\ell-1$ in
Eq. \ref{eq:FP_FP_e_corr} by $2\ell-3$, and $\gamma_{\rm{F}}$ by
$\tilde\gamma_{\rm{F}}$. The value of $\tilde\gamma_{\rm{F}}$ is
determined by requiring that $\nu_{\rm{FP}}=1$ when
$\nu_{\rm{QH}}=1/2$. This condition results from the fact that the
pairs are formed by two QHs separated by two filled CF states. The
resulting value of $\tilde\gamma_{\rm{F}}$ is 7, so Eq.
\ref{eq:nu_FP_nu} is replaced by:
\begin{equation}
\nu_{\rm{FP}}^{-1}=4\nu_{\rm{QH}}^{-1}-7~. \label{eq:nu_FP_nu_QH}
\end{equation}
The QH daughter states resulting from Laughlin correlated
$\rm{QH}_2$ (pairs of QHs of the $\nu=1/3$ state) and the electron
filling factor satisfying $\nu^{-1}=2+(1-\nu_{\rm{QH}})^{-1}$ are
given in Table \ref{tab:nu_QH_nu}.

\begin{table}
  \caption{Values of $\nu_{\rm{QH}}$ satisfying $1/3 > \nu_{\rm{QH}} \geq 1/5$ and
  the resulting electron filling factors $\nu$ for LC $\rm{QP}_2$s with $\nu_{\rm{FP}}^{-1}=7,9,11,13$}
  \label{tab:nu_QH_nu}
\begin{tabular}{|c|c|c|c|c|}
  \hline
  $\nu_{\rm{FP}}^{-1}$ & 7 & 9 & 11 & 13 \\
  \hline
  $\nu_{\rm{QH}}$ & 2/7 & 1/4 & 2/9 & 1/5 \\
  \hline
  $\nu$ & 5/17 & 3/10 & 7/23 & 4/13 \\
  \hline

\end{tabular}
\end{table}

\subsection{Numerical Studies of Spin Polarized QP States}


Standard numerical calculations for $N_e$ electrons are not useful
for studying such new states as $\nu=4/11$, because convincing
results require large values of $N_e$. Therefore we take advantage
of the knowledge \cite{WojsQuinnPRB00,SitkoYiYiQuinnPRL96,
LeeScarolaJainPRL01, LeeScarolaJainPRB02} of the dominant features
of the pseudopotential $V_{\rm{QE}}(\mathcal R)$ of the QE-QE
interaction (i.e, the QE-QE interaction energy $V_{\rm{QE}}$ as a
function of relative pair angular momentum $\mathcal R$), and
diagonalize the (much smaller) interaction Hamiltonian of
$N_{\rm{QE}}$ systems. This procedure was earlier
\cite{SitkoYiYiQuinnPRL96} shown to reproduce accurately the low
energy $N_e$-electron spectra at filling factors $\nu$ between $1/3$
and $2/5$. It was also used in a similar, many-QE calculation by Lee
et al. \cite{LeeScarolaJainPRL01,LeeScarolaJainPRB02} (who, however
have not found support for QE clustering).

\begin{figure}
\centerline{\includegraphics[width=
\linewidth]{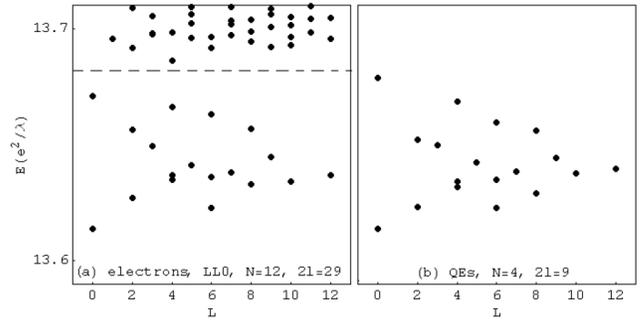}} \caption{Energy spectra for $N
= 12$ electrons in the lowest LL with $2\ell = 29$ and for $N = 4$
QEs in the CF LL1 with $2l = 9$. The energy scales are the same, but
the QE spectrum obtained using $V_{\rm{QE}}(\mathcal R)$ is
determined only up to an arbitrary constant.}
\label{Fig:12e2l29_4QE2l9}
\end{figure}

One might question whether using the pair pseudopotential for QPs
obtained by diagonalization of a finite system of $N$ electrons
(containing two QEs or two QHs) gives a reasonable accurate
description of systems containing more than a few QPs. We have
attempted to account for finite size effects
\cite{WojsQuinnPRB00,WojsPRB01_1,SzWojsQuinnPRB01_1,WojsWodzinskiQuinnPRB06}
by plotting the values of $V_{\rm{QP}}(\mathcal R)$ for each value
of $\mathcal R$ as a function of $N^{-1}$, where $N$ is the number
of electrons in the system that produced the two CFQPs \cite{He}. We
extrapolate $V_{\rm{QP}}(\mathcal R)$ to the macroscopic limit. In
addition, the low energy spectra of an $N$ electron system (obtained
by diagonalization of $V_0(\mathcal R)$) that contains $N_{\rm{QP}}$
quasiparticles is compared with the spectrum of $N_{\rm{QP}}$
quasiparticles [obtained by diagonalization of $V_{\rm{QP}}(\mathcal
R)$]. The results for $(N,2\ell)=(12, 29)$ and $(N_{\rm{QE}},2
\ell_{\rm{QE}})=(4,9)$ are shown in Fig. \ref{Fig:12e2l29_4QE2l9}.

\begin{figure}
\centerline{\includegraphics[width=
\linewidth]{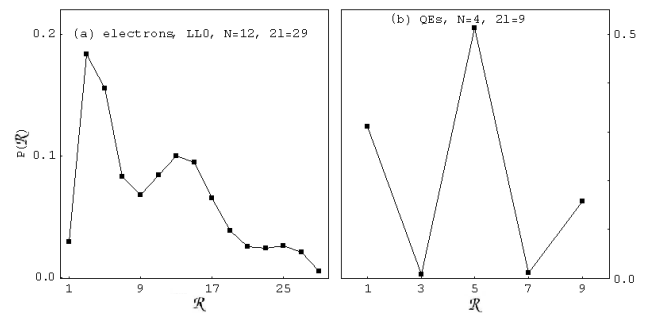}} \caption{Pair probability
functions $P(\mathcal R)$ for the two ground states shown in Fig.
\ref{Fig:12e2l29_4QE2l9}.} \label{Fig:12e2l29_4QE2l9_pr}
\end{figure}

A CF transformation on the $(N,2\ell)=(12,29)$ electron system gives
an effective CF angular momentum $\ell^{\ast}$ satisfying
$2\ell^{\ast}=2\ell-2(N-1)=7$. Eight of the $12$ CFs fill the shell
$\ell^{\ast}=7/2$, leaving the four CFQEs in the shell of
$\ell^{\ast}=9/2$. Although the four QE spectrum is not identical to
the low energy band of the $12$ electron spectrum, it is clearly a
rather good approximation. Both spectra have an $L=0$ ground state,
but the gaps are somewhat different in size. The fact that
$(2\ell_{\rm{QE}}, N_{\rm{QE}})$ system has an $L=0$ ground state at
$2\ell_{\rm{QE}}=3N_{\rm{QE}}-3$ led a number of researchers
\cite{SmetNature,GoerbigPhysicaE,GoerbigPRB69,LopezFradkinPRB04,
ChangJainPRL04} to suggest that it represented a second generation
of CFs giving rise to a daughter state and resulting $\nu=4/11$ spin
polarized IQL state observed by Pan et al.
\cite{PanStormerTsuiPRL03}. This idea can't be correct because
$V_{\rm{QE}}(L^{\prime})$ is not superharmonic at $\mathcal R=1$ and
cannot cause a Laughlin correlated CF daughter state of spin
polarized QEs. In Fig. \ref{Fig:12e2l29_4QE2l9_pr}, we show the pair
probability $P(\mathcal R)$ versus $\mathcal R$ for the
$(N,2\ell)=(12,29)$ electron ground state and for the
$(N_{\rm{QE}},2\ell_{\rm{QE}})=(4,9)$ QE daughter state. Because the
later has $P(\mathcal R)$ with maxima at $\mathcal R=1$ and $5$, and
minima at $\mathcal R=3$ and $7$ it is certainly not a Laughlin
correlated state of QEs. In fact it is a $\nu_{\rm{QE}}=1/2$ state
at $2\ell_{\rm{QE}}=2N_{\rm{QE}}+1$ (the conjugate of
$2\ell_{\rm{QE}}=2N_{\rm{QE}}-3$).

The fact that the magnitude of $V_{\rm{QE}}(\mathcal R)$ is only
about one fifth as large as the energy necessary to create an
additional QE-QH pair in a Laughlin correlated state permits
diagonalization in the subspace of the partially filled QE Landau
level with reasonably accurate results (see e.g, Fig.
\ref{Fig:12e2l29_4QE2l9}). For situations in which the width of the
band of two QP states is closer to the energy needed to create a
QE-QH pair higher bands (or higher QP LL) cannot be neglected. In
such cases the pseudopotential $V(L_j\beta_j)$, describing the
interaction energy of a multiplet $|\ell^j;L_j\beta_j>$ containing
$j$ particles in a state of total angular momentum $L_j$, may be
useful in accounting for the large Hilbert space needed for a more
accurate diagonalization (and more accurate description of
correlations) of a many particle system.

The value of $2\ell$ at which the IQL state at filling factor $\nu$
occurs in the spherical geometry is given by
$2\ell=\nu^{-1}N+\gamma(\nu)$, where $N$ is the number of particles
and $\gamma(\nu)$ is a finite size effect shift \cite{HaldanePRL83}.
For Laughlin correlated electrons in LL0 at filling factor $\nu$
equal to the inverse of an odd integer, $\gamma(\nu)=-\nu^{-1}$, so
that the $\nu=1/3$ IQL states occur at $2\ell=3N-3$. For
quasielectrons of Laughlin $\nu=1/3$ state, an IQL state occurs at
$(N,2\ell)=(4,9)$.

As mentioned earlier, we believe that because QEs will not support
Laughlin correlations at $\nu=1/3$, it is an ``aliased" state
\cite{MorfPRL98,MorfdAmbrumenilDasSarmaPRB02} at $2\ell=2N+1$
(conjugate to $2\ell=2N-3$) that supports pairing correlations. By
"aliased" states we mean two states with the same values of $N$ and
$2\ell$ that belong to different sequences
$2\ell=\nu{^(-1)}N+\gamma(\nu)$. Different values of $\gamma(\nu)$
for IQL states of electrons in LL0 and QEs in CFLL1 suggest that the
QE correlations are different from the Laughlin correlations for
electrons in LL0. It also gives emphasis to how important is to
carry out numerical calculations for many different values of
$2\ell$ for each value of the particle number $N$, instead of
assuming the value of assuming the value of $\gamma(\nu)$ at which
the IQL state is expected.

The essential information about the interaction of particles
confined to some Hilbert space can be obtained by defining the value
of interaction energy for all allowed pair states. For charged
particles confined to an LL in the presence of a magnetic field, the
relative motion is strongly quantized. The orbital pair eigenstates
can be labeled with a single discrete quantum number, relative
angular momentum $\mathcal R$. This number is a non-negative
integer; it must be odd (even) for a pair of identical Fermions
(Bosons), and it increases with increasing average distance
$\sqrt{<r^2>}$ between the two particles.

The pair interaction energy of two QPs, the QP pseudopotential
$V_{\rm{QP}}(\mathcal R)$, determines the correlations between QPs.
On a spherical surface, $\mathcal R \leq 2 \ell$, where $\ell$ is
the angular momentum of the QP shell. Thus the number of
pseudopotential parameters is finite. However, even in an infinite
(planar) system, only those few leading parameters at the values of
$\mathcal R$ corresponding to the average distance not exceeding the
correlation length $\xi$ are of significance (provided that the
correlations are indeed characterized by finite $\xi \sim \lambda$).


\subsection{Numerical Spectra}

\begin{figure}
\centerline{\includegraphics[width= \linewidth]{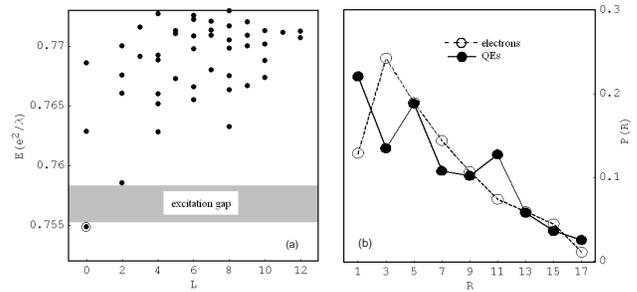}}
\caption{(a) Energy spectra as a function of total angular momentum
$L$ of 10 QEs at $2\ell=2N-3=17$ corresponding to
$\nu_{\rm{QE}}=1/2$ and $\nu=3/8$. It is obtained in exact
diagonalization in terms of individual QEs interacting through the
pseudopotential shown in Fig. \ref{fig:Pseudopot} (c) (triangles)
(b) Coefficient of $P(\mathcal R)$, the probability associated with
pair states of relative angular momentum $\mathcal R$, for the
lowest $L=0$ state. The solid dots are for 10 QEs of the
$\nu_{\rm{QE}}=1/2$ state in a shell of angular momentum
$\ell=17/2$. The open circles are for 10 electrons in the LL0 at
$\ell_0=17/2$ \cite{QuinnVarenna}.} \label{Fig:10QE_2l_17}
\end{figure}

We begin with numerical results for the spectrum $E_{\alpha}(L)$ and
the ground state pair probability $P(\mathcal R)$ of a system of $N$
quasielectrons in a shell of angular momentum $\ell=17/2$. $\mathcal
R$ is the relative angular momentum of a pair $\mathcal
R=2\ell-L^{\prime}$, where $L^{\prime}$ is the total angular
momentum of the pair. We observe in Fig. \ref{Fig:10QE_2l_17} (a)
that there is an $L=0$ ground state separated by a gap from the
lowest excited states.

The spectrum is obtained by exact diagonalization (within the
subspace of CF LL1) of the $N$ quasielectron system interacting
through the pseudopotential coefficients $V_{\rm{QE}}(\mathcal R)$
represented in Fig. \ref{fig:Pseudopot} (c)  by triangles. The solid
circles in Fig. \ref{Fig:10QE_2l_17} show $P(\mathcal R)$ the
probability that a QE pair has relative angular momentum $\mathcal
R$ in the $L=0$ ground state.  The solid circles in Fig.
\ref{Fig:10QE_2l_17} (b) show $P(\mathcal R)$, the probability that
a QE pair has relative angular momentum $\mathcal R$ in the $L=0$
ground state. The open circles in Fig. \ref{Fig:10QE_2l_17} (b)
show, for contrast, $P(\mathcal R)$ for ten electrons in LL0
interacting via the Coulomb interaction, The maxima in $P(\mathcal
R)$ at $\mathcal R=1$ and 5 and the minimum at $\mathcal R=3$ for
the QE system are in sharp contrast to the Laughlin correlated
$P(\mathcal R)$ of the ten electron system in LL0. The QE maximum at
$\mathcal R=1$ and minimum at $\mathcal R=3$ suggests formation of
QE pairs with $\ell_{\rm{P}}=2\ell-1$ and the avoidance of pairs
with $\mathcal R=2\ell-L^{\prime}=3$, the pair state with the
largest repulsion. This IQL ground state occurs at $2\ell=2N-3$ and
corresponds to $\nu_{\rm{QE}}=1/2$ and $\nu=3/8$.

For $\nu_{\rm{QP}}=1/2$ state should occur at the conjugate value of
$2\ell$ given by $2\ell=2N-3$ and $2N+1$. Thus Fig.
\ref{Fig:10QE_2l_17} can be thought of as $N_{\rm{QP}}=10$ or
$N_{\rm{QP}}=8$, the former corresponding to $2\ell=2N-3$ and the
latter to $2\ell=2N+1$. We have already mentioned that QEs in the
CFLL1 are Laughlin QEs of the $\nu=1/3$ IQL, while QHs in the CFLL1
are QHS of the Jain $\nu=2/5$ state. It seems reasonably to
diagonalize $V_{\rm{QP}}(\mathcal R)$ for QHs when CFLL1 is more
than half filled and for QEs when it is less than half filled. If
only $V_{\rm{QP}}(\mathcal R)$ for $\mathcal R \leq 5$ is important,
$V_{\rm{QE}}(\mathcal R)$ and $V_{\rm{QH}}(\mathcal R)$ are
qualitatively similar (but not identical). We should then expect the
same correlations independent of which $V_{\rm{QP}}(\mathcal R)$ is
used in the numerical diagonalization. This  would suggest that Fig
\ref{Fig:10QE_2l_17} be interpreted as containing $N_{\rm{QH}}=8$
and $2\ell=2N_{\rm{QH}}+1=17$ instead of as $N_{\rm{QE}}=10$ and
$2\ell=2N-3=17$.

\begin{figure}
\centerline{\includegraphics[width=
\linewidth]{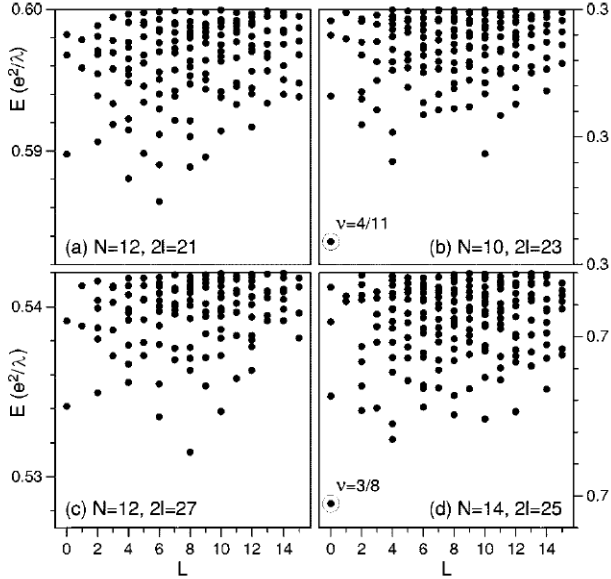}} \caption{Energy spectra
of up to $N=14$ QEs in LL shells with various degeneracies
$2\ell+1$, calculated using the pseudopotentials shown in Fig.
\ref{fig:Pseudopot} (c) \cite{WojsYiQuinnPRB04}.}
\label{Fig:En_spectra_quasiel_gen}
\end{figure}

In Fig. \ref{Fig:En_spectra_quasiel_gen} we display spectra for
$N=10,12$, and 14 QEs in angular momentum shells with various values
of $\ell$ ($2\ell =21,23,25$, and 27). In frames (a) and (c) the
ground states occur at total angular momentum $L=6$ and $L=8$
respectively. They are not IQL ground states. Frames (b) and (d)
have $L=0$ ground states separated from the low energy excitations
by a substantial gap. For frame (b) $2\ell=3N-7$, and for frame (d)
$2\ell=2N-3$. These are the values for which we find QE daughter
states with $\nu_{\rm{QE}}=1/3$ and $\nu_{\rm{QE}}=1/2$
respectively. Using $\nu^{-1}=2+(1+\nu_{\rm{QE}})^{-1}$ gives the
value $\nu=4/11$ and $\nu=3/8$ for these two states. It is worth
mentioning that the $N_{\rm{QE}}$ quasielectron systems used in
studying the energy spectra in Fig. \ref{Fig:En_spectra_quasiel_gen}
arise from much larger electron systems after a CF transformation
was applied. For example it can be seen that frame (d) results for
$(N_e, 2\ell_e)=(38,97)$ by noting that
$2\ell^{\ast}=2\ell_e-2(N_e-1)=23$. This lowest CF shell can hold
$2\ell^{\ast}+1=24$ of the CFs; the remaining 14 become CFQEs in CF
LL1 with angular momentum
$2\ell^{\ast}_{\rm{QE}}=2(\ell^{\ast}+1)=25$. This $(N_e,
2\ell_e)=(38,97)$ is far too large to study numerically, but
$(2\ell_{\rm{QE}},N_{\rm{QE}})=(25,14)$ can be handled without
difficulty.

We have calculated similar $(2\ell,N)$ spectra for up to 14 QEs at
filling factors $\nu_{\rm{QE}}\sim N/(2\ell+1)$ between $1/2$ and
$1/3$. Note that the assignment of the filling factor to a finite
system $(2\ell,N)$ is not trivial, and it depends on the form of
correlations. As discussed earlier, the $(2\ell,N)$ sequences that
correspond in the thermodynamic limit to a filling factor $\nu$ are
described by a linear relation,
\begin{equation}
2\ell=\nu^{-1} N -\gamma(\nu)~, \label{eq:N_2l_shift_gen}
\end{equation}
where the ``shift" $\gamma(\nu)$ depends on the microscopic nature
of the many body state causing incompressibility at this $\nu$. For
example, the sequence of finite-size nondegenerate ($L=0$) ground
states that extrapolates to $\nu=1/3$ occurs at $2\ell=3N-3$ for the
Laughlin state, at $2\ell=3N-5$ for the Laughlin correlated state of
Fermion pairs
\cite{WojsPRB01_1,QuinnWPhysE02,QuinnWojsYiPLA03,WojsQuinnPol03} and
at $2\ell=3N-7$ for the incompressible QE state identified below.

In Table \ref{tab:gap_QEs} we present the excitation gaps obtained
for the QE systems with various values of $N$ and $2\ell$. The table
should be symmetric under the replacement of $N$ by $2\ell+1-N$
which reflects the particle-hole symmetry in a partially filled QP
shell (i.e., in CF LL1). This symmetry is only approximate in real
systems. The largest of the gaps $\Delta$ (those shown in boldface)
occur for the following two sequences: $2\ell=3N-7$,  $2N-3$ (and
its $e-h$ conjugate $2\ell=2N+1$), corresponding to
$\nu_{\rm{QE}}=1/3$ and $1/2$. Using Eq. \ref{eq:nu_gen_CF}, these
values can be converted to the electron filling factors $\nu=3/8,
4/11$, and $5/13$ .

\begin{figure}
\centerline{\includegraphics[width=
\linewidth]{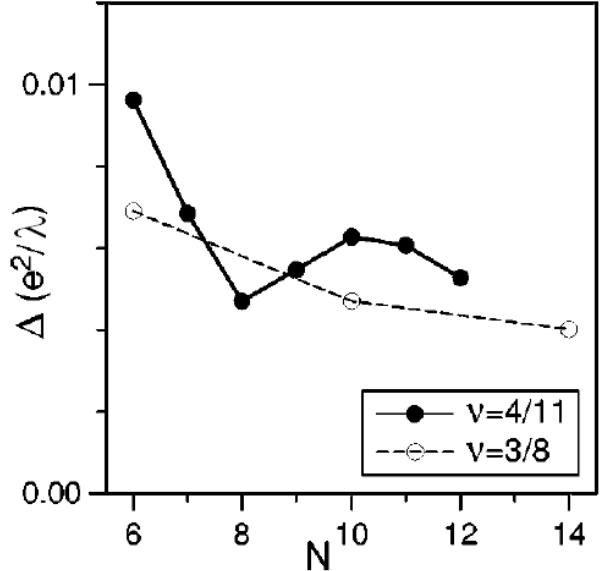}} \caption{Excitation gap $\Delta$
for the $\nu_{\rm{QE}}=\frac{1}{3}$ series of $N$ QE ground states
at $2\ell=3N-7$ (full dots) and for the $\nu_{\rm{QE}}=\frac{1}{2}$
series at $2\ell=2N-3$ (open circles), plotted as a function of the
QE number, $N$ \cite{WojsYiQuinnPRB04}.} \label{Fig:Gap_QE_states}
\end{figure}

The dependence of the excitation gaps $\Delta$ on the QE number $N$
for the $\nu_{\rm{QE}}=1/3$ series at $2\ell=3N-7$ (full dots) and
for the $\nu_{\rm{QE}}=1/2$ series at $2\ell=2N-3$ (open circles) is
plotted in Fig. \ref{Fig:Gap_QE_states}. It is difficult to
extrapolate accurately our finite-size data to the thermodynamic
limit to predict the magnitude of $\Delta$ in an infinite (planar)
system. However, we believe that the series of finite-size
nondegenerate ground states  occurring at $2\ell=2N+1$ (or
$2\ell=2N-3$) describe the FQH state observed experimentally at
$\nu=3/8$. The series at $2\ell=3N-7$ is less certain. It shows
large oscillations over the limited range of $N$ values for which we
can calculate, and we do not know if this series persists to the
thermodynamic limit.

\begin{table}
\caption{Excitation gaps $\Delta$ in units of $10^{-3}e^2/\lambda$,
above the nondegenerate ($L=0$) ground state of $N$ QEs each with
angular momentum $\ell$, interacting through pseudopotential in Fig.
\ref{fig:Pseudopot} (c). Circles $\circ$ mark degenerate ($L\neq 0$)
ground states. The value in the boldface are the largest; they all
belong to the three $(N,2\ell)$ sequences corresponding to
$\nu_{\rm{QE}}=\frac{1}{2},\frac{1}{3}$, and $\frac{2}{3}$.}
\label{tab:gap_QEs}
\begin{tabular}{c|c c c c c c c c c c c c c }
  \hline\hline
  $N^{2\ell}$ & 17 & 18 & 19 & 20 & 21 & 22 & 23 & 24 & 25 & 26 & 27 & 28 & 29 \\
  \hline
  8 & \bf{4.71} & $\circ$ & $\circ$ & $\circ$ & 0.01 &  &  &  &  &  &  &  &  \\
  9 & $\circ$ & $\circ$ & $\circ$ & \bf{5.47} & $\circ$ & $\circ$ & $\circ$ & 1.18 &  &  &  &  &  \\
  10 & \bf{4.71} & $\circ$ & $\circ$ & $\circ$ & $\circ$ & $\circ$ & \bf{6.29} & $\circ$ & 0.81 & $\circ$ & $\circ$ &  &  \\
  11 & $\circ$ & $\circ$ & $\circ$ & $\circ$ & $\circ$ & $\circ$ & $\circ$ & $\circ$ & $\circ$ & \bf{6.07} & $\circ$ & $\circ$ & $\circ$ \\
  12 &  & $\circ$ & $\circ$ & \bf{5.47} & $\circ$ & $\circ$ & 0.37 & $\circ$ & \bf{4.02} & $\circ$ & $\circ$ & $\circ$ & \bf{5.28} \\
  13 &  &  &  & $\circ$ & $\circ$ & $\circ$ & $\circ$ & $\circ$ & $\circ$ & $\circ$ & $\circ$ & $\circ$ & $\circ$ \\
  14 &  &  &  &  & 0.01 & $\circ$ & \bf{6.29} & $\circ$ & \bf{4.02} & $\circ$ & $\circ$ & $\circ$ & $\circ$ \\
  15 &  &  &  &  &  &  & $\circ$ & $\circ$ & $\circ$ & $\circ$ & $\circ$ & $\circ$ & $\circ$ \\
  16 &  &  &  &  &  &  &  & 1.18 & 0.81 & \bf{6.07} & $\circ$ & $\circ$ & $\circ$ \\
  17 &  &  &  &  &  &  &  &  &  & $\circ$ & $\circ$ & $\circ$ & $\circ$ \\
  18 &  &  &  &  &  &  &  &  &  &  & $\circ$ & $\circ$ & \bf{5.28} \\
  \hline\hline
\end{tabular}
\end{table}


In our numerical studies the $\nu_{\rm{QP}}=1/2$ state occurs only
when the number of QPs is even, suggesting that QP pairs are formed.
However, IQL states are formed only when the number of minority QPs
in CFLL1 is 8 or 12, but not when it is 10 or 14. This could
indicate that the CF pairs form quartets (i.e. pairs of CF pairs) in
the IQL state. This is completely speculative since we have very
little knowledge of the pseudopotential describing the interaction
between CF pairs. For $\nu_{\rm{QE}}=1/3$, a gap occurs at all
values of $N$ between 6 and 12. Even values of $N$ can be made up of
pairs; $N=8$ and 12 can give states containing quartets (pairs of
pairs); $N=6$, 9, and 12 could contain triplets. We have not yet
attempted to explore IQL states containing clusters of different
sizes (single QEs, CF pairs, triplets, etc.) that would be needed to
obtain IQLs at $N=7$ and $11$. The effect of different cluster sizes
might be responsible for large variations in the gap for
$\nu_{\rm{QE}}=1/3$ with $N$.

The ``shift" defined by Eq. \ref{eq:N_2l_shift_gen} and describing
the $2\ell =3N-7$ sequence identified here ($\gamma=7$) is different
not only from $\gamma=3$ describing a Laughlin state, but also from
$\gamma=5$ that results for a Laughlin state of Fermion pairs. This
precludes the interpretation of these finite-size
$\nu_{\rm{QE}}=1/3$ ground states found numerically (and also of the
experimentally observed $\nu=4/11$ FQH state) as a state of Laughlin
correlated pairs  of QEs (i.e., particles in the partially filled CF
LL1). However, it is far more surprising that paired state of QEs
turns out as an invalid description for these states as well.
Clearly, the correlations between the pairs of QEs at
$\nu_{\rm{QE}}=1/3$ must be of a different, non-Laughlin type, and
we do not have an alternative explanation for the incompressibility
of this state.

While we do not completely understand the correlations between QEs
at $\nu_{\rm{QE}}=1/3$, it may be noteworthy that the value of
$\gamma=7$ appropriate for the series of incompressible states found
here can be obtained for the Laughlin state of QE triplets
(QE$_3$s), each with the maximum allowed angular momentum,
$L=3\ell-3$, or of quartets (made up of pairs of pairs) with maximum
allowed angular momentum of the quartet $\ell_{\rm{Q}}=4\ell-10$.
The quartet state can be thought of as consisting of four filled
states ($\ell,\ell-1,\ell-4,\ell-5$) separated by two empty states
($\ell-2,\ell-3$). Both of these heuristic pictures give
$2\ell=3N-7$ for the $\nu=1/3$ state.


\subsection{Results from Model Interactions}


In this section we present the results of similar calculations,
obtained using the model pseudopotential given by Eq.
\ref{eq:model_pp}. It is known \cite{WojsPRB01_1,QuinnWPhysE02} that
the correlations characteristic of electrons in the partially filled
LL0 and LL1 are accurately reproduced by $V( \mathcal R_2)$ given by
Eq. \ref{eq:model_pp} with $\alpha \approx 0$ and $1/2$,
respectively. Similarly, by the comparison of pair amplitudes, we
have confirmed that this model pseudopotential with $\alpha \approx
1$ causes correlations characteristic of QEs in their partially
filled LL. We have repeated the diagonalization of a few finite
systems with $2\ell=2N-3$ (or $2N+1$) and $3N-7$, for $\alpha$
varying between 0 and 1, in order to answer the following two
questions. First, to what extent is the stability of the identified
$\nu=3/8$ and $4/11$ states affected by the (width dependent)
details of the QE-QE interaction? And second, does a phase
transition occur for values of $\alpha$ between $1/2$ and $1$,
indicating a different origin of the incompressibility of the $\nu
=3/8$ and $4/11$ states and their electron counterparts (in LL1) at
$\nu=5/2$ and $7/3$? The latter question is naturally motivated by
our observation that the $2\ell=2N+1$ sequence of nondegenerate
ground states occurs only for $ N=8$ and $12$, in contrast to the
situation in LL1 where they occurred for any integral value of
$N/2$.

\begin{figure}
\centerline{\includegraphics[width=
\linewidth]{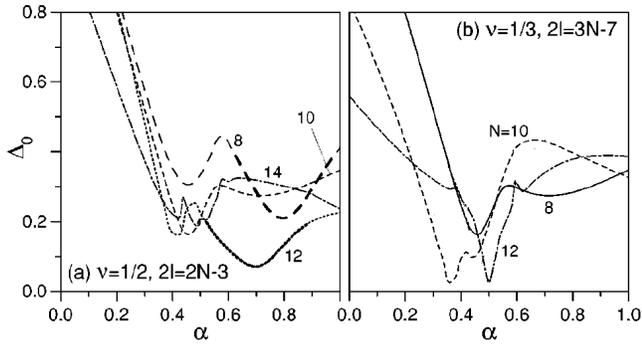}} \caption{The excitation gap
$\Delta_0$ between the lowest and the first excited states in the
$L=0$ subspace of $N$ particles on Haldane sphere with the values of
$2\ell$ corresponding to $\nu=\frac{1}{2}$ (a) and $\nu=\frac{1}{3}$
(b), plotted as a function of the interaction of parameter $\alpha$
defined by Eq. \ref{eq:model_pp} \cite{WojsYiQuinnPRB04}.}
\label{Fig:Gap_modelpp_L0}
\end{figure}


In Fig. \ref{Fig:Gap_modelpp_L0} we plot the $L=0$ excitation energy
gap $\Delta_0$ (difference between the two lowest energy levels at
L=0), as a function of $\alpha$. A minimum in $\Delta_0(\alpha)$
suggests a (forbidden) level crossing, i.e., a phase transition in
the $L=0$ subspace. Such minima occur near $\alpha=1/2$ for all
values of $N$ and for both $2\ell=2N-3$ and $3N-7$. They reveal
destruction of Laughlin correlations that occur for small $\alpha$
(e.g., for electrons in LL0) and formation of incompressible
$\nu=1/2$ and $1/3$ states of a different (paired) character that
occur for $\alpha \approx 1/2$ (e.g., for electrons in LL1). In Fig.
\ref{Fig:Gap_modelpp_L0} (a), similar strong minima occur at $\alpha
\approx 0.7$ for $N=8$ and $12$ (marked with thick lines). This is
consistent with our observation that the correlations between the
QEs and between the electrons in LL1 (both at the half filling) are
different. In Figs. \ref{Fig:Gap_modelpp_L0} (a) and (b), additional
weaker minima between $\alpha=1/2$ and 1 appear also for other
combinations of $N$ and $2\ell$. This confirms that the $\nu=1/2$
and $1/3$ incompressible states of QEs are generally different from
those of the electrons in LL1, despite the fact that they both
usually occur at the same values of $2\ell=2N+1$ and $3N-7$ in the
finite systems.

\begin{figure}
\centerline{\includegraphics[width=
\linewidth]{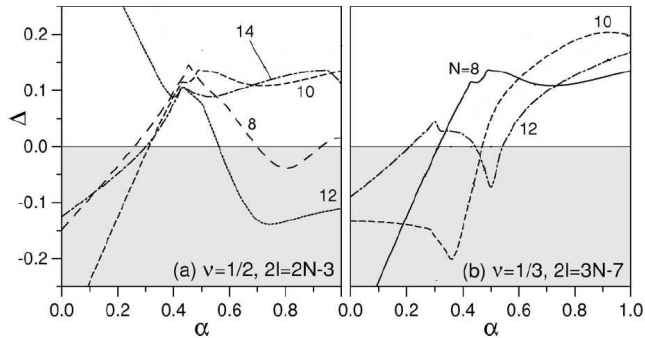}} \caption{The excitation gap
$\Delta$ from the lowest state with $L=0$ to the lowest state with
$L \neq 0$  for $N$ particles on Haldane sphere with values of
$2\ell$ corresponding to $\nu=\frac{1}{2}$ (a) and $\nu=\frac{1}{3}$
(b), plotted as a function of the interaction of parameter $\alpha$
defined by Eq. \ref{eq:model_pp} \cite{WojsYiQuinnPRB04}.}
\label{Fig:Gap_modelpp_gen}
\end{figure}

The absolute excitation gaps $\Delta(\alpha)$ of the $L=0$ ground
states (difference between the lowest energies at $L \neq 0$ and
$L=0$) are shown in Fig. \ref{Fig:Gap_modelpp_gen}. The negative
value of $\Delta$ means that the absolute ground state is degenerate
(~i.e., $L\neq 0$), and the abrupt changes in the slope of
$\Delta(\alpha)$ occur whenever level crossings occur for the lowest
$L\neq0$ state. Clearly, except for $N=8$ and $12$ with
$2\ell=2N-3$, the lowest $L=0$ states remain the absolute ground
states of the system in the whole range of $\alpha$ between $1/2$
and $1$. This shows that the incompressibility of the
$\nu_{\rm{QE}}=1/2$ and $1/3$ ground states will not be easily
destroyed in experimental systems by a minor deviation from the
model QE-QE pseudopotential used here in the numerical
diagonalization.

\begin{figure}
\centerline{\includegraphics[width=
\linewidth]{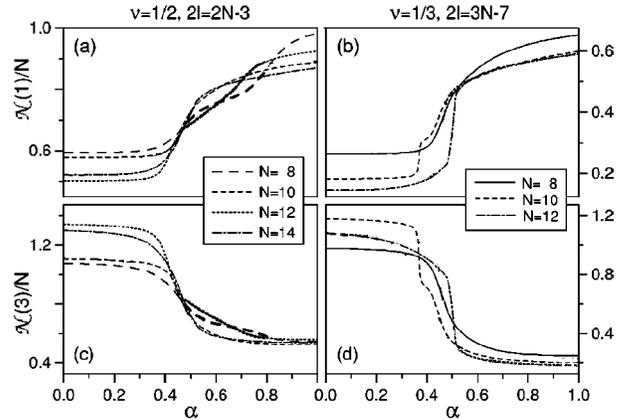}} \caption{The average number
of pairs with relative angular momentum $\mathcal R=1$ (a,b), and
$\mathcal R=3$ (c,d) per particle, $\mathcal N(\mathcal R)/N$,
calculated for the lowest state in the $L=0$ subspace of $N$
particles on Haldane sphere with values of $2\ell$ corresponding to
$\nu=1/2$ (a,c), and $\nu=1/3$ (b,d) plotted as a function of the
interaction of parameter $\alpha$ defined by Eq. \ref{eq:model_pp}
\cite{WojsYiQuinnPRB04}.} \label{Fig:Pairs_model_pp}
\end{figure}

Let us finally examine the dependence of the leading pair
amplitudes, $P(1)$ and $P(3)$, on $\alpha$. In Fig.
\ref{Fig:Pairs_model_pp} we plot the number of pairs, $\mathcal
N(\mathcal R)=\frac{1}{2} N(N-1) P(\mathcal R)$ divided by $N$. A
transition from Laughlin correlations at $\alpha=0$ to pairing at
$\alpha=1/2$ and possibly grouping into larger clusters at $\alpha
\sim 1$ is clearly visible in each curve. It is also confirmed that
just as the Laughlin ground state remains virtually insensitive to
the exact form of the interaction pseudopotential $V_e$ as long as
it is strongly superharmonic at short range, the correlations in the
$\nu_{\rm{QE}}=1/2$ and $1/3$ states  are quite independent of the
details of the QE-QE interaction, as long as $V_{\rm{QE}}$ is
strongly subharmonic at short range. This result supports our
expectation that the incompressible QE ground states found here
numerically indeed describe the FQH $\nu=3/8$ and $4/11$ electron
states observed in experiment. On the other hand, correlations at
$\alpha \approx 1/2$ (electrons in LL1), characterized by having
$P(1)\approx P(3)$, are quite different from those at $\alpha \sim
1$ (QEs), characterized by having the minimum possible $P(3)$, much
smaller than $P(1)$. Finally, with thick lines in Fig.
\ref{Fig:Pairs_model_pp} (a) we have marked the curves for $N=8$ and
$12$ in the vicinity of $\alpha=0.7$ at which the forbidden
crossings were found in Fig. \ref{Fig:Gap_modelpp_L0} (a). A
different behavior of $\mathcal N(1)/N$ and $\mathcal N(3)/N$ for
these two values of $N$ is clearly visible.

\subsection{Unresolved Questions}

We have demonstrated by direct calculation of the pair amplitudes
$P(\mathcal R)$ that, at sufficiently large filling factor
($\nu_{\rm{QE}} \geq 1/3$), the QEs form pairs or larger clusters,
with a significant occupation of the minimum relative pair angular
momentum, $\mathcal R=1$. The QE (and analogous QH) clustering is an
opposite behavior to Laughlin correlations characterizing, e.g.,
electrons partially filling LL0. Therefore it invalidates the
reapplication of the CF picture to the individual QEs or QHs (and
thus also the equivalent multiflavor CF model) and precludes the
simple hierarchy interpretation of any incompressible states at $1/3
\leq \nu_{\rm{QP}} \leq 2/3$. The series of finite-size
nondegenerate ground states at QE filling factors $\nu_{\rm{QE}}=1/2
, 1/3,$ and $2/3$ have been identified. These values correspond to
the electronic filling factors $\nu=3/8 , 4/11 ,$ and $5/13$, at
which the FQH effect has been experimentally discovered
\cite{PanStormerTsuiPRL03}. Due to a discussed similarity between
the QE-QE and QH-QH interactions, these three QE states have their
QH counterparts at $\nu_{\rm{QH}}=1/4 , 1/5,$ and $2/7$,
corresponding to $\nu=3/10, 4/13$, and $5/17$, all of which have
also been experimentally observed \cite{PanStormerTsuiPRL03}.

The finite-size $\nu_{\rm{QE}}=1/2$ states of QEs (CFs in LL1) are
found at the same values of $2\ell=2N-3$  (and its conjugate) as the
$\nu=5/2$ \cite{MooreReadNuclPhys91,RezayiHaldanePRL00,
GreiterWenWilczekPRL91, GreiterWenWilczekNuclPhys92,
MorfPRL98,MorfdAmbrumenilPRL95,MorfdAmbrumenilDasSarmaPRB02}),
despite the different electron and CF pseudopotentials. This is also
true for the $\nu_{\rm{QE}}=1/3$ state at $2\ell=3N-7$ and the
$\nu=7/3$ IQL in LL1. Therefore we have studied the dependence of
the wavefunctions and stability of the novel FQH states on the exact
form of interaction at short range. We found several indications
that the novel QE states are distinctly different from the electron
states in LL1: (i) the $\nu_{\rm{QE}}=1/2$ state appears
incompressible only for the even values of $N/2$, where $N$ is the
number of minority QPs; (ii) the pair-correlation functions
$P(\mathcal R)$ are quite different; (iii) although they remain
incompressible, the ground states appear to undergo phase
transitions when the QE-QE pseudopotential is continuously
transformed into that of electrons in LL1. However, further studies
are needed to understand these transitions. On the other hand, weak
dependence of the wave functions and excitation gaps of the novel
FQH states on the details of the QE-QE interaction, as long as it
remains strongly subharmonic at short range, justifies the use of a
model pseudopotential in the realistic numerical calculation.

We have also explored an idea \cite{HalperinHPA83,QuinnWojsYiPLA03,
WojsYiQuinnPRB04} of the formation of Laughlin states of QE pairs
(QE$_2$s). An appropriate composite Fermion model has been
formulated and shown to predict a family of novel FQH states at a
series of fractions including all those observed in experiment.
However, several observations strongly point against this simple
model: (i) Our best estimate of the QE$_2$-QE$_2$ interaction
pseudopotential is not superharmonic to support Laughlin
correlations of QE$_2$ (except possibly for $\nu_{\rm{QE}}=1/2$);
(ii) the values of $2\ell$ predicted for finite $N$ are different
from these obtained from the numerical diagonalization (except for
$\nu_{\rm{QE}}=1/2$ ); (iii) the numerical results do not confirm
the significance of parity of the number of QEs in finite systems
(the $\nu_{\rm{QE}}=1/2$ states occur only for $N=8$ and $12$ at
$2\ell=2N-3$, and the $\nu_{\rm{QE}}=1/3$ states occur for both even
and odd values of N).

\section{Partially Spin Polarized Systems}
\label{sec:part_spin_pol}

\subsection{Introduction and Model}


The experiment of Pan et al. \cite{PanStormerTsuiPRL03} has
suggested some of the novel IQL states (e.g. $\nu=4/11$) are fully
spin polarized and that other states could be partially spin
polarized. Sitko et al. \cite{SitkoYiYiQuinnPRL96,SitkoYiQuinnPRB97}
found that the $\nu=4/11$ state did not occur in the CF hierarchy of
spin polarized IQL sates. They suggested that the reason for this
was that the pseudopotential $V_{\rm{QE}}(L^{\prime})$ was not
sufficiently similar to $V_0(L^{\prime})$, the pseudopotential for
electrons in LL0, to support the same kind of correlations. It was
shown \cite{QuinnWoysJPhys00,WojsPRB01} that because
$V_{\rm{QE}}(L^{\prime})$ was not ``superharmonic" at $\mathcal R=
2\ell -L^{\prime}=1$, the QEs could not support Laughlin
correlations and no second generation of CFs could occur at
$\nu_{\rm{QE}}=1/3$ producing a completely spin polarized IQL state
at $\nu^{-1}=2+(1+\nu_{\rm{QE}})^{-1}=11/4$ (although an IQL with
some other, non-Laughlin, form of QE-QE correlations was not
excluded by this argument). This led to the suggestion
\cite{ParkJainPRB00} that the QE excitations would have to have
reversed spin in order to produce a daughter state at
$\nu_{\rm{QE}}=1/3$.

Until now, we have concentrated on fully spin polarized systems with
total spin $S=N/2$ (each electron having projection $s_z=-1/2$). In
this section we describe the numerical calculations for systems with
total spin $S=N/2-K$, where $K$ is the number of electrons with
reversed spin. The spin excitations of a fully spin polarized system
are evaluated for both integral \cite{RezayiPRB87} and fractional
IQL states. Reversed spin quasielectrons (QERs), skyrmions (SK)
\cite{SondhiPRB47}, and spin waves are found, and their properties
are discussed. The goal of this section is to present enough
information about spin excitations to be able to compare fully spin
polarized and partially spin polarized states in the FQH hierarchy.

We perform numerical diagonalization of the Coulomb interaction for
a system of $N$ electrons in a shell of angular momentum $\ell$,
specifying the total $z$-component of angular momentum for $N-K$
electrons of spin $\downarrow$ and $K$ electrons of spin $\uparrow$.
There are four conserved quantum numbers: $L$, the total angular
momentum; $S$, the total spin, and their projections $L_z$ and
$S_z$. The energy eigenvalues depend only on $L$ and $S$, and they
are therefore $(2L+1)(2S+1)$-fold degenerate. For more realistic
results, finite well width effects are accounted for by replacing
$e^2/r$ (where $r$ is in-plane separation) by
$V_{\xi}(r)=e^2\int{dzdz^{\prime}\xi^2(z)\xi^2(z^{\prime})
[r^2+(z-z^{\prime})^2]^{-1/2} } $, where $\xi(r)$ is the envelope
function for the lowest subband. The basis functions
$|m_1\sigma_1,m_2\sigma_2,\ldots, m_N\sigma_N
\rangle=c_{m_1\sigma_1}^{\dag}\ldots
c_{m_N\sigma_N}^{\dag}|vac\rangle$, where $|vac\rangle$ stands for
the vacuum state, have $L_z=\sum_im_i$ and $S_z=\sum_i\sigma_i$ as
good quantum numbers. The total angular momentum $L$ and total spin
$S$ are resolved numerically in the diagonalization of each
appropriate $(L_z,S_z)$ Hilbert subspace.


\subsection{Integral Filling}


\begin{figure}
\centerline{\includegraphics[width=
\linewidth]{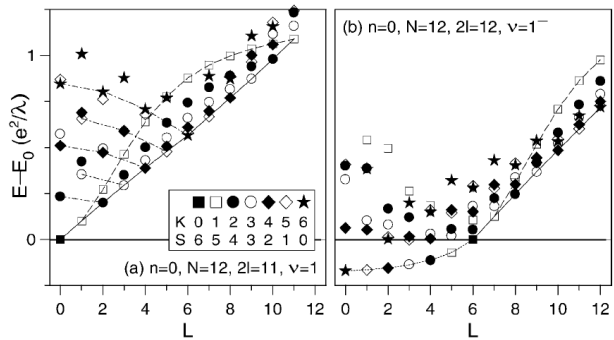}} \caption{The energy
spectra of 12 electrons in the LL0 calculated on Haldane sphere with
$2\ell=11$ (a) and 12 (b) \cite{QuinnQuinnSSC02sk}.}
\label{Fig:En_Spectra_12_el_spin}
\end{figure}


In Fig. \ref{Fig:En_Spectra_12_el_spin} (a) and (b) we present the
low energy spectra of the $\nu = 1$ and $1^-$ (a single hole in $\nu
= 1$) states, respectively. In this and all other spectra, only the
lowest state at each $L$ and $S$ is shown. $E_0$ is the energy of
the lowest maximally polarized state (K = 0), and the Zeeman energy
$E_Z$ is omitted. The ferromagnetic ground state of Fig.
\ref{Fig:En_Spectra_12_el_spin} (a) at $L = 0$ and $S = N/2 = 6$
results from the Coulomb interaction even when $E_Z = 0$. States
with different values of $S$ are indicated by the different symbols
shown in the inset. The lowest excited state is a spin wave (SW)
\cite{KallinHalperin} consisting of a hole in the spin $\downarrow$
level and an electron in the spin $\uparrow$ level with $L=K=1$. A
dashed line marks the entire single SW band at $1\leq L\leq 11$
(resulting from $\vec L = \vec \ell_e + \vec \ell_h$ with $\ell_e=
\ell_h = \ell = 11/2$). The lowest energy excitation for a given
value of either $L$ or $K$ occurs at $L=K$ where $K =(1/2)N-S$ is
the number of spin flips away from the fully polarized ground state.
The (near) linearity of $E(K)$ for this band of states (denoted by
$W_K$) suggests that it consists of $K$ SWs, each with $L=1$; which
are (nearly) noninteracting. As shown with the dot-dash lines
connecting different states of the same number $K$ of $L = 1$ SWs,
only the $L = K$ state (in which the SWs have parallel angular
momenta) is noninteracting, and all others (at $L < K$) are
repulsive.

We have compared the linear $W_K$ energy bands calculated for
different electron numbers $N \leq 14$, and found that they all have
the same slope $u \approx 1.15 e^2/\lambda$ when plotted as a
function of the `relative' spin polarization $\zeta =K/N$. The fact
that $E - E_0 = u\zeta$ for the $W_K$ band for every value of $N$
has two noteworthy consequences in the $N \rightarrow \infty$
limit.(i) For any value of $E_Z \neq 0$, the interaction energy of
each $W_K$ state, $E - E_0 \propto K/N$, is negligible compared to
its total Zeeman energy, $KE_Z$. (ii) The gap for spin excitations
at $\nu = 1$ equals $E_Z$; if this gap can be closed (e.g. by
applying hydrostatic pressure), the $\nu = 1$ ferromagnet becomes
gapless and the density of states for the $W_K$ excitations becomes
continuous.

Because of the exact particle-hole symmetry in the lowest LL, the
$\nu=1^{-} $ state whose spectrum appears in Fig.
\ref{Fig:En_Spectra_12_el_spin} (b) can be viewed as containing
either one hole or one reversed spin electron in a $\nu = 1$ ground
state. The band of states with $0 \leq L \leq 5$ and $S = L$ (dotted
line) is the skyrmion band denoted by $S_K$. Its energy increases
monotonically with $S$ and $L$. For $6\leq L\leq 12$, the single SW
band (dashed line) and band of $K$ SWs each with $L = 1$ (solid
line) resemble similar bands in Fig. \ref{Fig:En_Spectra_12_el_spin}
(a), except that their angular momenta are added to that of the hole
which has $\ell_h = \ell = 6$.

Fig. \ref{Fig:En_Spectra_12_el_spin} completely ignores the Zeeman
energy. The total Zeeman energy measured from the fully polarized
state is proportional to $K$. The total energy of the skyrmion band
is $E(K)= E_S(K) + KE_Z$ and the lowest $S_K$ state occurs when
$E(K)$ has its minimum. If we very roughly approximate the skyrmion
energy in a finite system by $E_S(K) \approx E_S(N/ 2) + \beta S^2$,
where $\beta \geq 0$ is a constant, this minimum occurs at $K = 1/(N
-E_Z/\beta)$ spin flips. This vanishes when $E_Z = \beta N$,
defining the critical value, $\tilde E_Z$, and it reaches its
maximum value $K= N/2$ (or complete depolarization) when  $E_Z = 0$.
At such $E_Z$ the ground state at $\nu = 1^{\pm}$ is a finite size
skyrmion, its gap for spin excitations (`internal' spin excitations
introduced by Fertig et al.\cite{FertigBreyCoteMacDonaldPRL96}) is
much smaller than (and largely independent of) $E_Z$. This is in
contrast to the exact $\nu = 1$ filling and allows spin coupling of
the electron system to the magnetic ions, nuclei, or charged
excitons.

The only difference between the filling factors $\nu = 3, 5,\ldots $
and $1$ is that the monopole harmonics $|Q; \ell =Q + n, m\rangle$
correspond to the excited LL instead of the lowest. Matrix elements
of the Coulomb interaction $e^2 /r$ between these higher monopole
harmonics give a different pseudopotential $V_n(\mathcal R)$ from
that for $n =0$. Though one might expect skyrmions to be the lowest
energy charged excitations in this case, the change in the
pseudopotential from $V_0 (\mathcal R)$ to $V_n(\mathcal R)$ with $n
\geq 1$ causes the charged spin flip excitations to have higher
energy than the single hole or reversed spin electron
\cite{WuJain94}.


\subsection{Fractional Filling}

\begin{figure}
\centerline{\includegraphics[width=
\linewidth]{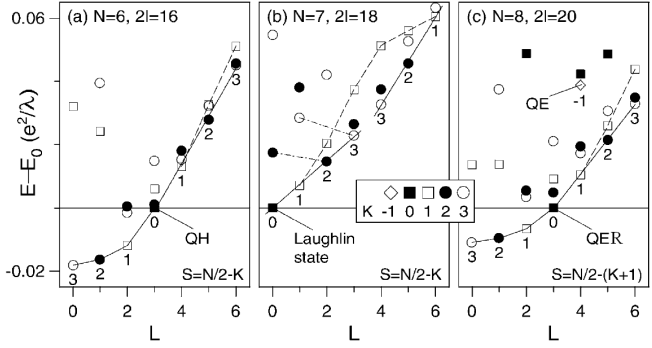}} \caption{The energy
spectra of $N=6-8$ electrons on Haldane sphere at values of $2\ell$
corresponding to $\nu=(1/3)^-$(a), $\nu=1/3 $(b), and $\nu=(1/3)^+$
(c) \cite{QuinnQuinnSSC02sk}}. \label{Fig:En_spectra_el_1_3_spin}
\end{figure}

Since the CF picture \cite{JainPRL89} describes the FQH effect in
terms of integral filling of effective CF levels, it is interesting
to ask \cite{KamillaWuJainSSC96} if spin excitations analogous to
the SWs and SKs occur at Laughlin fractional fillings $\nu =(2p
+1)^{-1}$ (where $p =1, 2, \ldots$). In Fig.
\ref{Fig:En_spectra_el_1_3_spin} we display numerical results for
$\nu \approx 1/3$.

The values of $N$ and $2\ell$ in frames (b), (a), and (c) correspond
to a Laughlin $\nu = 1/3$ condensed state, Laughlin quasihole (QH),
and Laughlin quasielectron (QE) or reversed spin quasielectron
(QER), respectively. For each of these cases, the lowest CF LL has a
degeneracy of seven. Clearly the single SW dispersion (dashed line)
and the linear $W_K$ band (solid line) both appear in Fig.
\ref{Fig:En_spectra_el_1_3_spin}(b). The SK bands beginning at $L
=0$ lie below the single QH state (a) and below the single
${\rm{QE_R}}$ state (c). The solid and dashed lines at $3\leq L\leq
6$ in Fig. \ref{Fig:En_spectra_el_1_3_spin} (a) and (c) are
completely analogous to those in Fig.
\ref{Fig:En_Spectra_12_el_spin} (b), and correspond to the single SW
band and the $W_K$ band, except that their angular momenta are added
to $\ell_{\rm{QH}}=3$ or $\ell_{\rm{QE_R}} =3$. What is clearly
different from the $\nu =1$ case is the smaller energy scale, and a
noticeable difference between the $\nu=(1/3)^{-}$ (QH) and
$\nu=(1/3)^{+}$ (QER) spectra. Since the QH$-$QH and QER$-$QER
interactions are known to be different \cite{SzWojsQuinnPRB01_1},
this lack of QH$-$QER symmetry is not unexpected. It implies a lack
of symmetry between the CF skyrmion (QER + $K$ SW) and CF
antiskyrmion (QH + $K$ SW) states in contrast to the
skyrmion-antiskyrmion symmetry of $\nu =1$. Because the CF skyrmion
energy scale is so much smaller than $E_{\rm{C}}$ at $\nu = 1$; the
critical $E_Z$ at which skyrmions are stable is correspondingly
smaller \cite{gmeasPRL97}.

\subsection{Spin-Reversed Quasielectrons}

\begin{figure}
\centerline{\includegraphics[width= \linewidth]
{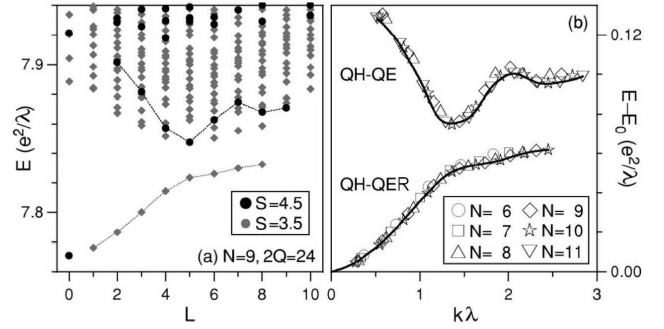}} \caption{(a) The energy spectrum of
the system of $N=9$ electrons on Haldane sphere at monopole strength
$2Q=3(N-1)=24$. Black dots and gray diamonds mark states with the
total spin $S=\frac{N}{2}=\frac{9}{2}$ (maximum polarization) and
$S=\frac{N}{2}-1=\frac{7}{2}$ (one reversed spin), respectively.
Ground state is the Laughlin $\nu=\frac{1}{3}$ state. Lines connect
states containing one QE-QH ($S=\frac{9}{2}$) or QER$-$QH
($S=\frac{7}{2}$) pair. (b) The dispersion curves (excitation energy
$\mathcal E_{K}=E-E_0$ vs. wavevector $k$) for the $K=0$
charge-density wave (QE-QH pair) and the $K=1$ spin-density wave
(QE$-$QER pair) in the Laughlin $\nu=\frac{1}{3}$ ground state,
calculated in the systems of $N \leq 11$ electrons on Haldane sphere
\cite{SzWojsQuinnPRB01_1}.} \label{Fig:En_1_3_1_spin_9el}
\end{figure}

It is well known that even in the absence of the Zeeman energy gap,
$E_Z=0$, the ground state of the 2DEG in the lowest LL is completely
spin-polarized at the precise values of the Laughlin filling factor
$\nu=(2p+1)^{-1}$, with $p=0,1,2,\ldots$. There are two types of
elementary charge-neutral excitations of Laughlin $\nu=(2p+1)^{-1}$
ground states, carrying spin $S=0$ or 1, respectively. Their
dispersion curves $E_{\rm{S}}(k)$ have been studied for all
combinations of $p$ and $S$. In Fig. \ref{Fig:En_1_3_1_spin_9el} we
present the exact numerical results for $\nu=1/3$ obtained from our
exact diagonalization of up to $N=11$ electrons on Haldane sphere
\cite{SzWojsQuinnPRB01_1}. As an example, in Fig.
\ref{Fig:En_1_3_1_spin_9el} (a), we show the entire low-energy
spectrum of an $N=9$ system with all spins polarized and with one
reversed spin (Hilbert subspaces of total spin $S=N/2-K=9/2$ and 7/2
for $K=0$ and 1, respectively), from which the dispersion curves
$E_S(k)$ are obtained. The energy $E$ is plotted as a function of
angular momentum $L$, and $2Q=3(N-1)=24$ is the strength of the
magnetic monopole inside Haldane sphere corresponding to the LL
degeneracy $g=2Q+1=25$ and the Laughlin filling factor
$\nu=(N-1)/(g-1)=1/3$ (for the details of Haldane spherical geometry
see Refs. \cite{HaldanePRL83,FanoOrtolaniColomboPRB86,
MonoploeHarmonics1,MonopoleHarmonics2,WojsQuinnPRB07} and Section
\ref{sec:num_diag}. The energy $E$ does not include the Zeeman term
$E_Z$, which scales differently from the plotted Coulomb energy with
the magnetic field $B$. The excitation energies $\mathcal E_K=E-E_0$
(where $E_0$ is the Laughlin ground state energy) have been
calculated for the states identified in the finite-size spectra as
the $S=0$ charge-density wave and the $K=1$ spin-density wave. These
states are marked with dotted lines in Fig.
\ref{Fig:En_1_3_1_spin_9el} (a). The values of $\mathcal E_K$
obtained for different $N \leq 11$ have been plotted together in
Fig. \ref{Fig:En_1_3_1_spin_9el} (b) as a function of the wave
vector $k=L/R =(L/\sqrt{S})\lambda^{-1}$. Clearly, using the
appropriate units of $\lambda^{-1}$ for wave vector and
$e^2/\lambda$ for excitation energy in Fig.
\ref{Fig:En_1_3_1_spin_9el} (b) results in the quick convergence of
the curves with increasing $N$, and allows an accurate prediction of
the dispersion curves in an infinite system, as marked with thick
lines. The most significant features of these curves are (i) the
finite gap $\Delta_0 \approx 0.076 e^2/\lambda$ and the magnetoroton
minimum $k=1.5\lambda^{-1}$ in $E_0(k)$ and (ii) the vanishing of
$\mathcal E_1(k)$ in the $k\rightarrow 0$ limit (for $E_Z=0$).

The similar nature of the charge and spin waves in the $\nu=1/3$
state to those at $\nu=1$ lies at the heart of the composite Fermion
picture \cite{JainPRL89,LopezFradkinPRB91,HalperinLeeReadPRB93} in
which these excitations correspond to promoting one CF from a
completely filled lowest $(n=0)$ spin $\downarrow$ CF LL either to
the first excited ($n=1$) CF LL of the same spin ($\downarrow$) or
to the same CF LL ($n=0$) but with the reversed spin ($\uparrow$).
The three constituent QPs of which the charge and spin waves are
composed (a hole in the $n=0$ spin $\downarrow$ CFLL and particles
in the $n=1$ spin $\downarrow$ and $n=0$ spin $\uparrow$ CFLLs) are
analogous to those in the electron LLs from which the charge and
spin waves at $\nu=1$ are built. Independently of the CF picture,
one can define three types of QPs (elementary excitations) of the
Laughlin $\nu=1/3$ fluid. They are Laughlin quasiholes and
quasielectrons and Rezayi spin-reversed quasielectrons (QER). The
excitations in Fig. \ref{Fig:En_1_3_1_spin_9el} are more complex in
a sense that they consist of a (neutral) pair of QH and either QE
($K=0$) or QE$_{R}$ ($K=1$). Each of the QPs is characterized by
such single-particle quantities as (fractional) electric charge
($\mathcal Q_{\rm{QH}}=+e/3$ and $\mathcal Q_{\rm{QE}}= \mathcal
Q_{{\rm{QER}}}=-e/3$), energy $\varepsilon_{\rm{QP}}$ , or
degeneracy $g_{\rm{QP}}$ of the single-particle Hilbert space. On
Haldane sphere, the degeneracy $g_{\rm{QP}}$ is related to the
angular momentum $\ell_{\rm{QP}}$ by
$g_{\rm{QP}}=2\ell_{\rm{QP}}+1$, with
$\ell_{\rm{QH}}=\ell_{{\rm{QER}}}=Q^{\ast}$ and
$\ell_{\rm{QE}}=Q^{\ast}+1$ and $2Q^{\ast}=2Q-2(N-1)$ being the
effective monopole strength in the CF model.

\begin{figure}
\centerline{\includegraphics[width=
\linewidth]{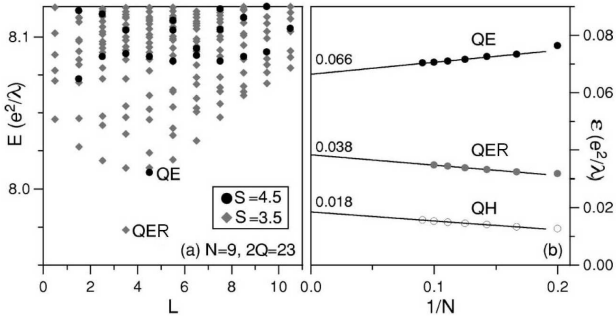}} \caption{(a) The energy spectrum
of the system of $N=9$ electrons on Haldane sphere at monopole
strength $2Q=3(N-1)-1=23$. Black dots and gray diamonds mark states
with the total spin $S=\frac{N}{2}=\frac{9}{2}$ (maximum
polarization) and $S=\frac{N}{2}-1=\frac{7}{2}$ (one reversed spin),
respectively. Ground state at $S=\frac{7}{2}$ is the QER of the
Laughlin $\nu=\frac{1}{3}$ state and the lowest energy state at
$S=\frac{9}{2}$ is the Laughlin QE. (b) The energies $\varepsilon$
of all three types of quasiparticles of Laughlin $\nu=\frac{1}{3}$
ground state (QH, QE, and QER) in the systems of $N \leq 11$
electrons on Haldane sphere and plotted as a function of $N^{-1}$.
The numbers give the results of linear extrapolation to an infinite
(planar) system.} \label{Fig:1_QER_9_el}
\end{figure}

The energies $\varepsilon_{\rm{QP}}$ to create an isolated QP of
each type in the Laughlin ground state have been previously
estimated in a number of ways. Here, we present our results of exact
diagonalization for $N\leq 11$ ($\varepsilon_{\rm{QE}}$ and
$\varepsilon_{\rm{QH}}$) and $N \leq 10$
($\varepsilon_{{\rm{QER}}}$) \cite{SzWojsQuinnPRB01_1}. In Fig.
\ref{Fig:1_QER_9_el} (a) we show an example of the numerical energy
spectrum for the system of $N=9$ electrons, in which an isolated QE
or QER occurs at $2Q=3(N-1)-1 =23$ in the subspace of $S=N/2=9/2$
and $S=N/2-1=7/2$, respectively. Both of these states have been
identified in Fig.\ref{Fig:1_QER_9_el} (a). To estimate
$\varepsilon_{\rm{QE}}$ and $\varepsilon_{{\rm{QER}}}$, we use the
standard procedure \cite{HaldaneRezayiPRL85,HaldaneQHE,
FanoOrtolaniColomboPRB86,WojsQuinnPhilMag,WojsQuinnPRB00} to take
into account the finite-size effects (the dependence of $\lambda$ on
$2Q$, $Q\lambda^2=R^2$), and express the energies $E$ of Fig.
\ref{Fig:1_QER_9_el} (a)  in units of $e^2/\lambda$ with $\lambda$
appropriate for $\nu=1/3$, before subtracting from them the Laughlin
ground state energy of Fig. \ref{Fig:En_1_3_1_spin_9el} (a).
Plotting the results for different values of $N$ in Fig.
\ref{Fig:1_QER_9_el} (b) as a function of $N^{-1}$ allows the
extrapolation to an infinite system, with the limiting values of
$\varepsilon_{\rm{QE}}=0.0664 e^2/\lambda$ and
$\varepsilon_{{\rm{QER}}} = 0.0383 e^2/\lambda$ (with the difference
$\varepsilon_{\rm{QE}}-\varepsilon_{{\rm{QER}}} =0.0281 e^2/\lambda$
in remarkable agreement with Rezayi's original estimate
\cite{RezayiPRB87,RezayiPRB91} based on his numerics for $N\leq6$).
For completeness, we have also plotted the QH energies, which
extrapolate to $\varepsilon_{\rm{QH}}=0.0185 e^2/\lambda$. Note that
to obtain the so-called ``proper" QP energies $\tilde
\varepsilon_{\rm{ QP}}(N)$ in a finite system
\cite{HaldaneRezayiPRL85,FanoOrtolaniColomboPRB86,WojsQuinnPhilMag},
the term $\mathcal Q_{\rm{QP}}^ 2 /2R$ must be added to each value
in Fig. \ref{Fig:1_QER_9_el} (b). The linear extrapolation of
$\tilde \varepsilon_{\rm{ QP}}(N)$ to $N^{-1} \rightarrow 0$ gives
$\tilde \varepsilon_{\rm{ QE}}=0.0737 e^2/\lambda$, $\tilde
\varepsilon_{\rm{ QER}} =0.0457 e^2/\lambda$, and $\tilde
\varepsilon_{\rm{ QH}}=0.0258 e^2/\lambda$. The energies of
spatially separated QE-QH and QER-QH pairs (activation energies in
transport experiments) are hence equal to $E_0(\infty)=\tilde
\varepsilon_{\rm{ QE}}+ \tilde \varepsilon_{\rm{ QH}} = 0.0995
e^2/\lambda$ and $E_1(\infty)=\tilde \varepsilon_{\rm{ QER}}+ \tilde
\varepsilon_{\rm{ QH}}=0.0715 e^2/\lambda$. While the QHs are the
only types of QPs that occur in low-energy states at $\nu
<(2p+1)^{-1}$, the QEs and QERs are two competing excitations at
$\nu>(2p+1)^{-1}$. As pointed out by Rezayi
\cite{RezayiPRB87,RezayiPRB91} and Chakraborty et al.
\cite{ChakrabortyPRL86}, whether QEs or QERs will occur at low
energy depends on the relation between their energies including the
Zeeman term, $\varepsilon_{\rm{QE}}$ and $\varepsilon_{{\rm{QER}}}
+E_Z$. Although it is difficult to accurately estimate the value of
$E_Z$ in an experimental sample because of its dependence on a
number of factors (material parameters, well width $w$, density
$\rho$, magnetic field $B$, etc.), it seems that both scenarios with
QEs and QERs being lowest-energy QPs are possible. For example,
using the bulk value for the effective $g^{\ast}$ factor in GaAs
($dE_Z /dB=0.03$ meV/T) results in the QER-QE crossing at $B=18 T$,
while including the dependence of $g^{\ast}$ on $w$ and $B$ as
described by W\'ojs et al. \cite{WojsQuinnHawrylakPRB00} makes QER
more stable than QE up to $B\sim 100$ T.

\begin{figure}
\centerline{\includegraphics[width= \linewidth]{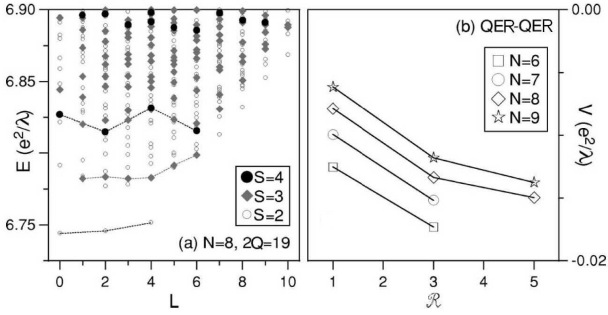}}
\caption{(a) The energy spectrum of the system of $N=8$ electrons on
Haldane sphere at monopole strength $2Q=3(N-1)-2=19$. Black dots and
gray diamonds mark states with the total spin $S=\frac{N}{2}=4$
(maximum polarization), $S=\frac{N}{2}-1=3$ (one reversed spin), and
$S=\frac{N}{2}-2=2$ (two reversed spin), respectively. Lines connect
states containing one QE-QE ($S=4$), QE$_{R}$-QE ($S=3$), or
QE$_{R}$-QE$_{R}$ ($S=2$) pair. (b) The pseudopotential (pair energy
$V$ vs. relative angular momentum $\mathcal R$) of the QER-QER
interaction calculated in the systems of $N\leq 9$ electrons on
Haldane sphere \cite{SzWojsQuinnPRB01_1}.} \label{Fig:2_QER}
\end{figure}

Once it is established which of the QPs occur at low energy in a
particular system (defined by $\rho, w, B, \nu,$ etc.), their
correlations can be understood by studying the appropriate pair
interaction pseudopotentials \cite{HaldaneQHE,WojsQuinnPhilMag,
QuinnWoysJPhys00, WojsPRB01}. The pseudopotential $V(\mathcal R)$ is
defined \cite{HaldaneQHE} as the dependence of pair interaction
energy $V$ on relative orbital angular momentum $\mathcal R$. On a
plane, $\mathcal R$ for a pair of particles $ab$ is the angular
momentum associated with the (complex) relative coordinate $z=z_a
-z_b$. On Haldane sphere, the compatible definition of $\mathcal R$
depends on the sign of $\mathcal Q_a\mathcal Q_b$ : for a pair of
opposite charges, $R$ is the length of total pair angular momentum,
$L=|\vec\ell_a +\vec\ell_b|$, while for two charges of the same
sign, $\mathcal R=|\ell_a+\ell_b -L|$. In all cases, $\mathcal R>0$
and larger $\mathcal R$ corresponds to a larger average $ab$
separation \cite{WojsQuinnPhilMag,QuinnWoysJPhys00}. Furthermore,
only odd values of $\mathcal R$ are allowed for indistinguishable
$(a=b)$ Fermions.

Since the QE-QH and QER-QH pseudopotentials have been plotted in
Fig. \ref{Fig:En_1_3_1_spin_9el} ($V_{\rm{QE}-\rm{QH}}=E_0$ and
$V_{\rm{QER-QH}}=E_1$), and the QE-QE and QH-QH pseudopotentials can
be found, for example, in \textcite{WojsQuinnPRB00}, we only need to
discuss $V_{{\rm{QER}}-{\rm{QER}}}$ and $V_{{\rm{QE}}-{\rm{QER}}}$.
Two QERs occur in an $N$-electron system with at least two reversed
spins ($S<(N/2)-1$) and at $2Q=3(N-1)-2$ (i.e., at $g=g_0-2$ where
$g_0$ corresponds to the Laughlin state). An example of the energy
spectrum is shown in Fig. \ref{Fig:2_QER} (a) for $N=8$ at $2Q=19$.
The lowest-energy states in the subspaces of $S=N/2=4, N/2-1=3,$ and
$N/2-2=2$ are connected with dashed lines and contain a QE-QE,
QE-QER , and QER-QER pair, respectively. The angular momenta $L$
that occur in these bands result from addition of $\vec
\ell_{\rm{QE}}$ and/or $\vec \ell_{{\rm{QER}}}$ (with
$\ell_{\rm{QE}}=Q^{\ast}+1=7/2$ and $\ell_{{\rm{QER}}}
=Q^{\ast}=5/2$ ). For identical Fermions, the addition must be
followed by antisymmetrization that picks out only odd values of
$\mathcal R$ for the QE-QE and QER-QER pairs. An immediate
conclusion from Fig. \ref{Fig:2_QER} (a) is that the maximally
spin-polarized ($S=N/2$) system is unstable at the filling factor
close but not equal to the Laughlin value of $\nu =1/3$ (the actual
spin polarization decreases with decreasing $E_Z$ , and $S=0$ for
$E_Z=0$). This was first pointed out by Rezayi
\cite{RezayiPRB87,RezayiPRB91} and interpreted in terms of an
effective attraction between $S=1$ spin waves; in this paper we
prefer to use charged QPs as the most elementary excitations and
explain the observed ordering of different $S$ bands by the fact
that $\varepsilon_{\rm{QE}} \neq \varepsilon_{{\rm{QER}}}$ (at
$E_Z=0$, $\varepsilon_{\rm{QE}}-\varepsilon_{{\rm{QER}}} =0.0281
e^2/\lambda$) and the particular form of involved interaction
pseudopotentials.

We have calculated the QE-QER and QER-QER pseudopotentials from the
energy spectra as that in Fig. \ref{Fig:2_QER} (a) by converting $L$
into $\mathcal R$ and subtracting the Laughlin ground state energy
and the energy of two appropriate QPs from the total $N$-electron
energy, $V_{AB}(\mathcal R)=E(L)-E_0- \varepsilon_A-\varepsilon_B$.
To minimize the finite-size effects, all subtracted energies are
given in the same units of $e^2/\lambda_0$, where
$\lambda_0=R/\sqrt{Q_0}$ corresponds to $2Q_0=3(N-1)$, i.e., to
$\nu=1/3$. The result for $V_{{\rm{QER}}-{\rm{QER}}}$ and $N\leq 9$
is shown in Fig. \ref{Fig:2_QER}(b). Clearly, obtained values of
$V_{{\rm{QER}}-{\rm{QER}}} (\mathcal R)$ still depend on $N$ and,
for example, the positive sign characteristic of repulsion between
equally charged particles is only restored in the $N^{-1}
\rightarrow 0$ limit with $V_{{\rm{QER}}-{\rm{QER}}} (1)$ of the
order of $0.01 e^2/\lambda$ (compare with discussion of the signs of
$V_{\rm{QE}-\rm{QE}}$ and $V_{\rm{QH}-\rm{QH}}$ in Ref.
\textcite{WojsPRB01_1} and Sec. \ref{sec:res_int} ). However, it
seems that the monotonic character of $V_{{\rm{QER}}-{\rm{QER}}}
(\mathcal R)$ is independent of $N$. More importantly,
$V_{{\rm{QER}}-{\rm{QER}}} (\mathcal R)$ is also a superlinear
function of $L(L+1)$. This implies \cite{WojsQuinnPhilMag,
QuinnWoysJPhys00,WojsPRB01} Laughlin correlations and
incompressibility at $\nu_{{\rm{QER}}}=(2p+1)^{-1}$, in analogy to
the spin-polarized Laughlin states of QEs or QHs in Haldane's
hierarchy picture \cite{HaldanePRL83,WojsQuinnPRB00}. The most
prominent of QER Laughlin states, $\nu_{{\rm{QER}}} =1/3$,
corresponds to the electronic filling factor of $\nu= 4/11$ and the
$75\%$ spin polarization ($S=N/4$). This state has been first
suggested by Beran and Morf \cite{BeranMorfPRB91}. The expected
critical dependence of the excitation gap at $\nu=4/11$ on the
Zeeman gap $E_Z$ might be revealed in tilted-field experiments. This
dependence will be very different from that at some other fractions.



\section{Spin Polarization Transition of the $\nu=4/11$ State}
\label{sec:4_11_polarization}

\subsection{Possible Incompressible Quantum Liquid States}


In Sec. \ref{sec:QE} we mentioned that there were at least two
candidates for the $\nu=4/11$ state observed by Pan et al.
\cite{PanStormerTsuiPRL03}. The fully spin polarized state (for
which there is some experimental support) cannot be a second
generation CF state resulting from Laughlin correlated QEs at
filling factor $\nu_{\rm{QE}}=1/3$. The QE pseudopotentials
$V_{\rm{QE}}(\mathcal R)$ is strongly subharmonic at $\mathcal R=1$
and cannot support Laughlin correlations. A state with pairs of
electrons of total angular momentum $\ell_{\rm{P}}=2\ell-1$ (where
$\ell$ is the QE angular momentum), or with larger clusters can
cause a totally spin polarized state at $\nu=4/11$. However, a
partially spin polarized state in which CF quasielectrons have
reversed spin (QERs) relative to those in the IQL state above which
they reside, could give rise to an IQL state with
$\nu_{{\rm{QER}}}=1/3$ \cite{ParkJainPRB00}. This is possible
because $V_{{\rm{QER}}}(\mathcal R)$ is superharmonic at $\mathcal
R=1$ \cite{SzWojsQuinnPRB01_1} allowing the QERs to form a Laughlin
state.

Which of these has a lower energy? The total energy depends on (i)
the energies of the quasielectrons, $\varepsilon_{\rm{QE}}$ and
$\varepsilon_{{\rm{QER}}}$, (ii) the interaction energy of these
quasiparticles, which depends on their pseudopotentials
$V_{\rm{QE}}(\mathcal R)$ and $V_{\rm{QER}}(\mathcal R)$, and
finally on (iii) the Zeeman energy $E_Z$ due to the total spin $S$
in the applied magnetic field $B$. In real samples each of these
energies depends upon the width of the quantum well in which the
electrons are confined. The QE energies and their interactions
depend on well-width primarily because the interactions of the
electrons in the systems that give rise to QEs involve form factors
resulting from integration over squares of the subband wavefunctions
$\chi(z)$ for the quantum well. The Zeeman energy $E_Z=g^{\ast}\mu_B
B$ depends on the effective $g$-factor of the electrons, observed
experimentally to increase from $g^{\ast}=-0.44$ for wide wells to
zero for well width of roughly 6 nm.


\subsection{Quasielectron Energies}


As we have seen in earlier sections, in the mean-field CF
transformation, the liquid of correlated electrons at filling factor
$\nu_e= 4/11$, is converted to the system of CFs with an effective
filling factor $\nu_{\rm{CF}}= 4/3$ . Approximately $3N_e/4$ of the
CFs fill the lowest CF energy level CF LL0 $\uparrow$, with angular
momentum $\ell^*=\ell-(N_e-1)$. The remaining of $N \sim N_e/4$ CFs
go into the lowest ($\simeq 1/3$-filled) excited CF energy level
(either CF LL0$\downarrow$ or CF LL1$\uparrow$) depending on the
relative magnitude of electron Zeeman energy $E_Z$ and the
``effective" CF cyclotron gap (proportional to $e^2 / \lambda$).
Each CF in the partially filled $1 \uparrow$ or $0 \downarrow$ CF
LLs represent a ``normal" QE \cite{LaughlinPRL83} or QER
\cite{RezayiPRB87,RezayiPRB91} of the underlying incompressible
Laughlin liquid, respectively.

\begin{figure}
\centerline{\includegraphics[width=
\linewidth]{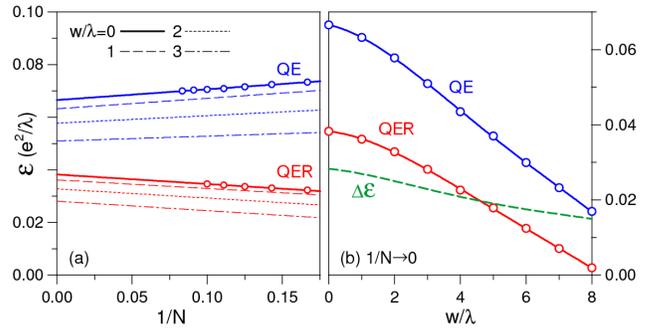}} \caption{(Color online)
Dependence of the QE and QER energies $\varepsilon$ on (a) the
inverse electron number $N_e ^{-1}$ in a finite-size calculation and
(b) the electron layer effective width $w$.  $\lambda$ is the
magnetic length \cite{wojs2006}.} \label{Fig:QP_energies_4_11}
\end{figure}

The Coulomb energies $\varepsilon_{\rm{QE}}$ and
$\varepsilon_{{\rm{QER}}}$ of these two QPs can be extracted
\cite{SzWojsQuinnPRB01_1,FanoOrtolaniColomboPRB86} from exact
diagonalization of finite systems of $N_e$ electrons in the lowest
LL with the appropriate degeneracy $g$. The Laughlin ground state
occurs at $g=3N_e-2\equiv g_L$; it is nondegenerate ($L=0$) and
spin-polarized ($S=N_e/2$). A single QE or QER appears in the
Laughlin liquid in the lowest states at $g=g_L-1$ and either $S=
N_e/2$ or $(N_e/2)-1$, respectively. The QE and QER energies defined
relative to the underlying Laughlin liquid are obtained from the
comparison of the $N_e$-electron energies at $g=g_L$ and $g_L-1$.
The numerical procedure and the result for an ideal 2D electron
layer \cite{SzWojsQuinnPRB01_1,WojsPRB01_1} were presented earlier
in Sections \ref{sec:res_int} and \ref{sec:part_spin_pol}. In Fig.
\ref{Fig:QP_energies_4_11}, we compare the QE/QER energies
calculated for quasi-2D layers of finite width $w$. Here, $w$ is the
effective width of the electron wavefunction in the normal $z$
direction, approximated by $\chi(z) \propto \cos(z \pi /w)$
\cite{wojs2006}. It is slightly larger than the quantum well width
$W$; e.g., for symmetric GaAs/Al$_{0.35}$Ga$_{0.65}$As wells,
$w=W+3$ nm over a wide range of $W\geq$10 nm. The regular dependence
on system size in Fig. \ref{Fig:QP_energies_4_11} (a) allows
reliable extrapolation to ($N_e ^{-1}\rightarrow 0$) planar
geometry. From the comparison of $\varepsilon_{\rm{QE}}(w)$ and
$\varepsilon_{{\rm{QER}}}(w)$ in Fig. \ref{Fig:QP_energies_4_11}
(b), it is clear that their difference is less sensitive to the
width than any of the $\varepsilon_{{\rm{QER}}}(w)$ or
$\varepsilon_{\rm{QE}}(w)$. To put the shown width range in some
perspective, let us note that a (fairly narrow) W=12 nm well in a
(fairly low) field $B=10$ T corresponds to $w/\lambda=1.9$ and
$\Delta \varepsilon(w) / \Delta \varepsilon(0) =0.9$, justifying the
2D approximation. On the other hand, a wide $W=$40 nm well in a high
field $B=23$ T gives $w/\lambda=8.1$ and $\Delta \varepsilon(w) /
\Delta \varepsilon(0) =0.5$, i.e., a significant width effect.

\subsection{Quasiparticle Interactions and Correlation Energy}


The weak effective CF-CF interactions are known with some accuracy
from earlier studies \cite{WojsWodzinskiQuinnPRB06, WojsQuinnPRB00,
SzWojsQuinnPRB01_1, LeeScarolaJainPRL01, LeeScarolaJainPRB02,
SitkoYiQuinnPRB97}. At least at sufficiently low CF fillings factors
$\nu \leq 1/3$, they can be well approximated by fixed Haldane
pseudopotentials independent of the CF LL filling or spin
polarization. The short-range QE-QE, QER-QER, and QE-QER
pseudopotentials can be obtained from finite-size diagonalization
for $N_e$ electrons with up to two reversed spins $S= N_e/2-2$ at
$g=g_L-2$.

\begin{figure}
\centerline{\includegraphics[width= \linewidth]
{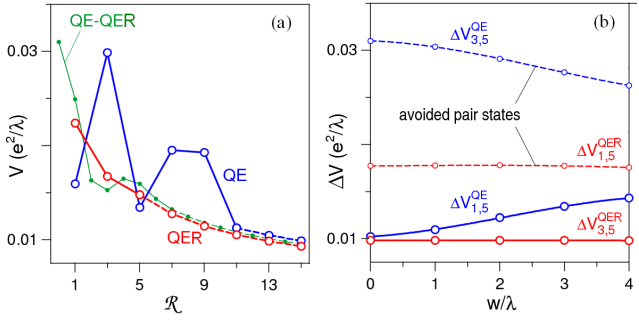}} \caption{(Color online) (a) Haldane
pseudopotential pair interaction energy $V$ as a function of
relative angular momentum $\mathcal R$ for QEs and QERs in an ideal
2D ($w=0$) electron layer. (b) Dependence of pseudopotential
increments $\Delta V_{\mathcal R \mathcal R^{\prime}}=V_{\mathcal
R}-V_{\mathcal R^{\prime}}$ on the electron layer effective width
$w$. $\lambda$ is the magnetic length \cite{wojs2006}.}
\label{Fig:Pseudopot_4_11.eps}
\end{figure}

The result is a reliable account of the relative values $\Delta
V_{\mathcal R \mathcal R^{\prime}}=V_{\mathcal R}-V_{\mathcal
R^{\prime}}$ at small neighboring $\mathcal R$ and $\mathcal
R^{\prime}$, but the absolute values are not estimated very
accurately. Fortunately, since vertical correction of $V(\mathcal
R)$ by a constant does not affect the many-CF wavefunctions and only
rigidly shifts the entire energy spectrum \cite{WojsQuinnPRB00}, a
few leading values of $\Delta V$ completely determine the
short-range CF correlations at a given $\nu$. Therefore, the
knowledge of those few approximate values of $\Delta V_{{\rm{QER}}}$
and $\Delta V_{\rm{QE}}$ was sufficient to establish that: (i) the
QERs form a Laughlin $\nu= 1/3$ liquid \cite{WojsQuinnPRB02_1,
MandalJainPRB02, ParkJainPRB00} which in finite $N-{\rm{QER}}$
systems on a sphere occurs at $g=3N-2$, and (ii) in contrast, the
QEs form a different (probably paired) state
\cite{WojsYiQuinnPRB04,WojsWodzinskiQuinnPRB05,
WojsWodzinskiQuinnPRB06} at the same $\nu= 1/3$, which, on a sphere,
occurs at $g=3N-6$.

However, the relative strength of QE-QE and QER-QER pseudopotentials
$V_{{\rm{QE}}}$ and $V_{\rm{QER}}$ must also be known in addition to
$\Delta V$ to compare the energies of many-QER and many-QE states
(i.e., of the spin-polarized and unpolarized electron states at
$\nu_e= 4/11$). The absolute values of $V_{{\rm{QER}}}$ and
$V_{\rm{QE}}$ can be obtained by matching \cite{HaldaneQHE} the
short-range behavior from exact diagonalization of small systems
with the long-range behavior predicted for a pair of charges
$q=-e/3$. Specifically, the short-range part of
$V_{{\rm{QER}}}(\mathcal R)$, which describes a pair of CFs in the
$1\downarrow$ CF LL, is shifted to match $\eta V_0(\mathcal R)$, the
electron pseudopotential in the lowest LL rescaled by $\eta \equiv
(q^2\lambda_q^{-1}) /(e^2\lambda_e^{-1})=(q/e)^{5/2}$. Similarly,
the short-range part of $V_{{\rm{QER}}}$, related to the $1\uparrow$
CF LL, is shifted to match $\eta V_1(\mathcal R)$.

The result in Fig. \ref{Fig:Pseudopot_4_11.eps} (a) for an ideal 2D
layer was reported earlier \cite{SzWojsQuinnPRB01_1}. In Fig.
\ref{Fig:Pseudopot_4_11.eps} (b), the width dependence of the
leading parameters $\Delta V$ has been plotted. It is noteworthy
that $V_{\rm{QE}}$ is much more sensitive to the electron layer
width $w$ than $V_{{\rm{QER}}}$. This is explained by stronger
oscillations in $V_{\rm{QE}}(\mathcal R)$ at $w =0$, which tend to
weaken in wider wells (when the characteristic in-plane distances
decrease relative to $w$). The curves involving $V_{{\rm{QER}}}(1)$
and $V_{\rm{QE}}(3)$ have been drawn with dashed lines, since the
QER-QER and QE-QE pair states associated with these dominant
pseudopotential parameters will be avoided
\cite{WojsYiQuinnPRB04,WojsWodzinskiQuinnPRB05} in the unpolarized
and polarized $\nu= 1/3$ CF ground states, respectively.

\begin{figure}
\centerline{\includegraphics[width= 0.532\linewidth]
{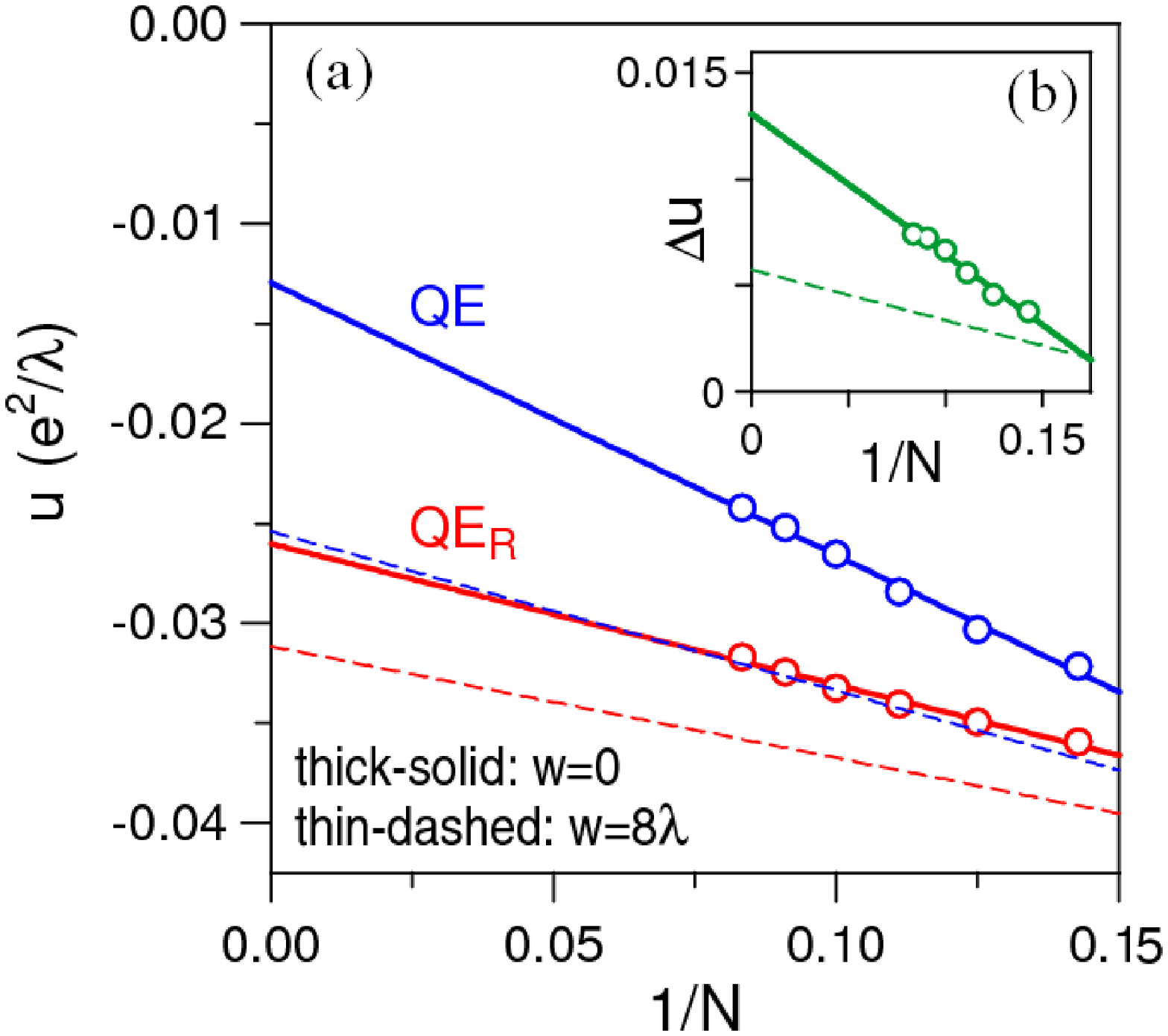}~\includegraphics[width=
0.500\linewidth]{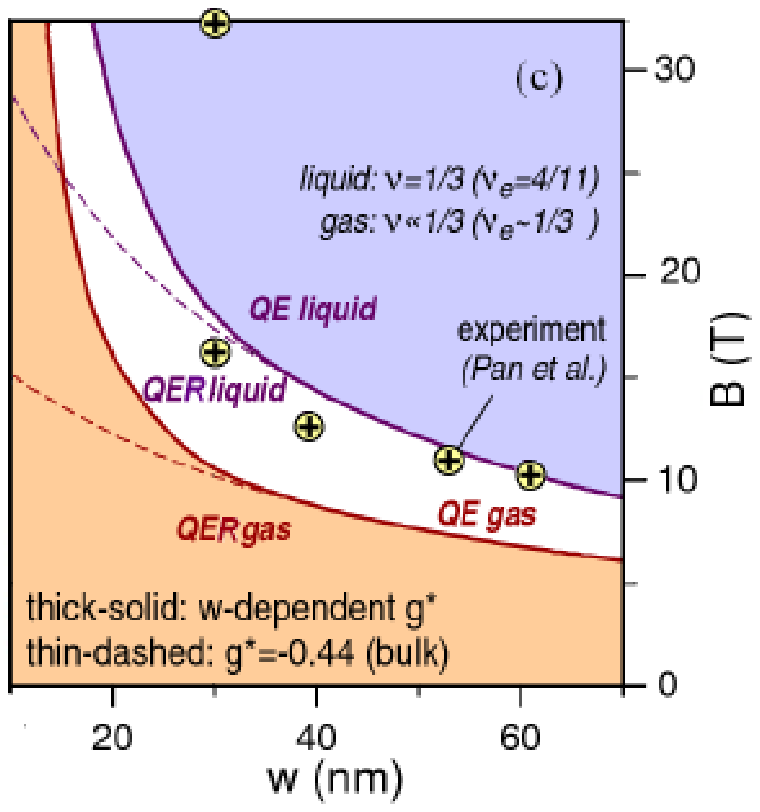}} \caption{(Color
online) (a) Correlation energy $u$ in the $\nu=1/3$ incompressible
liquid of QE or QER as a function of their inverse number $N^{-1}$
for two different widths $w$ of the quasi-2D electron layer;
$\lambda$ is the magnetic length. (b) Difference $\Delta
u=u_{\rm{QE}}-u_{{\rm{QER}}}$ as a function of $N^{-1}$. (c) Phase
diagram critical layer width $w$ vs magnetic field $B$ for the
QE-QER spin transition at $\nu=1/3$ i.e., at $\nu_e=4/11$, assuming
the effective electron Land\'e $g^{\ast}$ factor for GaAs. Dashed
line is for the bulk value $g^{\ast}=-0.44$, ignoring dependence on
the layer width $w$. The experimental points were taken after Pan et
al \cite{PanStormerTsuiPRL03}.} \label{Fig:Corr_Phse_diag_4_11}
\end{figure}

As mentioned above, due to the strong QER-QER repulsion at short
range ($\mathcal R=1$), the QERs form a Laughlin $\nu=1/3$ state
similarly to the electrons in LL0 at $\nu_e= 1/3$. The corresponding
series of nondegenerate $N$-${\rm{QER}}$ ground states on a sphere
occurs at the Laughlin sequence of $g=3N-2$. In Fig.
\ref{Fig:Corr_Phse_diag_4_11} (a), we plot the size dependence of
their correlation energy $u$ (per particle), defined as
\begin{equation}
u = \frac{E + U_{\rm{bckg}}}{ N } \zeta~. \label{eq:u_corr_4_11}
\end{equation}
Here, $E$ is the interaction energy of the ground state of $N$ QERs
and $U_{\rm{bckg}}=-(Nq)^2/2R$ is a correction due to interaction
with the charge-compensating background with the sphere radius
$R=\lambda\sqrt{Q}$ taken for $2Q+1=g$, in analogy to the relation
for electrons in the lowest LL. Factor $\zeta=\sqrt{Q(Q-1)^{-1}}$ is
used to rescale the energy unit $e^2/\lambda=\sqrt{Q}e^2 /R$ from
that corresponding to $g_{{\rm{QER}}} =3N-2$ to that of an average
$\overline{g}= 1/2(g_{{\rm{QER}}}+g_{\rm{QE}})=3N-4$, to allow for a
later comparison of $u$ calculated for QERs and QEs at different
values of $g$ and thus, at different magnetic lengths corresponding
to the same area $4 \pi R^2$. The correlation energies $u$ were
calculated for $N \leq 12$ and extrapolated to $N^{-1} \rightarrow
0$ to eliminate the finite-size effects.

Let us turn to the QEs. The dominant QE-QE repulsion at $\mathcal
R=3$ causes the QEs to form pairs \cite{WojsWodzinskiQuinnPRB06}
rather than a Laughlin state at $\nu=1/3$ although the exact wave
function of this incompressible state is still unknown. The
corresponding series of nondegenerate $N$-QE ground states on a
sphere was identified \cite{WojsYiQuinnPRB04,
WojsWodzinskiQuinnPRB05} at $g=3N-6$, different from the Laughlin
sequence. The QE correlation energy $u$ was calculated from the same
Eq. \ref{eq:u_corr_4_11} but with a different
$\zeta=\sqrt{Q(Q+1)^{-1}}$ (where $g =2Q+1$) also. By using
different $\zeta_{{\rm{QER}}}$ and $\zeta_{\rm{QE}}$, we removed the
discrepancy between $\lambda/R$ of finite $N$-QER and $N$-QE
systems, in order to improve size convergence of $\Delta
u=u_{\rm{QE}} -u_{{\rm{QER}}}$. In an ideal 2D system $w=0$, the
extrapolated value at $N^{-1} = 0$ is
$u_{\rm{QE}}=-0.013e^2/\lambda$, twice smaller in the absolute value
than $u_{\rm{QER}}$ of a Laughlin state. The difference is $\Delta u
=0.013e^2 /\lambda$. The accuracy of this estimate can be judged
from the extrapolation plot in Fig. \ref{Fig:Corr_Phse_diag_4_11}
(b).

The $u_{\rm{QER}}-u_{\rm{QE}}$ difference can be explained from the
comparison \cite{WojsWodzinskiQuinnPRB06} of QER and QE
charge-density profiles $\rho(r)$. The roughly Gaussian QER is (up
to normalization) very similar to $\rho_0$ of an electron in the
lowest LL, yielding similar QER and electron pseudopotentials
$V(\mathcal R)$ and correlation energies $u$ (in the $\eta$-rescaled
units). The ring-like $\rho_{\rm{QE}}$ is more complicated and has a
bigger radius, causing stronger (on the average) QE-QE repulsion.
The difference between $u_{\rm{QER}}$ and $u_{\rm{QE}}$ appears to
result primarily from the difference between QE and QER charge
densities.

\subsection{Spin Phase Diagram for $\nu=4/11$}

Whether QEs or QERs will form a $\nu= 1/3$ state at $\nu_e= 4/11$
depends on the competition of Coulomb and Zeeman energies. The
condition for the QE$\leftrightarrow$QER transition is:
\begin{equation}
\Delta \varepsilon + \Delta u = E_Z~. \label{eq:4_11_trans_cond}
\end{equation}
The competing phases differ in electron-spin polarization. They are
both incompressible but probably have different excitation gaps (and
thus might not show equally strong FQH effect). In an ideal 2D
electron layer, the excitation gap for neutral excitations of the
polarized state can be expected \cite{WojsYiQuinnPRB04,
WojsWodzinskiQuinnPRB05} below $0.005e^2 / \lambda$, and, for the
Laughlin state of QERs, it is estimated at $0.06 \eta e^2
/\lambda=0.004 e^2 /\lambda$. The nature of charged excitations and
the corresponding transport gaps (especially in more realistic
conditions, i.e., for $w>0$, including LL mixing and disorder, etc.)
are not known, and their prediction should require a much more
extensive calculation.

Let us concentrate on the question of stability of either QERs or
QEs at $\nu_e= 4/11$. In order to draw the phase diagram for GaAs
heterostructures in Fig. \ref{Fig:Corr_Phse_diag_4_11} (c), we
combined the estimated dependences of $\Delta \varepsilon/
(e^2\lambda^{-1})$ and $\Delta u/ (e^2\lambda^{-1})$ on $w/\lambda$
(where $e^2\lambda^{-1} /\sqrt{B}=4.49$ meV/T$^{1/2}$ and $\lambda
\sqrt{B}=25.6$ nm T$^{1/2}$) with published data \cite{WojsPRB01} on
width dependence of the effective Land\'e factor $g^{\ast}$,
governing the Zeeman splitting $E_Z=g^{\ast} \mu_B B$ (for $W \geq
30$ nm, it is $g^{\ast}=-0.44$ and $E_Z /B$=0.03 meV/T; in narrower
wells, $g^{\ast}$ increases, passing through zero at $W \approx5.5$
nm; recall that $w\approx W+3$ nm).

The most important phase boundary drawn in Fig.
\ref{Fig:Corr_Phse_diag_4_11} (c) divides the polarized and
unpolarized $\nu_e= 4/11$ states, i.e., the correlated QE and QER
liquids at a finite $\nu= 1/3$ . In the experiment of Pan et al.
\cite{PanStormerTsuiPRL03}, the polarized $\nu_e= 4/11$ state was
observed in a symmetric $W=50$ nm GaAs quantum well at $B=11$ T. The
corresponding experimental point $(w,B)$ marked with a plus lies
very close to the predicted phase boundary, suggesting that the
experimentally detected polarization depended critically on the
choice of a very wide well. \textcite{PanStormerTsuiPRL03} report
identification of an incompressible $\nu_e= 4/11$ state at a very
high field $B=33$ T, taken as an argument for spin polarization.
Indeed, the corresponding data point marked with a cross $W=30$ nm
lies deep inside the predicted ``QE liquid" phase area. However, no
clear evidence for an unpolarized $\nu_e= 4/11$ has yet been
reported. It is clear from Fig. \ref{Fig:Corr_Phse_diag_4_11} (c)
that the spin transition in narrower wells shifts quickly to higher
magnetic fields (i.e., to higher electron concentrations $\rho_e=
\nu_e(2\pi\lambda^2)^{-1}$), especially when the width dependence of
$g^{\ast}$ is taken into account. This suggests that the spin
transition at $\nu_e= 4/11$ might be confirmed in a similar
experiment, carried out in a sample with the same $W$ and $\rho_e$
but with the layer width $w$ tuned by the electric gates inducing a
controlled well asymmetry. The role of QP interaction in stabilizing
the QER phase is clear from the comparison of boundaries dividing
correlated QE/QER liquids and noninteracting QE/QER gases (the gas
occurs at $\nu \ll 1/3$ , with the critical equation $\Delta
\varepsilon=E_Z$; the CF gas $\leftrightarrow$ liquid transition was
recently demonstrated by inelastic light scattering
\cite{gallaisPRL06}. Additional boundaries (not shown here, but see
Fig. 13(b) in \textcite{WojsQuinnPRB02}) appear at even smaller $B$,
defining the areas of stability for a gas of CF skyrmions of
different sizes
\cite{KamillaWuJainSSC96,MacDonaldPalaciosPRB98,WojsQuinnPRB02,gmeasPRL97}.
Note also that $\Delta \varepsilon$ is determined more accurately
than $\Delta u$, possibly explaining the incorrect position of the
experimental point inside the predicted QE gas and/or QER liquid
area.

The spin polarization transition results from a competition between
the Zeeman energy which is proportional to $B$ and the interaction
energy which is proportional to $B^{1/2}$ (or $e^2/{\lambda}$).
Large Zeeman energy favors the totally spin polarized state. Large
quantum well width decreases the interaction energy relative to the
Zeeman energy, so that wide wells and large total magnetic field
(the perpendicular component of $B$ is fixed by the electron density
and the filling factor $\nu=4/11$) favor fully spin polarized state.
Our phase diagram is clearly qualitatively correct, but the
evaluation of the energy of each state involves a number of
approximations. The phase transition lines in the $w-B$ (well
width-applied magnetic field) plane is only approximate. Our
suggestion of using a back gate to change the quantum well size in a
single sample offers a conceptually simple way to test our simple
model. It should be noted that we have considered only the two
extreme polarizations
$P=(n_{\uparrow}-n_{\downarrow})/(n_{\uparrow}+n_{\downarrow})$
equal to 1 and $1/2$, omitting the possibility of intermediate $P$.


\section{Electron System Containing Valence Band Holes}
\label{sec:e_vh_syst}

The first observations of both the integral and fractional quantum
Hall effects were made in magnetotransport studies
\cite{vonKlitzingPRL80,TsuiPRL82}. Deep minima in the longitudinal
conductivity, $\sigma_{xx}$, and flat plateaus in the transverse
conductivity $\sigma_{xy}$, at special filling factors $\nu$ were
the signatures of the IQL states. Magnetotransport has continued to
be a very important technique for studying quantum Hall systems.
However, optical measurements, including infrared spectroscopy,
inelastic light scattering, and photoluminescence have been valuable
probes of quantum Hall systems
\cite{HeimanPinczukGossardPRL88,PinczukPRL93,
KukushkinHaugvonKlitzingPloogPRL94, HawrilakNature}. Many of the
optical processes involve valence band holes interacting with the
electrons confined in a quasi 2D system. A valence band hole can
bind one or two electrons to form a neutral or a negatively charged
exciton ($X$ or $X^-$). In this section we study the properties of a
quasi 2D system containing $N_e$ electrons interacting with $N_h$
valence band holes.

The electron-hole systems are of interest for several reasons. In
quasi 2D systems, neutral excitons and negatively charged excitonic
complexes can form in relatively stable bound states. The negatively
charged excitonic complexes are charged Fermions with LL structure
of their own. They have correlations with one another and with
electrons, just as unbound electrons have with one another. The
correlations between unbound electrons and negatively charged
excitonic complexes is another example of the usefulness of the
generalized CF picture. In some ways it is a simpler example because
the constituents ($e$, $X^-=e^2 h$, $X_2^-=e^3h^2, \cdots$) all have
the same total charge. However, it is more complicated because more
than two different kinds of Fermion can occur.

\subsection{Hidden Symmetry and Multiplicative States}

If the electrons and the valence band holes are confined to the same
2D plane, and if the magnetic field is sufficiently large that the
Landau level separations are large compared to the Coulomb
interaction energy of a pair of particles, only a single LL for
electrons and a single LL for holes need to be considered. In such
case, the magnitude of the interaction between a pair of particles
($e-e, e-h,h-h$) is the same. Then a ``hidden symmetry"
\cite{lernerlozovik81,dzyubenkolozovik,PaquetRiceUeda,
MacDonaldRezayiPRB90,MacDonaldRezayiKellerPRL92} results from the
fact that the commutator of the Hamiltonian $\hat H$ with the
operator $d^{\dag}(0)=N_{\phi}^{-1/2}\sum_{\vec k'}c^{\dag}_{\vec
k'}d^{\dag}_{-\vec k'}$, which creates a neutral exciton with
wavevector $k=0$, satisfies the relation:
\begin{equation}
\left[\hat H, d^{\dag}(0)\right]=E_X(0)d^{\dag}(0)~.
\label{eq:comm_hidden_sym}
\end{equation}
Here $E_X(0)$ is the energy of the exciton, $N_{\phi}=2Q+1$ is the
LL degeneracy and $c^{\dag}_{\vec k'}$ (or $d^{\dag}_{\vec k'}$)
creates an electron (or hole) in LL0 with wavenumber (in
$y$-direction for the Landau gauge) equal to $k$. Because of Eq.
\ref{eq:comm_hidden_sym}, if $|\Phi>$ is an eigenstate of $\hat H$
with energy $E_{\Phi}$, then $d^{\dag}(0)|\Phi>$ is an eigenstate of
$\hat H$ with energy $E_{\Phi}+E_X$. The neutral $k=0$ exciton is
essentially uncoupled from the electron system. States containing
$N_X$ such neutral excitons and $N_e$ free unbound electrons are
referred to as `` multiplicative states"
\cite{lernerlozovik81,dzyubenkolozovik,PaquetRiceUeda,
MacDonaldRezayiPRB90,MacDonaldRezayiKellerPRL92}. These low energy
multiplicative states are not necessarily the ground states of a
system of $N_h$ holes and $N_e$ ($>N_h$) electrons. For
multiplicative states, the photoluminescence (PL) results from the
recombination of the electron-hole pair bound in the ``uncoupled"
exciton ($X$). Since $X$ is not coupled to the background 2D system,
this PL spectrum contains no information about the correlations in
the fluid of free electrons. To obtain information about those
correlations, it is necessary to break the ``hidden symmetry". In
real systems, this does occur as a result of: (i) finite well width
giving different subband wavefunction for electrons and holes and
different $e-e$ and $e-h$ pseudopotentials, (ii) separation of the
centers of mass of the electron and hole layers due to asymmetry of
the quantum well (e.g. not symmetrically modulation doped), and
(iii) admixing of higher LLs when the Coulomb interaction is not
very small compared to the LL separations. For understanding the
qualitative aspects of PL spectrum, it is sufficient to remove the
``hidden symmetry" by introducing a separation $d$ between the 2D
planes on which the electrons and holes reside. Then
$V_{e-h}=e^2(r^2+d^2)^{-1/2}$ and $|V_{e-e}|=e^2/r$. This breaks the
hidden symmetry without the need of including subband wavefunction
or admixing higher LLs. For comparison with real experiments, a more
careful treatment of the $e-e$ and $e-h$ interactions and the
admixing of higher LLs is necessary.

In the next subsection we present spectra obtained by numerical
diagonalization of systems containing up to four electrons and two
valence holes. From the results we obtain binding energies and
angular momenta of the neutral exciton ($X$), the negatively charged
exciton ($X^-=e+X$), and the negatively charged biexciton
($X_2^-=X+X^-$). We also obtain the pseudopotentials describing the
interaction $V_{AB}$ of charged Fermion pairs where $A$ and $B$ can
be $e^-,X^-,X_2^-$, etc. Many of the results in the reminder of this
section have been summarized in \textcite{QWYRRDP}.

\subsection{Numerical Diagonalization}


\subsubsection{Numerical Results}

\begin{figure}
\centerline{\includegraphics[width= \linewidth]
{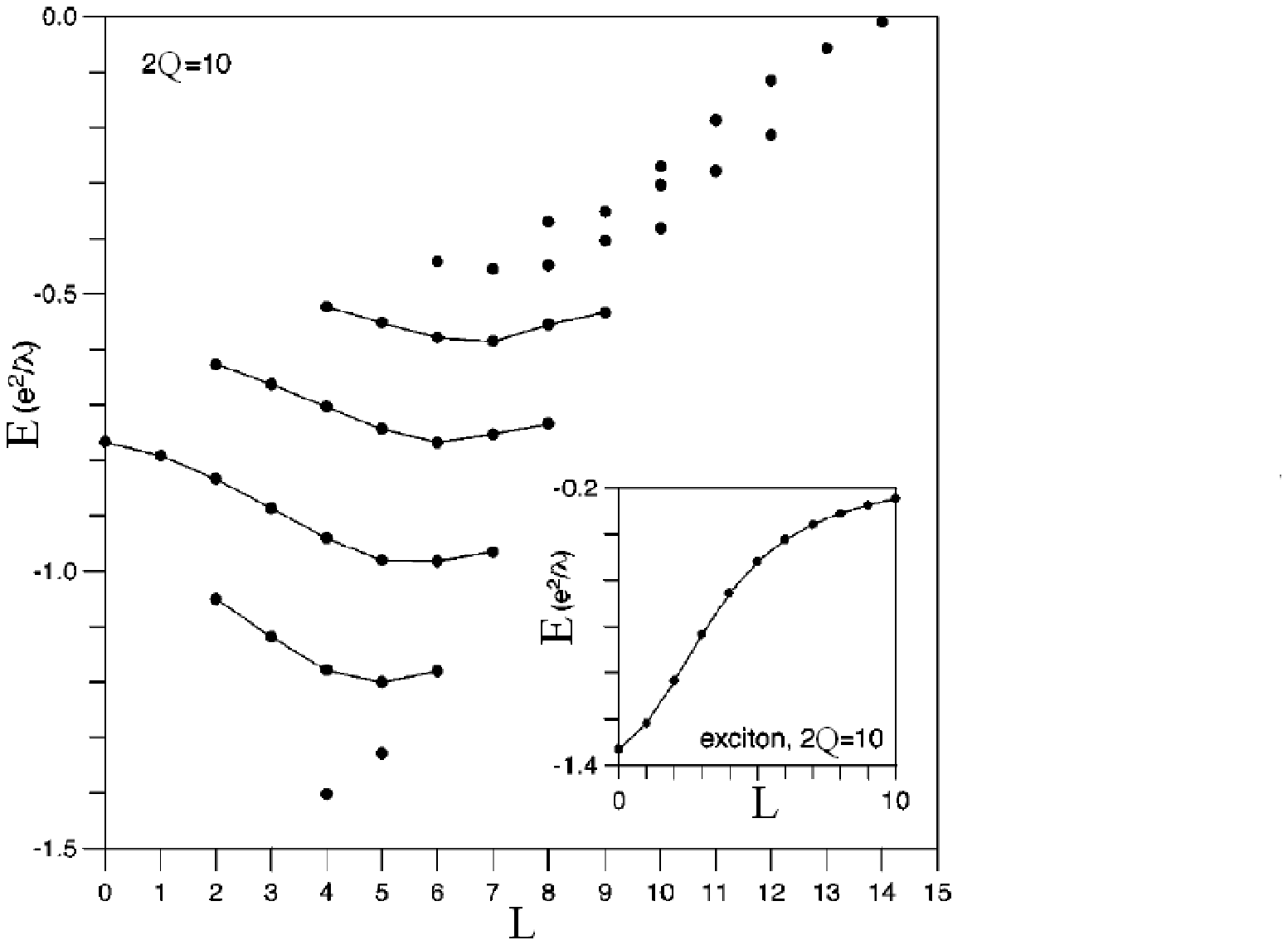}} \caption{The energy spectrum of two
electrons and one hole at $2Q=10$. Inset: the energy spectrum of an
electron-hole pair \cite{QWYRRDP}.} \label{Fig:En_spectrum_2el_1h}
\end{figure}

In Fig. \ref{Fig:En_spectrum_2el_1h}, we show the spectrum (in
magnetic units) of a system with two electrons and one hole at
$2Q=10$ as a function of the total angular momentum $L$
\cite{WojsSzYiQuinnPRB99}. The lowest energy state at $L=Q$ is the
multiplicative state with one neutral exciton in its $\ell_X=0$
ground state and one electron of angular momentum $\ell_e=Q$. Only
one state of lower energy occurs in the spectrum. It appears at
$L=Q-1$ and corresponds to the only bound state of the negatively
charged exciton $X^-$. The value of the $X^-$ angular momentum
$\ell_{X^-}=Q-1$, can be understood by noticing that the lowest
energy single particle configuration of the two electrons and one
hole is the ``compact droplet", in which the two electrons have
$z$-component of angular momentum $m= Q$ and $m = Q - 1$, and the
hole has $m= - Q$ giving $M = Q - 1$. As marked with lines in Fig.
\ref{Fig:En_spectrum_2el_1h} unbound states above the multiplicative
state form bands, which arise from the $e-h$ interaction and are
separated by gaps associated with the characteristic excitation
energies of an $e-h$ pair. (The $e-h$ pseudopotential, i.e., the
energy spectrum of an exciton, is shown in the inset). These bands
are rather well approximated by the expectation values of the total
($e-e$ and $e-h$) interaction energy, calculated in the eigenstates
of the $e-h$ interaction alone without $e-e$ interaction.

\begin{figure}
\centerline{\includegraphics[width= \linewidth]
{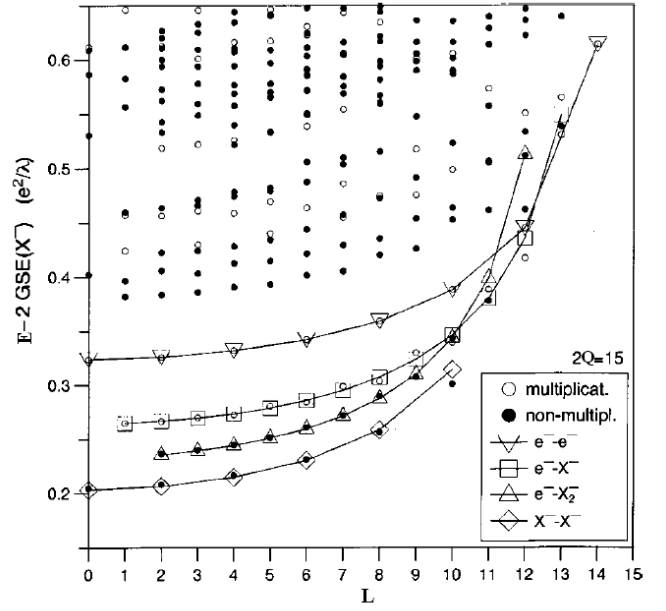}} \caption{The energy spectrum of four
electrons and two holes at $2Q=15$. Open circles: multiplicative
states; solid circles: non-multiplicative states; triangles, squares
and diamonds: approximate pseudopotentials \cite{QWYRRDP}. }
\label{Fig:En_spectrum_4el_2h}
\end{figure}

In Fig. \ref{Fig:En_spectrum_4el_2h}, we display the energy spectrum
obtained by numerical diagonalization of the Coulomb interaction of
a system of four electrons and two holes at $2Q = 15$
\cite{WojsSzYiQuinnPRB99}. The states marked by open and solid
circles are multiplicative (containing one or more decoupled $X$s)
and non-multiplicative states, respectively. For $L < 10$ there are
four rather well defined low lying bands. Two of them begin at $L =
0$. The lower of these consists of two $X^-$ ions interacting
through a pseudopotential $V_{X^--X^-}(L^{\prime})$. The upper band
consists of states containing two decoupled $X$s plus two electrons
interacting through $V_{e^--e^-}(L^{\prime})$. The band that begins
at $L = 1$ consists of one X plus an $X^-$ and an electron
interacting through $V_{e^--X^-}(L^{\prime})$, while the band which
starts at $L = 2$ consists of an $X_2^-$ interacting with a free
electron.

Knowing that the angular momentum of an electron is $\ell_e = Q$, we
can see that $\ell_{X^-_k}= Q - k$, and that decoupled excitons do
not carry angular momentum ($\ell_X = 0$). For a pair of identical
Fermions of angular momentum $\ell$ the allowed values of the pair
angular momentum are $L^{\prime} = 2\ell - j$, where j is an odd
integer. For a pair of distinguishable particles with angular
momenta $\ell_A$ and $\ell_B$, the total angular momentum satisfies
$|\ell_A-\ell_B| \leq L^{\prime} \leq \ell_A + \ell_B$. The states
containing two free electrons and two decoupled neutral excitons fit
exactly the pseudopotential for a pair of electrons at $2Q = 15$;
the maximum pair angular momentum is $L^{\prime MAX} = 14$ as
expected. By comparing this band of states with the band containing
two $X^-$s, we can obtain the binding energy of the neutral exciton
to the electron to form the $X^-$. The other binding energy, that of
a neutral exciton to an $X^-$ to form an $X^-_2$ can be obtained in
a similar way.

\subsubsection{Binding Energies}

We define $\varepsilon_0$ as the binding energy of a neutral
exciton, $\varepsilon_1$ as the binding energy of an $X$ to an
electron to form an $X^-$, and $\varepsilon_k$ as the binding energy
of an X to an $X_{k-1}^-$ to form an $X_k^-$. An estimate of these
binding energies (in magnetic energy units $e^2/\lambda$, where
$\lambda$ is the magnetic length) as a function of $2Q$ are given in
Table \ref{tab:binding_energies}.  We note clearly that
$\varepsilon_0 > \varepsilon_1 > \varepsilon_2 > \varepsilon_3$.

\begin{table}
\caption{Binding energies $\varepsilon_0,\varepsilon_1,
\varepsilon_2,\varepsilon_3$ of $X,X^{-},X^-_2$, and $X_3^-$,
respectively, in units of $e^2/\lambda$.}
\label{tab:binding_energies}
\begin{tabular}{|c||c|c|c|c|}
  \hline
  $2Q$ & $\varepsilon_0$ & $\varepsilon_1$ & $\varepsilon_2$ & $\varepsilon_3$ \\
  \hline\hline
  10 & 1.3295043 & 0.0728357 & 0.0411069 & 0.0252268 \\
  \hline
  15 & 1.3045679 & 0.0677108 & 0.0395282 & 0.0262927 \\
  \hline
  20 & 1.2919313 & 0.0647886 & 0.0381324 & 0.0260328 \\
  \hline
  \hline
\end{tabular}
\end{table}

\subsubsection{Pseudopotentials $V_{\rm{AB}}(L^{\prime})$ of Charged
Fermions}

In Fig. \ref{Fig:En_spectrum_4el_2h} the band of states containing
two $X^-$s terminates at $L^{\prime} = 10$. Since the $X^{-}$s are
Fermions, one would have expected a state at $L^{\prime
MAX}=2\ell_{X^-} - 1 = 12$. This state is missing in Fig.
\ref{Fig:En_spectrum_4el_2h}. We surmise that the state with
$L^{\prime} = L^{\prime MAX}$ does not occur because of the finite
size of the $X^-$. Large pair angular momentum corresponds to the
small average separation, and two $X^-$s in the state with
$L^{\prime MAX}$ would be too close to one another for the bound
$X^-$ to remain stable. We can think of this as a ``hard core"
repulsion for $L^{\prime} = L^{\prime MAX}$. Effectively, the
corresponding pseudopotential parameter, $V_{X^--X^-}(L^{\prime
MAX})$ is infinite. In a similar way, $V_{e^--X^-}(L^{\prime MAX})$
is infinite for $L^{\prime} = L^{\prime MAX}=14$ and
$V_{e^--X^-_2}(L^{\prime MAX})$ is infinite for $L^{\prime} =
L^{\prime MAX}=13$.

Once the maximum allowed angular momenta for all four pairings $AB$
are established, all four bands in Fig. \ref{Fig:En_spectrum_4el_2h}
can be roughly approximated by the pseudopotentials  of a pair of
point charges with angular momentum $\ell_A$ and $\ell_B$, shifted
by the binding energies of appropriate composite particles. For
example, the $X^- — X^-$ band is approximated by the $e^--e^-$
pseudopotential for  $\ell=\ell_{X^-} = Q - 1$ plus twice the $X^-$
energy. The agreement is demonstrated in Fig.
\ref{Fig:En_spectrum_4el_2h}, where the squares, diamonds, and two
kinds of triangles approximate the four bands in the
four-electron--two-hole spectrum. The fit of the diamonds to the
actual $X^- - X^-$ spectrum is quite good for $L^{\prime} < 10$. The
fit of the $e^- - X^-$ squares to the open circle multiplicative
states is reasonably good for $L^{\prime} < 12$, and the
$e^--X^-_{2}$ triangles fit their solid circle non-multiplicative
states rather well for $L^{\prime} < 11$. At sufficiently large
separation (low $L^{\prime}$), the repulsion between ions is weaker
than their binding, and the bands for distinct charge configurations
do not overlap.

There are two important differences between the pseudopotentials
$V_{AB}(L^{\prime})$ involving composite particles and those
involving point particles. The main difference is the hard core
discussed above. If we define the relative angular momentum
$\mathcal R= \ell_A+\ell_B-L^{\prime}$ for a pair of particles with
angular momentum $\ell_A$ and $\ell_B$ then the minimum allowed
relative angular momentum (which avoids the hard core) is found to
be given by
\begin{equation}
\mathcal R_{AB}^{min}=2\min (k_A,k_B)+1~, \label{eq:R_min_hard_core}
\end{equation}
where $A = X_{k_A}^-$ and $B = X_{k_B}^-$. The other difference
involves polarization of the composite particle. A dipole moment is
induced on the composite particle by the electric field of the
charged particles with which it is interacting. By associating an
"ionic polarizability" with the excitonic ion $X_k^-$, the
polarization contribution to the pseudopotential can easily be
estimated. When a number of charges interact with a given composite
particle, the polarization effect is reduced from that caused by a
single charge, because the total electric field at the position of
the excitonic ion is the vector sum of contributions from all the
other charges, and there is usually some cancelation. We will ignore
this effect in the present work and simply use the pseudopotential
$V_{AB}(L^{\prime})$ obtained from Fig. \ref{Fig:En_spectrum_4el_2h}
to describe the effective interaction.


\subsection{Generalized Composite Fermion Picture}


The electron, $X^-$, $X_2^-$,... are different types of Fermions,
all which have the same charge. These Fermions belong to different
classes $a,b,c \ldots$, distinguishable from one another. In a
system containing $N_{\alpha}$ Fermions of type $\alpha$ ($\alpha
\in a,b, \ldots$) the energy of interaction of a Fermion pair as a
function of pair angular momentum can be expressed as $V_{\alpha
\beta}(L')$. Here $\hat L' \equiv \hat \ell_i+\hat \ell_j$ is the
sum of the angular momentum $\hat \ell_i$ of the $i^{\rm{th}}$
particle of type $\alpha$ and $\hat \ell_j$ of the $j^{\rm{th}}$
particle of type $\beta$, and $\alpha, \beta$ can be in the same
class or in different classes. All of the pseudopotentials in Fig.
\ref{Fig:En_spectrum_4el_2h} are superharmonic. Because of this, the
lowest energy states in a system containing $N_{\alpha}$ Fermions of
type $\alpha$ ($\alpha \in a,b, \ldots$) will be Laughlin
correlated. We can describe the Laughlin correlations by introducing
an ``effective monopole strength" $2Q^*_{\alpha}$ seen by Fermions
of type $\alpha$ using the generalized CF picture introduced in Sec.
\ref{sec:LL1} F.

We write:
\begin{equation}
2Q_a^{\ast}=2Q-\sum_b(m_{ab}-\delta_{ab})(N_b-\delta_{ab})~.
\label{eq:Q_eff_gen_CF_transf}
\end{equation}
What we have done here is to attach to all type $a$ Fermions
$(m_{aa}- 1)$ flux quanta that couple only to the charges on all
other type $a$ Fermions and $m_{ab}$, flux quanta sensed only by
charges on the type $b$ Fermions. This is a straightforward
generalization of what we did in making in Sec. \ref{sec:LL1}F. The
coefficients $m_{ab}$ are the powers that occur in the generalized
Laughlin wavefunction, $\prod_{<i,j>}(z_i^{(a)} - z_j^{(b)}
)^{m_{ab}}$ where $z_i^{ a}$ is the complex coordinate of the
$i^{\rm{th}}$ Fermion of type $a$ and the product is over all pairs
$<i,j>$. For different multicomponent systems generalized Laughlin
incompressible states are expected to occur when (i) all the
hard-core pseudopotentials are avoided and (ii) each type of CF's
(i.e., CF$_a$s, CF$_b$s, ...) completely fills an integral number of
their angular momentum shells. In other cases, low lying multiplets
are expected to contain different kinds of CF quasiparticles
(QE$_{a}$s, QE$_b$s, $\ldots$, or QH$_a$s, QH$_b$s, $\ldots$) of the
incompressible generalized Laughlin states.

Correlations between different particles $a$ and $b$ result from
adiabatically adding $m_{ab}$ flux quanta (sensed by the charge
$q_a$ on particle $a$) to particle $b$. Because the charge $q_i$ is
the same for each of the negatively charged Fermions
($i=e,X^{-},X^{-}_2,\ldots$), $m_{ab}=m_{ba}$ produces the same
$a-b$ correlations by flux attachment to particle $a$ or to particle
$b$. In Section \ref{sec:LL1} we discuss Fermions of different
charge ($q_e=q_{\rm{FP}}/2$). In that case $q_a m_{ab}=q_b m_{ba}$
is required for the same $a-b$ correlations. This is why $2\gamma
N_e$ appeared in Eq. \ref{eq:FP_FP_e_corr} and $\gamma N_{\rm{P}}$
appeared in Eq. \ref{eq:e_FP_e_corr}.


\subsection{Low Lying Bands of $N_e$ Electron -- $N_h$ Hole Systems}


\subsubsection{Condensed States of Charged Excitons}

Consider for a moment a system containing $12$ electrons and six
holes on a Haldane spherical surface at monopole strength $2Q = 17$.
The charge configuration with the largest binding energy is that
containing six $X^-$ charged excitons. We will refer to it as (i);
its total binding energy $\varepsilon_i$ is equal to
$6(\varepsilon_0 + \varepsilon_1)$. If we make a CF transformation
on this system of $N_{X^-} = 6$ negatively charged excitons, we
obtain $2Q^{\ast}_{X^-} = 2Q - 2(N_{X-} - 1) = 7$. The angular
momentum of the $X^-$ is given by $\ell_{X^-} = Q - 1 = 15/2$ and
that of the CF $X^{-}$ by $\ell_{X^-}^{\ast} = Q _{X^-}^{\ast}- 1 =
5/2$. This means that the six CF $X^-$s completely fill the
$\ell_{X^-}^{\ast}=5/2$ shell giving a Laughlin $L=0$ incompressible
state at $\nu_{X^-}=1/3$. Note that $2\ell=\nu^{-1}(N-1)$ holds for
the quantum liquid of $X^-$s just as it did in the case of
electrons.

One point worth noting is that the generalized CF picture of a
multicomponent plasma can be thought of in terms of fictitious CF
fluxes and CF charges that have different ``colors" as discussed in
Section \ref{sec:LL1}. For example, electrons could have a red
Chern-Simons charge and $X^{-}$s a green charge. Then $m_{ee}-1$ red
and $m_{eX^-}$ green Chern-Simons fluxes would be attached to each
electron, while ($m_{X^-X^-}-1$) green and $m_{X^-e}$ red
Chern-Simons fluxes would be attached to each $X^-$.

Although $X^-$s have relatively long lifetimes for radiative
recombination of an electron-hole pair, it seems unlikely that the
Laughlin condensed state of negatively charged excitons can be
observed by standard experimental techniques used in case of
condensed states of an electron liquid. However, the PL spectrum
might give some indication of the correlations in the initial state.
For example, if the ground state of a twelve electron-six hole
system underwent $e-h$ recombination, the initial state would be an
$L=0$ IQL state of six $X^-$ excitons. Many different final states
of the eleven electron-five hole system would be possible.
Evaluating their eigenvalues and eigenfunction by numerical
diagonalization would allow one to identify the energies and
oscillator strength associated with different PL peaks. Selection
rules would depend on sample properties like quantum well width,
ratio of $e^2/\lambda$, Coulomb energy scale, to the LL separation,
impurity concentration, etc. Such PL processes have not yet been
studied in detail.

\subsubsection{Other Charge Configurations}

\begin{figure}
\centerline{\includegraphics[width= \linewidth]
{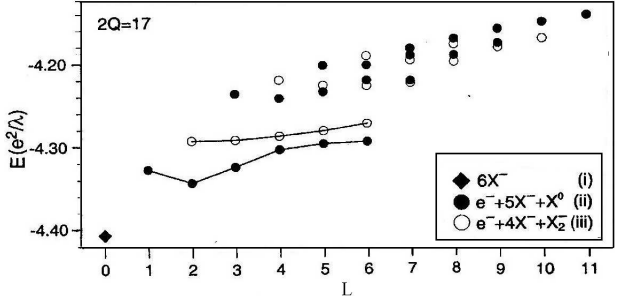}} \caption{Low energy spectra of different
charge configurations of $12e+6h$ on a Haldane sphere at $2Q=17$:
$6X^-$ (diamonds), $e^-+5X^-+X$ (solid circles), and
$e^-+4X^-+X_2^-$ (open circles) \cite{WojsSzYiQuinnPRB99}.}
\label{Fig:En_spectrum_12el_6h}
\end{figure}

For the 12-electron-6-hole system, other charge configurations
besides the six $X^-$s can occur as excited states. Among these are
(ii) $e^- + 5X^- + X$ with total binding energy $\varepsilon_{ii} =
6\varepsilon_0 + 5\varepsilon_1$, and (iii) $e^- + 4X^- + X_2^-$
with total energy $\varepsilon_{iii} = 6\varepsilon_0 + 5
\varepsilon_1 + \varepsilon_2$. The total energy of any state
depends on the interaction energy of the constituent charged
particles as well as the binding energy. The system of eighteen
particles (12 electrons and 6 holes) at $2Q = 17$ is too large for
us to diagonalize in terms of the electrons and holes and their
interactions. However we can obtain a reasonable approximation to
the low lying energy spectrum by considering the different charge
configurations denoted by (i) through (iii) each of which contains
only six charged Fermions. We make use of our knowledge of the
binding energies, angular momenta, and pseudopotentials
$V_{AB}(L^{\prime})$ where $A$ and $B$ can be $e^-$, $X^-$ or
$X_2^-$. The results of this simpler numerical calculation are
presented in Fig. \ref{Fig:En_spectrum_12el_6h}
\cite{WojsSzYiQuinnPRB99}. There is only one low lying state of the
six $X^-$ configuration, the $L = 0$ Laughlin $\nu_{X^-} = 1/3$
state. There are two bands of states in each of the charge
configurations (ii) and (iii). The results presented in Fig.
\ref{Fig:En_spectrum_12el_6h} can be understood from the generalized
CF model. The CF predictions are: (i) For the system of $N_{X^-} =
6$, we take $m_{X^-X^-} = 3$ and obtain the Laughlin $\nu_{X-} =
1/3$ state as discussed earlier. Because of the hard core of the
$X^- - X^-$ pseudopotential,  this is the only state of this charge
configuration. (ii) For the $e^- +5X^- +X$ configuration, we can
take $m_{X^-X^-} = 3$ and $m_{eX^-} = 1,2,$ or $3$. For $m_{eX^-} =
1$ we obtain $L=1\oplus 2 \oplus 3^2 \oplus 4^2 \oplus 5^3 \oplus
6^3 \oplus 7^3 \oplus 8^2 \oplus 9^2 \oplus 10 \oplus 11$. For
$m_{eX^-} = 2$ we obtain $L=1\oplus 2 \oplus 3 \oplus 4 \oplus 5
\oplus 6 $ and for $m_{eX^-} = 3$ we obtain $L=1$. (iii) For the
grouping $e^-+4X^-+X_2^-$, we set $m_{X^-X^-}=3$, $m_{eX_2^-}=1,
m_{X^-X_2^-}=3$ and $m_{eX^-}=1,2$ or $3$. For $m_{eX^-}=1$, we
obtain $L=2\oplus 3 \oplus 4^2 \oplus 5^2 \oplus 6^3 \oplus
7^2\oplus 8^2 \oplus 9 \oplus 10$. For $m_{eX^-}=2$, we obtain the
multiplets $L=2\oplus 3 \oplus 4\oplus 5\oplus 6$, and for
$m_{eX^-}=3$, we have $L=2$ \cite{QWYRRDP}. In the groupings (ii)
and (iii) the sets of multiplets obtained for higher values of
$m_{eX^-}$ are subsets of those obtained for lower values of
$m_{eX^-}$. We would expect them to form lower energy bands since
they avoid additional $\mathcal R_{eX^-}$. However, note that the
(ii) and (iii) states predicted for $m_{eX^-} = 3$ (at $L=1$ and
$2$, respectively) do not form separate bands in Fig.
\ref{Fig:En_spectrum_12el_6h}. This is because the $V_{eX^-}$
increases more slowly than linearly as a function of
$L^{\prime}(L^{\prime} + 1)$ in the vicinity of $\mathcal
R_{eX^-}=3$. In such case the CF picture fails
\cite{QuinnWoysSSC98,WojsQuinnSSC99}.

The agreement of our CF predictions with the data in  Fig.
\ref{Fig:En_spectrum_12el_6h} is really quite remarkable and
strongly indicates that the multicomponent CF picture is correct. We
were indeed able to confirm predicted Jastrow type correlations in
the low lying states by calculating their coefficients of fractional
parentage \cite{NuclearShellBook}. We have also verified the CF
predictions for other systems that we were able to treat
numerically. If exponents $m_{ab}$ are chosen correctly, the CF
picture works well in all cases.

\subsection{Spectra of $N_e$ Electron-Single Hole System}


In PL experiments the absorption of light creates a small number of
electron-valence hole pairs in a quantum well that already has a
concentration of conduction electrons. Because $N_h \ll N_e$, the
valence holes are rather far apart, and the PL spectrum is not
influenced by $h-h$ interactions. One can evaluate the energies and
wavefunctions for a single hole interacting with a gas of $N_e$
electrons, investigate the allowed final states of $N_e-1$
electrons, and calculate the energy and intensity of the PL spectrum
lines. For this reason, it is useful to study the eigenstates of a
$N_e$-electron -- 1-valence hole system. In order to remove the
``hidden symmetry" that decouples the PL spectrum from the
correlations in the underlying electron gas, we assume that the
electrons and the hole reside on  different 2D planes separated by a
distance $d$ between zero and four magnetic lengths, $\lambda=(\hbar
c/eB)^{1/2}$. We take the cyclotron energies to be large compared to
the Coulomb energy ($e^2/\lambda$) so that only a single LL for each
kind of carrier is necessary.

\begin{figure}
\centerline{\includegraphics[width= \linewidth] {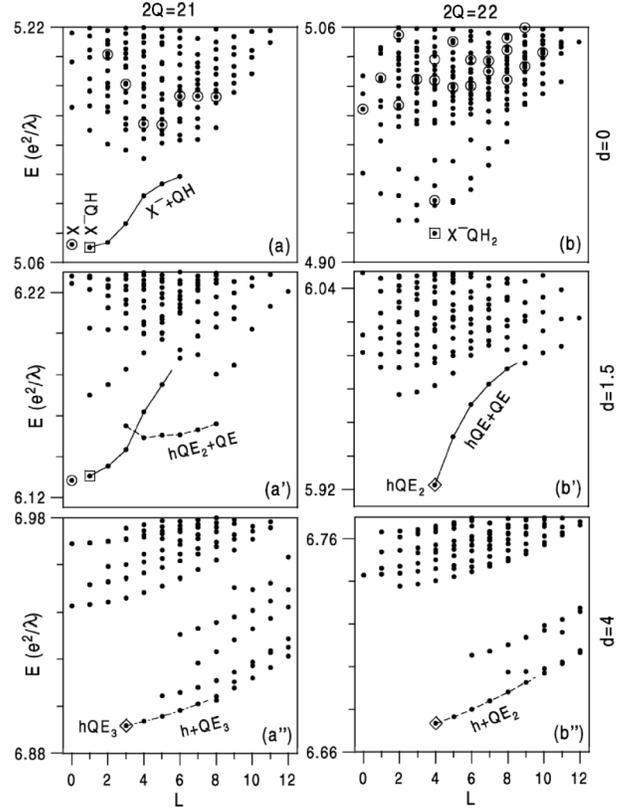}}
\caption{Energy spectra of nine-electron -- one-hole system for the
monopole strength $2Q=21,22$ (from left to right), and for the
interplane separation $d=0,1.5,4$ (from top to bottom). Lines and
open symbols mark the low states containing different bound
excitonic complexes \cite{QuinnWQuinnBPhysE01_1}.}
\label{Fig:E_9e_1h}
\end{figure}

In  Fig. \ref{Fig:E_9e_1h} we present the energy spectra for a
system of 9 electrons and 1 valence band hole at three different
values of separation $d$ between the 2D layers and two different
values of the monopole strength $2Q$ \cite{QuinnWQuinnBPhysE01_1}.

For $d\ll 1$ we have strong coupling between the electrons and the
hole. Neutral ($X$) and charged excitons $X^{-}$ are found. The
multiplicative states at $d = 0$ are shown as solid dots surrounded
by a small circle. Non-multiplicative states at $d = 0$ can have an
$X^-_{\rm{t}}$ exciton interacting with the remaining $N - 2$
electrons. For $d  \gg 1$ the valence hole interacts very weakly
with the $N$-electron system, and the spectra can be described in
terms of the eigenstates of the $N$-electron system multiplied by
the eigenfunction of the hole with total angular momentum $\hat L =
\hat L_e + \hat \ell_h$. For intermediate values of d ($d \simeq 2$)
the $e-h$ interaction is not a weak perturbation on the electronic
eigenstates, but it is not always strong enough to bind a full
electron to form an exciton.

For $d = 0$, $X$ and $X^-$ bound states occur. Due to the ``hidden
symmetry", the multiplicative states containing an $X$ have the same
spectrum as the eight electron system shifted by the $X$ binding
energy. The CF model \cite{WojsSzYiQuinnPRB99,QuinnWoysSSC98,
WojsQuinnSSC99} tells us the effective monopole strength seen by one
CF in a system of $N^{\prime} =N-1 = 8$ electrons near $\nu = 1/3$
is $2Q^{\ast} = 2Q-2(N^{\prime}-1)$. $Q^{\ast}$ plays the role of
the angular momentum of the lowest CF electron shell, therefore
$Q^{\ast}=7/2$ and $4$ for the multiplicative states in frame (a)
and (b) of Fig. \ref{Fig:E_9e_1h}, respectively. Since the lowest CF
shell can accommodate $2Q^{\ast}+1$ CFs, it is exactly filled in
frame (a), but there is one empty state in the $\ell^{\ast}=4$ CF
level, or one QH of angular momentum $\ell_{\rm{QH}}=4$ in frame
(b). Thus the lowest multiplicative states have $L=0$ in frame (a)
and $L=4$ in frame (b). The magnetoroton  band of multiplicative
states in frame (a) is clearly marked. It has $2 \leq L \leq 8$, and
it is contained within the quasicontinuum of non-multiplicative
states.

For the non-multiplicative states we have one $X^-$ and $N_e = N -
2$ remaining electrons. The generalized CF picture
\cite{WojsSzYiQuinnPRB99} allows us to predict the lowest energy
band in the spectrum in the following way. The effective monopole
strength seen by the electrons is $2Q^{\ast}= 2Q - 2(N_e - 1) -
2N_{X^-}$, while that seen by the $X^-$ is $2Q^{\ast}_{X-} = 2Q-
2N_e$. Here, we have attached to each Fermion (electron and $X^-$)
two fictitious flux quanta and used the mean field approximation to
describe the effective monopole strength seen by each particle (note
that a CF does not see its own flux). The angular momentum of the
lowest CF electron shell is $\ell_0^{\ast} = Q_e^{\ast}$, while that
of the CF $X^-$ shell is $\ell_{X^-}=Q^{\ast}_{X^-}-1$
\cite{WojsSzYiQuinnPRB99,QuinnWQuinnBPhysE01_1}. For the system with
$N_e=7$ and $N_{X^-}=1$ at $2Q=21$ and 22, the generalized CF
picture leads to: one QH with $\ell_{\rm{QH}}=7/2$ and one $X^-$
with $\ell_{X^-}^{\ast}=5/2$, giving a band at $1 \leq L \leq 6$ for
Fig. \ref{Fig:E_9e_1h} (a) and two QHs with $\ell_{QH}=4$ and one
$X^-$ with $\ell_{X^-}^{\ast}=3$ giving $L=0\oplus 1 \oplus 2^3
\oplus 3^3 \oplus 4^4 \oplus 5^3 \oplus 6^3 \oplus 7^2 \oplus 8^2
\oplus 9 \oplus 10$ for Fig. \ref{Fig:E_9e_1h} (b).

For $d \gg 1$, the electron-hole interaction is a weak perturbation
on the energies obtained for the $N$-electron system. The numerical
results can be understood by adding the angular momentum of a hole
$\ell_h=Q$, to the electron angular momentum obtained from the
simple CF model. The predictions are: for $2Q=21$ there are three
QEs each with $\ell_{\rm{QE}} = 7/2$ and the hole has $\ell_h =
21/2$; for $2Q = 22$ two QEs each with $\ell_{QE}=4$ and
$\ell_h=11$. Adding the angular momenta of the identical Fermion QEs
gives $L_e$, the electron angular momenta of the lowest band; adding
to $L_e$ the angular momentum $\ell_h$ gives the allowed  set of
allowed multiplets appearing in the low energy sector. For example,
in Fig. \ref{Fig:E_9e_1h} (b'') the allowed values of $L_e$ are $1
\oplus 3 \oplus 5 \oplus 7$, and the multiplets at 7 and 3 have
lower energy than at 1 and 5. Four low energy bands appear at $4
\leq L \leq 18$, $8 \leq L \leq 14$, $6 \leq L \leq 16$, and $10
\leq L \leq 12$, resulting from $L_e = 7, 3, 5$, and 1,
respectively.

For $d \approx 1$, the electron-hole interaction results in
formation of bound states of a hole and one or more QEs. In the
two-electron--one-hole system, the $X$ and $X^-$ unbind for $d
\approx 1$, but interaction with the surrounding unbound electrons
in a larger system can lead to persistence of these excitonic states
beyond d = 1. For example, the band of states at $d=0$ in Fig.
\ref{Fig:E_9e_1h} (a) that we associated with an $X^-$ interaction
with a QH persists at $d = 1.5$ in Fig. \ref{Fig:E_9e_1h} (a').
However, it appears to cross another low energy band that extends
from $L = 3$ to 8. This latter band can be interpreted in terms of
three QEs interacting with the hole as done in the weak-coupling
limit shown in Fig. \ref{Fig:E_9e_1h} (a''). The other bands of the
weak coupling regime (those beginning at $L = 5, 6, 7, 8,$ and 9)
have disappeared into the continuum of higher states as a result of
the increase of $V_{eh}$.

For $2Q = 22$, the lowest band can be interpreted in terms of one
$X^-$ interacting with two QHs of the generalized CF picture. The
$X^-$ has $\ell^{\ast}_{X^-} = 3$ and each QH  has $\ell_{\rm{QH}} =
4$. The allowed values of $L_{2\rm{QH}}$ are 7, 5, 3, and 1, and the
molecular state QH$_2$ which has the smallest average QH-QH distance
would have $\ell_{\rm{QH}_2} =7$. This gives a band of $X^- +
\rm{QH}_2$ states going from $L = \ell_{\rm{QH}_2} -
\ell^{\ast}_{X^-} = 4$ to $L = \ell_{\rm{QH}_2} + \ell^{\ast}_{X^-}
= 10$. A higher band beginning at $L = 2$ might be associated with a
2QH state at $L_{2QH} = 5$ interacting with an $X^-$. The origin of
the other bands is less certain.

It is worth noting that the $X^-{\rm{QH}}$ band in Fig.
\ref{Fig:E_9e_1h}(a) resembles the neutral exciton band shown in the
inset of Fig. \ref{Fig:En_spectrum_2el_1h}. The latter band begins
at $ L=0$ because $\ell_e=\ell_h=5$ and $|\ell_e-\ell_h|\leq L \leq
\ell_e+\ell_h$. For the $X^-QH$ band $\ell_{X^-}=Q^*-1=5/2$ and
$\ell_{\rm{QH}}=Q^*=7/2$ giving a band starting at
$L=\ell_{\rm{QE}}-\ell_{X^-}=1$ and ending at
$L=\ell_{\rm{QE}}+\ell_{\rm{QH}}=6$. The width of this band is
smaller than that of the neutral $X$ by at least an order of
magnitude. This reflects the fact that the magnitude of the
effective charges of the correlated $X^-$ (and of the QH) is one
third of the electron charge, and of the fact that the charge is
spread over a wider region. This makes it clear that the correlated
$X^-$ can be thought as a quasi-$X^-$ (${\rm{QX}}^-$), and just like
the Laughlin QH it has an effective charge of magnitude 1/3. This
allows us to call the ${\rm{QH}}-{\rm{Q}}X^-$ band state a neutral
quasiexciton ${\rm{Q}}X^0$



\section{Photoluminescence}

\subsection{General Considerations}

Exact numerical diagonalization gives both the eigenvalues and the
eigenfunctions. The low energy states $|i\rangle$ of the initial
$N$-electron--one-hole system have just been discussed. The final
states $|f\rangle$ contain $N^{\prime} = N-1$ electrons but no
holes. The recombination of an electron-hole pair is proportional to
the square of the matrix element of the photoluminescence operator
$\hat \mathcal L$, where $\hat \mathcal L = \int d^3 r \Psi_e(\vec
r) \Psi_h(\vec r)$ and $\Psi_e(\vec r)$ (or $\Psi_h(\vec r)$)
annihilates an electron (or hole). We have evaluated $|\langle f|
\hat \mathcal L|i\rangle|^2$ for all of the low-lying initial states
and have found the following results \cite{WojsQuinnPRB01_1}. (i)
Conservation of the total angular momentum $L$ is at most weakly
violated through the scattering of spectator particles (electrons
and quasiparticles) which do not participate directly in the
recombination process if the filling factor $\nu$ is less than
(approximately) $1/3$. (ii) In the strong coupling region, the
neutral $X$ line is the dominant feature of the PL spectrum. The
$X^-{\rm{QH}}_2$ state has very small oscillator strength for
radiative recombination. (iii) For intermediate coupling, the
$h{\rm{QE}}_2$ and an excited state of the $h$QE (which we denote by
$h$QE$^{\ast}$) are the only states with large oscillator strength
for photoluminescence.

At zero temperature ($T = 0$), all initial states must be ground
states of the $N$-electron--one-hole system. At finite but low
temperatures, excited initial states contribute to the PL spectrum.
The photoluminescence intensity is proportional to
\begin{equation}
w_{i\rightarrow f}=\frac{2\pi}{\hbar}\mathcal Z^{-1} \sum_{i,f}
e^{-\beta E_i} \left|\langle f|\hat \mathcal L|i\rangle\right|^2
\delta(E_i-E_f-\hbar \omega)~, \label{eq:PL_intensity}
\end{equation}
where $\beta=(kT)^{-1}$ and $\mathcal Z=\sum_i e^{-\beta E_i}$.

\subsection{Singlet and Triplet Charged Excitons: Photoluminescence
for Dilute ($\nu \ll 1$) Systems}

Only spin polarized charged excitons (with $S=1 $) are bound when
the ratio $(\hbar \omega_C) / (e^2 /\lambda)$ tends to infinity. In
real systems at finite values of this parameter, both singlet
$(S=0)$ and triplet $(S = 1)$ charged excitons occur. According to
the theory  \cite{WojsQuinnPRB01_1} the singlet $X_{\rm{s}}^-$ is
the ground state at low values of the magnetic field, while the
triplet $X_{\rm{t}}^-$ is the ground state at very high magnetic
fields. Numerical calculations of the ground states of both the
singlet and triplet charged excitons \cite{WojsQuinnPRB01_1}
indicated a crossing at roughly $B\approx 30~{\rm{T}}$ for a
symmetric GaAs quantum well, the width of which was about
$w=10~{\rm{nm}}$. Observation of PL by Hayne et al.
\cite{HayneetalPRB99} displaying three peaks that were interpreted
as the $X$, $X_{\rm{t}}^-$, and $X_{\rm{s}}^-$, showed no crossing
of the $X_{\rm{t}}^-$ and $X_{\rm{s}}^-$ up to the fields of 50
Tesla. This led the experimenters to question the validity of
variational calculations.

In this section we study very small systems (either two or three
electrons and one valence band hole) in narrow ($w \sim $11.5 nm)
symmetric GaAs quantum wells. We include the effects of Landau level
mixing caused by the interactions, and the effect of finite well
width on the effective interaction. Only a single subband is used in
the calculations, since the quantum well is relatively narrow. Both
electrons and holes are described in the effective mass
approximation, and interband coupling is partially accounted for by
a magnetic field dependence of the cyclotron mass of the hole (taken
from experimental data). The Zeeman energy depends on both the well
width and the magnetic field $B$. Five Landau levels for both the
electrons and holes were included in the calculation in order to
obtain satisfactory convergence. The energies obtained for different
values of the monopole strength $2Q$ were extrapolated to the large
$Q$ limit to eliminate finite-size effects.

\begin{figure}
\centerline{\includegraphics[width= \linewidth]
{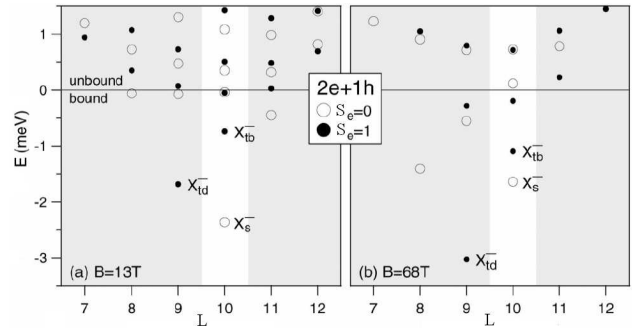}} \caption{Energy spectra (binding energies
vs. angular momentum) of the two-electron -- one hole system on a
Haldane sphere with the Landau level degeneracy $2Q+1=21$. $S_e$
denotes the total electron spin. The parameters are appropriate for
the 11.5 nm GaAs quantum well \cite{QuinnWPhysE00_1}.}
\label{Fig:_E_2e_1h_spin}
\end{figure}

The energy spectra of the two-electron--one-hole system calculated
for $2Q = 20$ are shown in Fig. \ref{Fig:_E_2e_1h_spin}. Open and
solid symbols mark singlet and triplet states ($S_e$ is the total
electron spin), and each state with $L> 0$ represents a degenerate
$L$ multiplet. Since the PL process (annihilation of an $e-h$ pair
and emission of a photon) occurs with conservation of angular
momentum, only states from the $L = Q$ channel are radiative
\cite{QuinnWPhysE00_1}. Recombination of other non-radiative states
requires breaking rotational symmetry (e.g., by collisions with
electrons). This result is independent of the chosen spherical
geometry and holds also for a planar quantum well, except that the
definition of the conserved momentum is different
\cite{DzyubenkoPRL00,DzyubenkoPRB94,DzyubenkoPRB93}.

The occurrence of a strict PL selection rule at finite $B$ may seem
surprising, since the hidden symmetry that forbids the
$X_{\rm{td}}^-$ recombination in the lowest LL does not hold when
the mixing with higher LLs is included. (The ``d" in $X_{\rm{td}}^-$
means ``dark" and $X_{\rm{td}}^{-}$ is called the dark triplet
because it is forbidden to decay radiatively.) However, it is both
the hidden symmetry and the above-mentioned angular momentum
conservation that independently forbid the $X_{\rm{td}}^{-}$
recombination, and the latter remains valid at finite $B$. Although
the hidden symmetry and resulting $N_X$ conservation law no longer
hold at finite $B$, the $X_{\rm{td}}^{-}$ recombination remains
strictly forbidden because of the independently conserved $L$.

We expect breaking of both symmetries for real experimental
situations. The presence of impurities and defects, and
$e-X_{\rm{td}}^{-}$ scattering during recombination in the presence
of excess electrons can relax the strict conservation of the $X^-$
angular momentum in the radiative decay. However, for narrow and
symmetric quantum wells containing a relatively small number of
excess electrons, the symmetries may only be weakly broken and some
remnant of the strict conservation laws may survive.

\begin{figure}
\centerline{\includegraphics[width= \linewidth]
{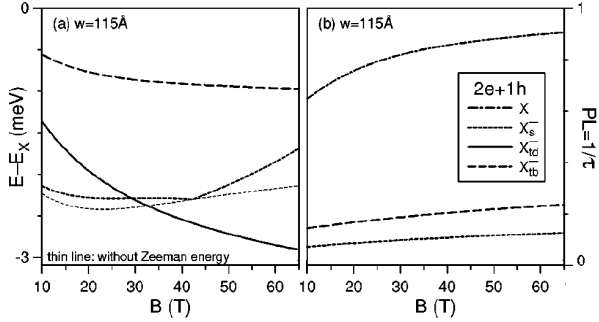}} \caption{The $X^-$ energies (a) and
oscillator strength (b) in the $11.5$ nm GaAs quantum well plotted
as a function of the magnetic field \cite{WojsQuinnHawrylakPRB00}.}
\label{Fig:pl_Strength}
\end{figure}

Three states marked in Fig. \ref{Fig:_E_2e_1h_spin} are of
particular importance: $X_{\rm{s}}^-$ and $X_{\rm{tb}}^-$ (``b"
stands for "bright") are the only strongly bound radiative states,
while $X_{\rm{td}}^{-}$ has by far the lowest energy of all
non-radiative states. The radiative triplet bound state
$X_{\rm{tb}}^{-}$ was identified for the first time by W\'{o}js et
al. \cite{WojsQuinnHawrylakPRB00, QuinnWPhysE00_1}. The binding
energies of all three $X^-$ states are extrapolated to $\lambda / R
\rightarrow 0$ and plotted in Fig \ref{Fig:pl_Strength} (a) as a
function of $B$. For the $X^-_{\rm{s}}$, the binding energy differs
from the PL energy (indicated by thin dotted line) by the Zeeman
energy needed to flip one electron's spin, and the cusp at $B
\approx 42$ T is due to the change in sign of the electron
$g$-factor. For the triplet states, the PL and binding energies are
equal. The energies of $X_{\rm{s}}^-$ and $X_{\rm{td}}^-$ behave as
expected: The binding of $X^-_{\rm{s}}$ weakens at higher $B$ and
eventually leads to its unbinding in the infinite field limit; the
binding energy of $X_{\rm{td}}^{-}$ changes as $e^2/\lambda \propto
\sqrt {B}$; and the predicted transition from the $X_{\rm{s}}^-$ to
the $X_{\rm{td}}^-$ ground state at $B \approx 30 T$ is confirmed.
The new $X_{\rm{tb}}^-$ state remains an excited triplet state at
all values of $B$, and its binding energy is smaller than that of
$X_{\rm{s}}^-$ by about 1.5 meV. The oscillator strengths
$\tau^{-1}$ of a neutral exciton X and the two radiative $X^{-}$
states are plotted in Fig. \ref{Fig:pl_Strength} (b). In the
two-electron--one-hole spectrum, the strongly bound $X_{\rm{s}}^-$
and $X_{\rm{tb}}^-$ states share a considerable fraction of the
total oscillator strength of one $X$, with $\tau_{\rm{tb}}^{-1}$
nearly twice larger than $\tau_{\rm{s}}^{-1}$.

The comparison of calculated magnitude and magnetic field dependence
of the $X^{-}$ binding energies with the experimental PL spectra, as
well as high oscillator strength of the $X_{\rm{tb}}^-$ , lead to
the conclusion that the three peaks observed in PL are the $X$,
$X_{\rm{s}}^-$,and $X^-_{\rm{tb}}$.

To understand why the $X_{\rm{td}}^{-}$ state remains optically
inactive even in the presence of collisions, the $e-X^-$ interaction
must be studied in greater detail. Our numerical results for a three
electron--one-hole system indicate that the lowest band of states
consists of a triplet $X^-$ and one unbound electron. Because the
$X_{\rm{t}}^- - e$ pseudopotential is superharmonic, in real
experimental systems at a low electron concentration ($\nu \leq
1/3$) Laughlin correlations between the electron and $X_{\rm{t}}^-$
will effectively isolate the $X_{\rm{t}}^-$ from the surrounding 2D
electron system. This prevents close collisions of the
$X_{\rm{t}}^-$ and the spectator electron during the $e-h$
recombination. Although the $X_{\rm{td}}^-$ is no longer forbidden
to decay radiatively since the spectator electron can change its
angular momentum in the recombination process, this scattering
process is weak for  $\nu < 1/3$. The oscillator strength for
radiative decay of the $X_{\rm{td}}^-$ is found to be more than an
order of magnitude smaller than those of the $X_{\rm{s}}^-$ and
$X_{\rm{tb}}^-$. These results support the interpretation that the
three peaks observed in many experiments correspond to the $X$,
$X_{\rm{s}}^-$, and $X_{\rm{tb}}^-$. The $X_{\rm{td}}^-$ is not
observed because of its small oscillator strength. The
$X_{\rm{td}}^-$ recombination line has been observed
\cite{YusaShtrikmanBarJosephPRL01}, when special care (very low
temperatures and high quality samples) was taken to detect its weak
signal. Even more convincing is the comparison with infrared
absorption at very low temperature where only the $X_{\rm{td}}^-$
state is heavily occupied. Absorption spectra show only one strong
peak in contrast to PL spectra which shows three, because the higher
population of the $X_{\rm{td}}^-$ compensates for its lower
oscillator strength for photon absorbtion compared to the
$X_{\rm{s}}^-$ and $X_{\rm{tb}}^-$ \cite{HamburgPRB02,HamburgPRL03}.


\subsection{$X^{-}$ in an Incompressible Quantum Liquid of Electrons:
Fractionally Charged Quasiexcitons}

In Fig. \ref{Fig:E_9e_1h} (a) and (a') we observed both a low energy
multiplicative state (consisting of a neutral exciton effectively
decoupled from the remaining $N_e^{\prime}=N_e-1$ unbound electrons)
at $L=0$, and a band of non-multiplicative states extending from
$L=1$ to $L=6$. This band could be identified  (using a generalized
CF picture) as a QH of the Laughlin IQL state coupled to the $QX^-$
(which has Laughlin correlations with the $N_e^{\prime}=N_e-2$
unbound electrons). In Section \ref{sec:e_vh_syst} we discussed how
it could be thought of as a neutral quasiexciton $QX^0$. We will
sometimes use the symbols ($\chi^-$, $\chi$, $\chi^+$) for the three
different possible quasielectrons in place of ($QX^-$, $QX$,
$QX^{+}$). $\chi$ is the neutral $QX$ (a bound state of $\chi^-$ and
a QH), while $\chi^+$ is a positively charged $QX^{+}$ ( a bound
state of $\chi$ and a QH).

In this subsection we review the many-body correlations associated
with negatively charged excitons (or trions) immersed in a Laughlin
IQL state, and predict a discontinuity of the PL spectrum at
$\nu=1/3$ \cite{PinczukPRL90,HawrilakNature,WoysGladyQuinnPRB06} and
for spin-polarized systems, we elucidate the earlier theory
\cite{ApalkovRashbaPRB92,ApalkovRashbaPRB93,ZangBirmanPRB95} by
identifying the ``dressed exciton" with $\chi$, its suppressed
dispersion with the $\chi^{-1/3}$-QH pseudopotential of interaction
among two Laughlin charge quanta, and the ``magnetoroton-assisted
emission" with the $\chi^{-1/3}$ recombination.

Photoluminescence from systems containing a small number of
quasiexcitons ($\chi^-$, $\chi$, $\chi^+$) reflect the properties of
these quasiexcitons in the initial state. The $\chi^-$ is formed
when a valence band hole binds two electrons to form an $X^-$, which
then becomes Laughlin correlated with the remaining unbound
electrons (in, for example, an IQL $\nu=1/3$ state). If several
$\chi^-$ quasielectrons are present,they repel one another. Then the
radiative $e-h$ recombination is essentially that of an isolated
$\chi^-$ in the IQL state of the remaining electrons. If the
magnetic field is increased to values that make $\nu>1/3$, Laughlin
QHs will be present. The $\chi^-$ attracts the Laughlin QHs, and can
form a neutral ($\chi^0$) or positively charged ($\chi^+$)
quasiexcitons. The resulting PL spectrum would be expected to
reflect the properties of the initial $\chi^-$ or $\chi^+$ for
$\nu>1/3$ and $\nu<1/3$ respectively. For $\nu$ very close to the
IQL value ($\nu=1/3$) an initial $\chi^0$ state might also be
observed. The PL from different initial state could have different
energies and different intensities, so observing a charge in the PL
spectrum as $\nu$ passes through an IQL value like $\nu=1/3$ is not
surprising \cite{HamburgPRB02,HamburgPRL03,PinczukPRL90}

We illustrate these concepts by use of exact numerical
diagonalization for $N \leq 10$ electrons and one valence hole on a
Haldane sphere \cite{HaldanePRL83} with radius $R$, magnetic
monopole strength $2Q=4 \pi R^2Be/hc$, and magnetic length
$\lambda=R/\sqrt{Q}$. The second-quantization Hamiltonian reads $H =
\sum_i U_i c_i^{\dag} c_i + \sum_{ijkl} V_{ijkl} c_i^{\dag}
c_j^{\dag} c_k c_l $.  Here, $c_i^{\dag}$ and $c_i$ are operators
creating and annihilating an electron in the conduction band or a
hole in the valence band, in the state labeled by a composite index
$i$ containing all relevant single-particle quantum numbers (band,
subband, and LL indices, angular momentum, and spin). The single
particle energies are measured from the ground states in conduction
and valence bands, respectively. The Coulomb interaction matrix
elements $V$ were integrated in 3D by taking the actual electron and
hole subband wavefunctions $\phi(z)$ calculated self-consistently
\cite{TanSniderChangHuJAP90} for $w=10$ and $20$ nm GaAs quantum
wells, doped on one side to $n=2 \times 10^{11}$ cm$^{-2}$ (yielding
$\nu= 1/3$ at $B=25$ T). The diagonalization was carried out in
configuration-interaction basis, $|i_1 , \cdots , i_N;
i_h>=c_{i_1}^{\dag} \cdots c_{i_N}^{\dag} c_{i_h}^{\dag} |vac>$,
where indices $i_1 \cdots i_N$ denote the occupied electron states,
and $i_h$ describes the hole. Finite size and surface curvature
errors were minimized by extrapolation to the $\lambda/R \rightarrow
0$ limit. The combination of closed geometry, used as an alternative
to periodic boundary conditions for modeling in-plane dynamics, with
exact treatment of the single-particle motion in the normal
direction allowed for quantitative estimates of binding energies
characterizing extended experimental systems.

\begin{figure}
\centerline{\includegraphics[width= \linewidth]
{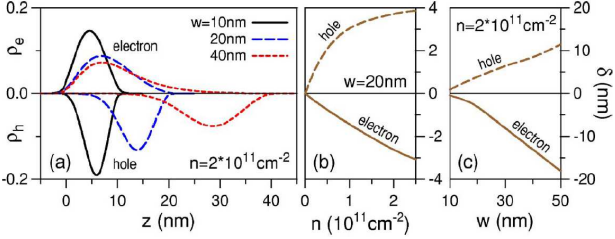}} \caption{(color online) (a) Lowest
subband electron and heavy-hole charge density profiles in the
normal direction $\rho(z)$ for one-sided doped GaAS quantum wells
(b), (c) Displacements $\delta$ of the density maxima from the
center of the quantum well as a function of electron concentration
$n$ and well width $w$ \cite{WoysGladyQuinnPRB06}.}
\label{Fig:density_profile}.
\end{figure}
We begin with the calculation of $X^-$ Coulomb binding energies
$\Delta$ using $\phi(z)$, i.e., in the mean normal electric field
due to a doping layer, but ignoring in-plane $X^--$IQL coupling. We
included five LLs and two $\phi$-subbands for both $e$ and (heavy)
$h$. The lowest-subband $e$ and $h$ density profiles for $w$=10, 20,
and 40 nm are plotted in Fig. \ref{Fig:density_profile} (a). The
effect of charge separation in wider wells is evident. The shifts of
the density maxima as a function of $n$ and $w$ are shown in Figs.
\ref{Fig:density_profile} (b) and \ref{Fig:density_profile} (c). For
the cyclotron energies $\omega_c$ (at $B=25$ T; after experiment of
\textcite{ColeetalPRB97}) and intersubband gaps $\Omega_s$ (from own
calculations) we took $\omega_{ce}=44.5$ meV, $\omega_{ch}=7.7$ meV,
$\Omega_{se}=29.6$ meV; $\Omega_{sh} =10.0$ meV for $w=20$ nm, and
$\omega_{ce}=44.5$ meV, $\omega_{ch} =8.1$ meV, $\Omega_{se}=89.8$
meV; $\Omega_{sh}=24.5$ meV for $w=10$ nm. The valence subband
mixing was neglected. The result for $w=10$ nm is $\Delta_s=2.3$ meV
and $\Delta_t=1.5$ meV, in qualitative agreement with earlier work.
\cite{StebeAinane,WhittakerShieldsPRB97, RivaPeetersVargaPRB01,
WojsQuinnHawrylakPRB00}, which also predicted the $X_{\rm{s}}^-$
ground state. For $w=20$ nm, neither symmetric-well nor
lowest-subband approximation works well (e.g., the latter
exaggerates charge separation in $X/X^-$ which mostly affects the
$X_{\rm{s}}^-$ and predicts its breakup at $B \geq 22$ T). Our best
estimates are $\Delta_s=1.5$ meV and $\Delta_t=1.2$ meV. They are
rather sensitive to the parameters, making prediction of the $X^-$
ground state in real samples difficult and somewhat pointless.
However, we expect that the $X_{\rm{t}}^-$s, additionally favored by
the Zeeman energy, could at least coexist with the $X_{\rm{s}}^-$s
at finite temperatures.

Consider a trion (either $X_{\rm{s}}^-$ or $X_{\rm{t}}^-$, whichever
state occurs at given $w$, $n$, and $B$) immersed in an IQL state.
Effective $e-X^-$ pseudopotentials are similar
\cite{WojsQuinnHawrylakPRB00} to the $e-e$ one \cite{HaldaneQHE}. In
the lowest LL, this causes similar $e-e$ and $e-X^-$ correlations,
described in a generalized two-component
\cite{WojsSzYiQuinnPRB99,WojsHawrylakQuinnPRB99} CF picture
\cite{JainPRL89}. At Laughlin-Jain fillings $\nu_{\rm{IQL}}=s /
(2ps+1)$, electrons converted to CF$_e$s fill the lowest $s$ LLs in
an effective magnetic field $B^{\ast}=B -2pn(hc/e)=B/ (2ps+1)$. At
$\nu \neq \nu_{\rm{IQL}}$, QEs in the $(s+1)$st or QHs in the $s$th
CF LL occur, carrying effective charge $\tilde q = \pm e/( 2ps+1)$.
We find that, similarly, an $X^-$ which is Laughlin correlated with
surrounding electrons can be converted to a CF$_{X^-}$ with charge
$\mathcal Q=-\tilde q$.

This value can be obtained, e.g., by noting that when an $X^-$
recombines, it leaves behind an indistinguishable electron which
becomes a CF$_e$ that either fills a QH in the $s$th CF$_e$ LL or it
appears as an additional QE in the $(s+1)$st CF$_e$ LL. More
importantly, partial screening of the trion's charge is independent
of either the particular $X^-$ state or the filling factor, as long
as correlations are described by the CF model. The same value
$\mathcal Q=-\tilde q $ results for any other distinguishable charge
$-e$ immersed in an IQL, if it induces Laughlin correlations around
itself (e.g., an impurity \cite{HaldaneRezayiPRL85} or a
reversed-spin electron).

A trion coupled to an IQL and carrying reduced charge is a many body
excitation. To distinguish it from an isolated $2e+h$ state, we call
it a charged $QX$ and denote it by $\chi^-\equiv \chi^{-\tilde q}$.
Being negatively charged, an $\chi^-$ interacts with IQL QPs. At
$\nu < \nu_{\rm{IQL}}$, the $\chi^-$ binds to a QH to become a
neutral $\chi^-{\rm{QH}}=\chi$, with a binding energy called
$\Delta_0$. Depending on sample parameters and spin of the trion,
$\chi$ may bind an additional QH to form a positively charged
$\chi^-{\rm{QH}}_2=\chi^+$, with binding energy $\Delta^+$. At
$\nu>\nu_{\rm{IQL}}$, the $\chi^+$ attracts and annihilates a QE:
$\chi^++QE \rightarrow \chi$; this process releases energy
$\Delta_{\rm{IQL}}-\Delta^+$ (where
$\Delta_{\rm{IQL}}=\varepsilon_{\rm{QE}}+\varepsilon_{\rm{QH}}$ is
the IQL gap). The $\chi$ may annihilate another QE: $\chi+QE
\rightarrow \chi^-$, with energy gain
\begin{equation}
\Delta^- =\Delta_{\rm{ IQL}} - \Delta^0 \label{eq:gain_annihiltion}
\end{equation}
that can be interpreted as $\chi-$ binding energy. The $\chi$ and
$\chi ^{\pm}$ are different states in which a hole can exist in an
IQL. If $\Delta^{\pm}>0$, then depending on $\nu$, either $\chi^-$
or $\chi^+$ is the most strongly bound state. If $\Delta^- \neq
\Delta^+ $, the PL spectrum will be discontinuous at
$\nu_{\rm{IQL}}$. For long-lived $\chi^{\pm}$ (made of a dark
$X_{\rm{t}}^-$), recombination of the $\chi$ is also possible,
especially at $\nu\approx \nu_{\rm{IQL}}$ (within a Hall plateau),
when QP localization impedes $\chi^{\pm}$ formation. The $QX$s
resemble normal excitons in $n$- or $p$-type systems, except that
the concentration of their constituent QPs can be varied (in the
same sample) by a magnetic field. Also, their kinetics ($\chi
\leftrightarrow \chi^{\pm}$) are more complicated because of the
involved QE-QH annihilation.

\begin{figure}
\centerline{\includegraphics[width= \linewidth]
{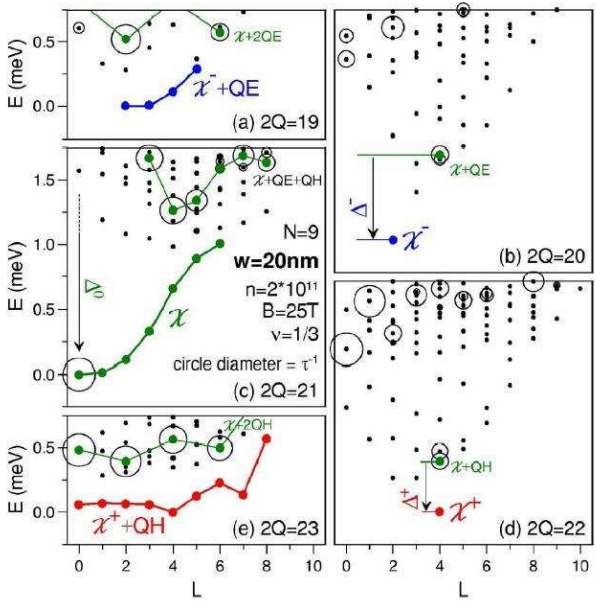}} \caption{(color online)
Excitation energy spectra (energy $E$ as a function of total angular
momentum $L$) of $9e+h$ systems on a sphere, with up to two QEs or
QHs in Laughlin $\nu=1/3$ IQL. Oscillator strengths $\tau^{-1}$ are
indicated by the area of the open circles
\cite{WoysGladyQuinnPRB06}.} \label{Fig:Excitation_spectrum_9e_1h_w}
\end{figure}

\begin{figure}
\centerline{\includegraphics[width= \linewidth]
{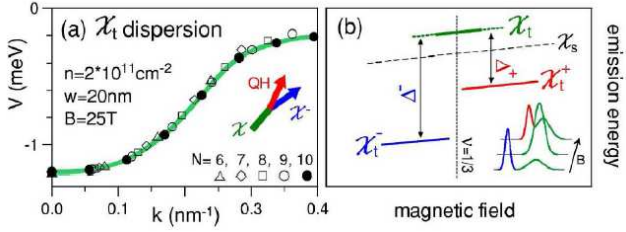}} \caption{(color online) Dispersion of
neutral quasiexciton $\chi_{\rm{t}}$ in Laughlin $\nu=1/3$ IQL;
$\chi_{\rm{t}}$ splits into $\chi_{\rm{t}}^-$ and QH at $k>0$. (b)
Schematic PL discontinuity due to $\chi_{\rm{t}}^{\pm}$ emissions
\cite{WoysGladyQuinnPRB06}.} \label{Fig:chi_dispersion}
\end{figure}

We have tested the $QX$ idea numerically for Laughlin  $\nu= 1/3$
IQL. First, we calculated spin-polarized $Ne+h$ energy spectra for
$w=20$ nm, in search of the $QX_{\rm{t}}$s. The $X_{\rm{t}}^-$ has
94$\%$ squared projection onto the lowest LL, so we ignored LL
mixing in the $Ne+$h calculation (direct tests confirmed that it is
negligible). The low-lying states in Fig.
\ref{Fig:Excitation_spectrum_9e_1h_w} are understood using the CF
picture \cite{JainPRL89, WojsSzYiQuinnPRB99, WojsHawrylakQuinnPRB99}
and addition rules for angular momentum. On a sphere, the CF
transformation introduces an effective monopole strength
$2Q^{\ast}=2Q-2(K-1)$, where $K =N-1$ is the total number of free
electrons and $X^-$s. The angular momenta of constituent QPs are
$\ell_{\rm{QH}}=Q^{\ast}$, $\ell_{\rm{QE}}=Q^{\ast} +1$, and
$\ell_{\chi^-}=Q^{\ast}-1$. The $\chi^{-}$ is a dark ground state in
(b) at $L=\ell_{\chi^-} =2$, and $\chi^+$ is found in (d) at
$L=\ell_{\chi^+}=|(2\ell_{\rm{QH}}-1)-\ell_{\chi^-}| =4$. Bands of
$\chi^--$QE and $\chi^+-$QH pairs are marked in (a) and (e). In (c)
the radiative $L=0$ ground state is a multiplicative state, opening
a $X=X^--$QH band \cite{ChenQuinnPRB93,WojsQuinnPRB01_1,
WojsQuinnPRB01_2}, earlier called a ``dressed exciton" and
identified \cite{ApalkovRashbaPRB92,ApalkovRashbaPRB93,
ZangBirmanPRB95} as responsible for the doublet structure in PL. The
continuous $\chi$ dispersion shown in Fig. \ref{Fig:chi_dispersion}
(a) results \cite{ApalkovRashbaPRB92,ApalkovRashbaPRB93,
ZangBirmanPRB95} from the in-plane dipole moment being proportional
to the wave vector $k=\ell /R$. It is suppressed (compared to $X$)
because of the reduced charge of the $\chi$ constituents,
$\chi_{\rm{t}}^-$ and QH. In the absence of an IQL the center of
mass of the two charges are separated and the cyclotron motion of
each charge together with their Coulomb attraction causes them to
move with a momentum proportional to their separation ($d \propto
k$). In an IQL, the charge quantum is reduced to $\tilde q$. This
has no consequence at $k=0$, and the $\chi$ is equivalent to an $X$
decoupled from the remaining electrons. A moving $\chi$ has a dipole
moment proportional to its wavevector but it is smaller because the
charges are $\pm \tilde q =\pm e/3$. The $\chi$ and $X$ dispersions
become similar in size, when $\chi$ acquires dipole moment in a
different way than $X$, by splitting into $\chi^-$ and QH, each
carrying only one small quantum $\pm \tilde q$. Indeed, the $\chi$
and $X$ dispersions become similar when energy and length scales are
rescaled in account of the $\tilde q \rightarrow e$ charge
reduction. Note that we also explain the emission from $\chi$ at
$k\lambda \sim 1.5$, proposed \cite{ApalkovRashbaPRB92,
ApalkovRashbaPRB93, ZangBirmanPRB95} for the lower peak in PL, as
the $\chi^- \rightarrow$ QE recombination assisted by QH scattering.
However, a small $dV/dk$ and a large $\tau^{-1}$ at $k \lambda \sim
1.5$ needed for this emission requires significant well widths, $w >
20$ nm.

By identifying the multiplicative states containing an $\chi$ with
$k=0$, one can estimate $\Delta^{\pm}$ and $\Delta^0$ as marked in
Figs. \ref{Fig:Excitation_spectrum_9e_1h_w} (b) and (d). More
accurate values were obtained by comparing the appropriate energies
identified in the spectra obtained at different values of $2Q$, in
which either $\chi^{\pm}$, $\chi$, or QP is alone in the IQL,
followed by extrapolation to $N\rightarrow \infty$. Our best
estimates, whose reasonable accuracy of under 0.05 meV is confirmed
by Eq. \ref{eq:gain_annihiltion}, are $\mathcal E _{\rm{QH}}=0.73$
meV, $\mathcal E_{\rm{QE}} =1.05$ meV, $\Delta^0=1.20$ meV,
$\Delta^-=0.52$ meV, and $\Delta^+ =0.27$ meV. Depending on $\chi^0
/\chi^{\pm}$ kinetics, either $\Delta^+ \neq \Delta^-$ or $\Delta ^0
\neq \Delta^{\pm}$ asymmetry will make PL energy jump at $\nu= 1/3$,
as sketched in Fig. \ref{Fig:chi_dispersion} (b). Similar behavior
has been observed \cite{PinczukPRL90,HawrilakNature}. The
$\chi^{\pm}$ discontinuity is different from that due to anyon
excitons \cite{RashbaPortnoiPRL93, PortnoiRashbaPRB96,
ParfittPortnoiPRB, ChenQuinnPRB94, WojsQuinnPRB01_1,
WojsQuinnPRB01_2} anticipated in much wider wells (e.g., for $w \geq
40$ nm at $n=2 \times 10^{11}$ cm$^{-2}$). The two effects can be
distinguished by different magnitude ($\sim \Delta_{\rm{IQL}}$ vs
$\Delta^{\pm}$) and opposite direction of the jump of emission
energy when passing through $\nu= 1/3$. In the present case, the
small ratio of $\chi^{\pm}$ and $X^{\pm}$ binding energies is the
signature of the fractional charge of the IQL excitations--directly
observable as splittings in PL. The $QX$s are defined through a
sequence of gedanken processes: (i) trion binding: $2e+h \rightarrow
X^-$, (ii) Laughlin correlation: $X^- \rightarrow \chi^-$, (iii) QH
capture: $\chi^- \rightarrow \chi/\chi^+$. Hence, $\chi$ and
$\chi^{\pm}$ are in fact the same $X^-$, only differently separated
from the surrounding electrons.

\begin{figure}
\centerline{\includegraphics[width= \linewidth]
{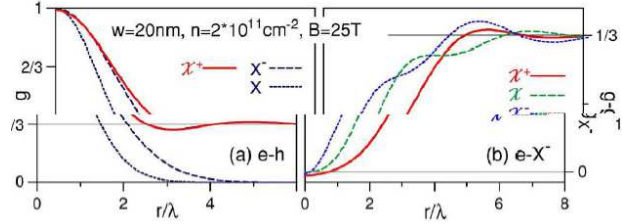}} \caption{(color online) (a) The
$e-h$ pair-distribution function (PDF) of quasiexciton
$\chi_{\rm{t}}^+$ and isolated $X_{\rm{t}}^-$ and $X$, normalized to
measure the local filling factor. (b) The $e-X^-$ PDF for different
$QX$s; curve for $\chi^+$ resembles $e-e$ PDF of Laughlin liquid;
shoulders for $\chi$ and $\chi^+$ reflect additional charge quanta
pushed onto the hole \cite{WoysGladyQuinnPRB06}.}
\label{Fig:e_h_pair_distribution}
\end{figure}

This is evident in the $e-h$ pair-distribution functions $g(r)$
shown in Fig. \ref{Fig:e_h_pair_distribution} (a) and normalized so
as to measure electron concentration near the hole in units of
$\nu$. The $\chi^+$ curve calculated for $N=10$ is compared with
$g_{X^-}(r)=\exp(-r^2/4)$ which accurately describes an
$X_{\rm{t}}^-$. The similarity at short range proves that the
$\chi^+$ is an $X^-$ well separated from the 2D electron gas. In
Fig. \ref{Fig:e_h_pair_distribution} (b) we plotted $\delta
g=g-g_{X^-}$ which measures the $e-X^-$ correlations in different
$QX$ states. Clearly, $\delta g_{\chi^+}$ resembles the $e-e$
pair-distribution function of a Laughlin  $\nu= 1/3$ liquid, while
shoulders in $\delta g_{\chi}$ and $\delta g_{\chi^-}$ reflect
additional charge quanta pushed onto the hole in $\chi$ and
$\chi^-$. Let us add that integration of $[g(r)- 1/3]$ directly
confirms fractional electron charge of $-(4/3)e$, $-e$, and
$-(2/3)e$ bound to the hole in the $\chi^-$, $\chi$, and $\chi^+$
states.

\begin{figure}
\centerline{\includegraphics[width= \linewidth]
{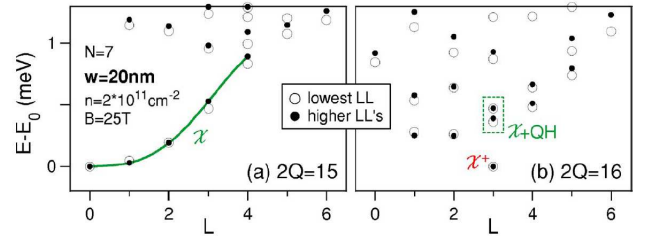}} \caption{(color online)
Excitation spectra similar to Fig.
\ref{Fig:Excitation_spectrum_9e_1h_w}, but for $7e+h$ system with
and without LL mixing \cite{WoysGladyQuinnPRB06}.}
\label{Fig:Excit_sp_7e_1h_w_wo_LL_mixing}
\end{figure}

The accuracy of the lowest LL approximation is demonstrated in Fig.
\ref{Fig:Excit_sp_7e_1h_w_wo_LL_mixing}, in which we compare the
excitation energy spectra similar to Figs.
\ref{Fig:Excitation_spectrum_9e_1h_w} (a) and (d), but calculated
for the $7e+h$ systems, with and without inclusion of one higher $e$
and $h$ LL. Evidently, neither the $\chi$ dispersion nor the
$\chi^+$ binding energy appear sensitive to the LL mixing. This is
in contrast to the behavior of $X$ or $X^-$, and the difference
obviously reflects weaker interactions among the fractional $QX$
constituents (compared to the same cyclotron energy scale).

The quasiexcitons formed by the singlet $X_s^-$  have been studied
numerically for a quantum well of width $w=10$ nm by considering an
$8e+h$ system with spin $S=3$ (i.e., one spin flipped). In contrast
to the results for the quasiparticles formed by the triplet $X_t^-$,
the $\chi_s^{\pm}$ charged singlet quasiexcitons are excited states.
The $X_s^-$ charge distribution is more compact than that of
$X_t^-$, leading to stronger dispersion of neutral $\chi_s^0$ and a
different coupling of the $X_s^-$ to the Laughlin quasiparticles.
The neutral $\chi_s^0$ is the most strongly bound state regardless
of the presence of Laughlin QEs and QHs. This may result in a
continuous PL peak for $\chi_s^0$, but precludes PL discontinuity in
narrow wells with a strong $X_{\rm{s}}^-$ ground state. The
$\chi_{\rm{s}}$ peak splits into a $\sigma_{\pm}$ doublet due to
spin $\downarrow$ and $\uparrow$ recombination involving either QEs
or ``reversed-spin" QERs \cite{RezayiPRB87}, but
temperature-activated emission at $k>0$ is not expected. The $QX$
idea can be extended to other IQLs (e.g., $\nu= 2/3$ or $2/5$).
However, different behavior of $QX_{\rm{t}}$s and $QX_{\rm{s}}$s at
$\nu=1/3$ is an example that PL discontinuity is not guaranteed. Via
Eq. \ref{eq:gain_annihiltion}, it is governed by sample and
$\nu$-dependent $\Delta_{\rm{IQL}}$ and $\Delta^0$ which must be
recalculated.


\section{Summary and Conclusions}


The fractional quantum Hall effect is a paradigm for all strongly
interacting systems, containing, at high magnetic field $B$, only a
single energy scale, the Coulomb scale $e^2/\lambda$, where
$\lambda$ is the magnetic length. Understanding all of the observed
IQL states may well give insight into a number of strongly
interacting system of great current interest.

In this paper we have reviewed exact numerical diagonalization of
small systems within the Hilbert subspace of a single partially
occupied LL. The numerical results are thought as ``numerical
experiment", and simple intuitive models fitting the numerical data
are sought, to better understand the underlying correlations. We
describe calculations for $N$ electrons confined to a Haldane
spherical surface, and present simple results at different values of
the LL degeneracy $g=2\ell+1$. We demonstrate that Jain's remarkable
CF picture predicts not only the values of $2\ell$ at which IQL
ground states occur for different values of $N$, but also predicts
the angular momenta $L$ of the lowest band of multiplets for any
value of $2\ell$ in a very simple way. We emphasize that Jain's CF
picture is valid, not because of some magical cancelations of
Coulomb and Chern-Simons gauge interactions beyond mean-field, but
because it introduces Laughlin correlations by avoiding pair states
with the lowest allowed relative angular momentum $\mathcal
R=2\ell-L^{\prime}$. The allowed angular momentum multiplets which
avoid pair states with $\mathcal R=1$ form a subset of the set of
multiplets $G_{N\ell}(L)$ that can be formed from $N$ Fermions in a
shell of angular momentum $\ell$. This subset avoids the largest
repulsion and has the lowest energy. Our adiabatic addition of
Chern-Simons flux introduces Laughlin correlations without the
necessity of introducing an irrelevant mean field energy scale
$\hbar \omega_c^{\ast}=\nu \hbar \omega_c$.

Jain's sequence of filled CF shells does not require an interaction
between CF quasiparticles. The incompressibility results from the
energy required to create a QE-QH pair in the integrally filled CF
state. Haldane's hierarchy of IQL states was based on the implicit
assumption that the residual interaction between QPs was
sufficiently similar to the Coulomb interaction between electrons in
LL0 that the QPs would form their own Laughlin correlated daughter
states.

The experiment of Pan et al. showed that neither Jain's CF picture
nor Haldane's hierarchy was the whole story. Residual pair
interactions between QPs had been determined
\cite{SitkoYiYiQuinnPRL96} up to an overall constant (unimportant
for QP correlations). This pseudopotential $V_{\rm{QP}}(L')$ could
be used to determine the spectrum of daughter states containing
$N_{\rm{QP}}$ quasiparticles in a partially filled QP shell.
Qualitatively correct results can be expected when $V_{\rm{QP}}(L')$
is small compared to the energy necessary to create a QE-QH pair in
the IQL state. When the CF picture was reapplied to the QPs, the
Haldane hierarchy of all odd denominator fractions resulted
\cite{SitkoYiQuinnPRB97}. Numerical calculations demonstrates that
this CF hierarchy scheme of Laughlin correlated QPs at each level
didn't always work, probably because $V_{\rm{QP}}(L')$ was not
sufficiently similar to $V_0(L')$, the pseudopotential for electrons
in LL0.

The energy of a multiplet $|\ell^N;L\alpha>$ formed from $N$
electrons in a shell of angular momentum $\ell$ is given by Eq.
\ref{eq:energyElalpha}. W\'{o}js and Quinn proved a simple theorem,
Eq. \ref{eq:L_identity} that led to the conclusion that a
pseudopotential of the form $V_H(\hat L')=A+B\hat L ^{'2}$ (referred
to as a `` harmonic" pseudopotential) failed to lift the degeneracy
of the multiplets $\alpha$ that had the same total angular momentum
$L$. Correlations (removal of this degeneracy) were caused only by
the anharmonic part of $V(L')$, i.e. by $\Delta
V(L')=V(L')-V_H(L')$. For $\Delta V(L')=k\delta (\mathcal R ,1)$,
($\mathcal R=2\ell-L'$ is referred to as the relative pair angular
momentum), where $k>0$, the lowest energy state for each value of
$L$ is the multiplet for which $P_{L\alpha}(\mathcal R=1)$is a
minimum. Here $P_{L\alpha}(\mathcal R=1)$ is the probability that
$|\ell^N;L\alpha>$ has pairs with pair angular momentum
$L'=2\ell-1$. This is exactly the condition for Laughlin
correlations at $\nu=(2\mathcal R+1)^{-1}=1/3$. If the anharmonic
part of $V(L')$ is negative (i.e. $k<0$), then the lowest energy for
each angular momentum $L$ occurs for the multiplet with
$P_{L\alpha}(\mathcal R=1)$ equal to a maximum, indicating a
tendency to form pairs with $\mathcal R=1$.

Because the pseudopotentials for electrons in LL0 and LL1 are
well-known, and for QEs and QHs of the Laughlin $\nu=1/3$ (and other
IQL states) can be evaluated, we can attempt to interpret the
numerical diagonalization results in terms of simple intuitive
pictures of the correlations expected from $V_0(\mathcal R)$,
$V_1(\mathcal R)$, and $V_{\rm{QP}}(\mathcal R)$.

For LL0 Laughlin correlations among the electrons are expected and
found. For LL1, pairing correlations are found for $1/2\geq \nu_1 >
1/3$, and Laughlin correlation are found $1/3 > \nu_1 \geq 1/5$
($\nu_1=\nu-2$). The strongest IQL states are found at
$\nu_1=1/2,3/3$, and their $e-h$ conjugate states. Laughlin
correlations with four Chern-Simons fluxes (CF$^4$) are expected for
$1/3 \geq \nu_1 >1/5$. The $N_{\rm{P}}=N/2$ pairs are Laughlin
correlated and give an IQL state at $2\ell=2N+1$ and its conjugate
at $2\ell=2N-3$. The elementary excitations can also be interpreted
in terms of a generalized CF picture described by Eqs.
\ref{eq:FP_FP_e_corr} and \ref{eq:e_FP_e_corr}. The $\nu_1=1/3$
state is found to $2\ell=3N-7$, not at $2\ell=3N-3$ of the Laughlin
state in LL0. We do not completely understand correlations at
$\nu_1=1/3$, but they could arise from triplets or from forming
pairs of pairs.

We investigate the possibility of a spin phase transition in the
$\nu=4/11$ IQL state observed by Pan et al. The two spin states are
daughter states of the Laughlin $\nu=1/3$ IQL state, each of which
has a QE filling factor $\nu_{\rm{QE}}=1/3$. For the fully spin
polarized state, the QEs partially fill CFLL1 and have the same spin
as the filled CFLL0$\uparrow$. For the partially spin polarized
state the quasiparticles are QERs, and they partially fill
CFLL0$\downarrow$.

By numerical diagonalization of $N$ electron systems with different
values of the total electronic spin, we determine the QP energies
$\varepsilon_{\rm{QP}}$ and their interactions $V_{\rm{QP}}(\mathcal
R)$ (for QP=QE and QER) as a function of the width $w$ of the
quantum well. The total energy is the sum of the QP energies, their
interaction energy, and the Zeeman energy. Wide wells weaken
electron-electron interactions and favor partially spin polarized
states. Large Zeeman energy favors fully spin polarized states. We
sketch a phase diagram in the well-width vs. Zeeman energy plane and
show a rough estimate of the phase boundary between the two states.

Finally, a system containing electrons and valence band holes is
studied. Neutral excitons $X=(eh)$, and charged excitonic complexes
$X^-=e(eh)$, $X_2^-=e(eh)^2$, etc. are found and their angular
momenta, binding energies and interactions with one another are
evaluated. In dilute systems with $\nu \ll 1/3$, the singlet $X_s^-$
and triplet $X_t^-$ electron spin states are the ground states at
low and high magnetic field respectively. The $X_s^-$ ground state
and an excited triplet state are shown to be the only strongly
radiative states. The latter state is called the bright triplet
$X_{tb}^-$, while the triplet ground state is called the dark
triplet $X_{td}^-$.

For systems with filling factor $\nu$ close to an IQL value (e.g
$\nu\simeq 1/3$), the $X^-$ becomes Laughlin correlated with the
electrons and has effective charge $-e/3$, the same as that of
Laughlin QEs. This $QX^{-}$ can bind one or two Laughlin QHs to form
a $QX^0$ or a $QX^-$. The spectra of these systems and their PL
intensities can be evaluated numerically, and they agree quite well
with the predictions of the CF picture.

The unified thread connecting the work included in this manuscript
is the generalized CF picture. By knowing the behavior of the
appropriate electron or QP pseudopotential, one can make an educated
guess at the nature of the ground state correlations. Laughlin
correlations in LL0 are the simplest type. Pairing or formation of
larger clusters when $V(\mathcal R)$ is subharmonic is more
complicated. However, the generalized CF picture (built on the ideas
of Laughlin, Haldane, Halperin and Jain) can be applied to pairs of
electrons in LL1 or to pairs of QHs in CFLL0 and pairs of QEs in
CFLL1. This simple model seems to give qualitatively correct results
not just for when an IQL ground state occurs, but often for the
spectrum of low energy excitations. We don't totally understand the
correlations in every case (e.g. at $\nu=7/3$ and at
$\nu_{\rm{QE}}=1/3$ for spin polarized systems). However, we are
certain that the full hierarchy of FQH states involves other types
of correlations in addition to Laughlin. The nature of the
correlations at each level of the hierarchy will depend on the
appropriate pseudopotential, as will the path through the hierarchy
levels that results.


\section*{Acknowledgments}

The authors acknowledge the important contribution of Jennifer J.
Quinn, Piotr Sitko, X. M. Chen, Daniel Wodzinski, Isabela
Szlufarska, Anna G{\l}adysiewicz and Pawel Hawrylak to parts of the
work presented in this paper. JQ and GS acknowledge the partial
support from Basic Science Program of DOE and KS from
KRF-2006-005-J02804.



\end{document}